%% file: ms_2017_astroph.tex
\documentclass[trackchanges]{aastex61}
\usepackage{subfigure,rotating,moresize}
\usepackage{wrapfig}
\pdfoutput=1

\newcommand\aastex{AAS\TeX}

\newcommand{\barm}{\mbox{$\bar{m}$}}
\newcommand{\barM}{\mbox{$\bar{M}$}}

\newcommand{\samename}{\vrule height0.4pt depth0.0pt width1.0in \thinspace.}

\received{2017 October 20}
\revised{2018 February 20}
\accepted{2018 February 21}
\submitjournal{ApJ}

\shorttitle{\aastex\ sample article}
\shortauthors{Gonz\'alez-L\'opezlira}

\begin{document}

\title{Tracers of Stellar Mass-loss - II. Mid-IR Colors and Surface Brightness Fluctuations}

\correspondingauthor{Rosa A.\ Gonz\'alez-L\'opezlira}
\email{r.gonzalez@irya.unam.mx}

\author[0000-0003-1557-4931]{Rosa A.\ Gonz\'alez-L\'opezlira}
\affiliation{Instituto de Radioastronomia y Astrofisica, UNAM, Campus Morelia,
     Michoacan, Mexico, C.P.\ 58089}

\begin{abstract}
I present integrated colors and surface brightness fluctuation magnitudes in the mid-IR,
derived from stellar population synthesis models that
include the effects of the dusty envelopes around thermally pulsing asymptotic
giant branch (TP-AGB) stars.
The models are based on the Bruzual \& Charlot CB$^*$ isochrones; they are single-burst, range in age from a few
Myr to 14 Gyr, and comprise metallicities between $Z =$ 0.0001 and $Z =$ 0.04.
I compare these models to mid-IR data of AGB stars and star clusters in the 
Magellanic Clouds, and study the effects of varying self-consistently the mass-loss rate,
the stellar parameters, and the output spectra of the stars plus their dusty envelopes.
I find that models with a higher than fiducial mass-loss rate are needed to fit the mid-IR colors of
``extreme" single AGB stars in the Large Magellanic Cloud. Surface brightness fluctuation 
magnitudes are quite sensitive to metallicity for 4.5 $\mu$m and longer wavelengths at all stellar 
population ages, and powerful diagnostics of mass-loss rate in the TP-AGB for intermediater-age populations, 
between 100 Myr and 2-3 Gyr.  
\end{abstract}


\section{Introduction.} \label{sec:intro}

Asymptotic giant branch (AGB) stars are central to the chemical evolution of galaxies, and understanding the 
contribution of these evolved stars to the spectral energy distribution (SED) of galaxies is essential 
for the interpretation of galactic emission in the near and mid-infrared (IR). 
Thermally pulsing AGB (TP-AGB) evolution is very complex, however, on account of a large number of physical processes 
at work, and the difficulties in constraining them (see, for a brief recent
summary, Marigo et al.\ 2013). 
A common tool to this end has been the
comparison of TP-AGB lifetimes and luminosity functions with observations. 
While several processes and parameters ---dredge-up efficiency, mixing-length, hot-bottom burning, pulsations---
are degenerate on their effects on both TP-AGB lifetimes and luminosity functions,
there is no doubt that mass-loss is the
most important parameter determining the duration of the phase \citep[e.g.,][]{rosen14,rosen16}.
Even at the beginning of the TP-AGB phase, especially for low-mass stars, 
 mass-loss can produce an early envelope ejection \citep{lg10} and hence a premature termination of the phase, 
resulting in a drastic reduction of the number of TP-AGB stars \citep[see, for example,][]{raim09,rosen14}.
In turn, the luminosity functions, integrated light, broadband colors, and surface brightness fluctuation (SBF) 
amplitudes of stellar populations will be impacted, as already stated by, e.g., 
Maraston (1998), Lan\c con \& Mouhcine (2002), Cantiello et al.\ (2003), Maraston (2005), 
Raimondo et al.\ (2005), Lee et al.\ (2010).
%\citet{mara98,lm02,cant03,mara05,raim05,lee10}.

In a previous work \citep{gonz10}, 
we compared star and stellar cluster data of the Large and Small Magellanic Clouds
(LMC and SMC, respectively)
with model colors and SBFs in the optical and near-IR.
The conclusion of that research was that broadband colors and SBFs at those wavelengths
cannot discern global variations in mass-loss rate, but that different mass-loss rates should
leave detectable imprints on mid-IR models and data. This prediction is the subject of this paper. 

Here, I study the impact of mass-loss in 
stars undergoing the TP-AGB phase,  
in particular during the superwind phase, on
the mid-IR integrated colors and SBFs of stellar populations.   

\section{Stellar Population Synthesis Models.} \label{sec:2013mod}

I explore the contribution of intermediate- and low-mass stars in the TP-AGB phases of their evolution
to the mid-IR light of simple stellar populations (SSPs), in particular through their mass-loss.
The treatment of this stellar phase in stellar population synthesis models determines the predicted SED 
of stellar populations in this wavelength range at ages from about 1 to 2 Gyr.
For this purpose I use an updated version of the Bruzual \& Charlot (2003; BC03 hereafter) models, dubbed CB$^*$ by these authors.
The CB$^*$ models are based on the stellar evolution models computed by \citet{bertl08}.
Evolutionary tracks are available for metallicities $Z =$ 0.0001, 0.0005, 0.001, 0.002, 0.004, 0.008, 0.017 ($Z_\odot$), and 0.04. {Table~\ref{cbstar} lists the spectral libraries used for different stellar types in the CB$^*$ models.
Models computed with the \citet{Chabrier2003} IMF have been used throughout this paper.
The CB$^*$ models have been employed by, e.g., \citet{labarbera12} and \citet{bruz13}.

\input{table1.tex}

In these models the evolution of the TP-AGB stars
follows the results of \citet{Marigo2013}, who used the COLIBRI code incorporating as much detailed
physics as possible into the calculation of this evolutionary phase.
This is a big improvement over previous treatments of the TP-AGB, which just followed a semi-empirical
prescription to describe the lifetimes, luminosities, and effective temperatures of these stars (e.g., BC03).
To check the calibration of the \citet{Marigo2013} results with a fiducial mass-loss rate,
computed from the difference in the stellar mass along the evolutionary track,
Bruzual et al.\ (2013) modeled the distribution of TP-AGB stars in the color-magnitude diagram (CMD) in various optical and near-IR bands for a stellar population with $Z$ = 0.008, close to the LMC metallicity, by means of Monte Carlo simulations (see Bruzual 2002; 2010).
At each time step the mass formed in stars was derived from the LMC star formation history \citep[SFH;][]{hz09}.
The stars were distributed in the CMD according to the isochrones computed with the CB$^*$ models.
Figure 2 of \citet{bruz13} shows a comparison between the theoretical luminosity function (LF) derived from their simulations, and the observed Spitzer Space Telescope Surveying the Agents of a Galaxy's Evolution (SAGE) AGB data set \citep{srin09} in the Infrared Array Camera (IRAC) [4.5] $\mu$m band; their Figure 3 compares the model and observed $[3.6 -4.5]\ \mu$m color distributions for the same data set. (For simplicity, from now on
the symbol $\mu$m will be omitted from filter names in the text and figure captions.) 
Using the same procedure and the SFH of the SMC from \citet{hz04}, \citet{bruz13} modeled the TP-AGB stellar
population in the SMC galaxy (see their Fig.\ 4). In the case of the SMC, the chemical evolution 
indicated by \citet{hz04} was included in the simulations.
Inspection of these results shows that the LFs computed with the CB$^*$ models are in closer
agreement with the observations than those obtained with previous models. Furthermore, they are consistent with the findings by \citet{mk10}, Melbourne et al.\ (2012), and \citet{sz12}, all of which support the treatment of TP-AGB stars in the CB$^*$ models.

\subsection{Mass-loss, Stellar Parameters, and Dusty Envelopes.} \label{dotMnstarpars}

Following the roadmap laid out in \citet{gonz10}, 
I use the CB$^*$ isochrones to calculate models with TP-AGB stars in the superwind phase whose  
dusty envelopes have been produced by a mass-loss rate $\dot M$ one order of magnitude above and
below fiducial; twice and half fiducial; and five times and one-fifth fiducial. The superwind phase
has three different stages in the CB$^*$ isochrones.  
Figure~\ref{mlrs} shows fiducial mass-loss rate, $\dot M$, versus initial (zero-age main-sequence) stellar mass, $M_{\rm i}$, 
both at the onset or first stage (left), and at the peak or third stage (right) of the superwind phase. Different 
metallicities are indicated by line type and color ($Z=$ 0.0005, black dotted; $Z=$ 0.004, blue short-dashed; $Z=$ 0.008,
cyan solid; $Z=$ 0.017, red long-dashed; $Z=$ 0.04, magenta dotted-short-dashed).

\begin{figure*}
\begin{tabular}{lll}
\hspace*{0.3cm}\includegraphics[scale=0.30]{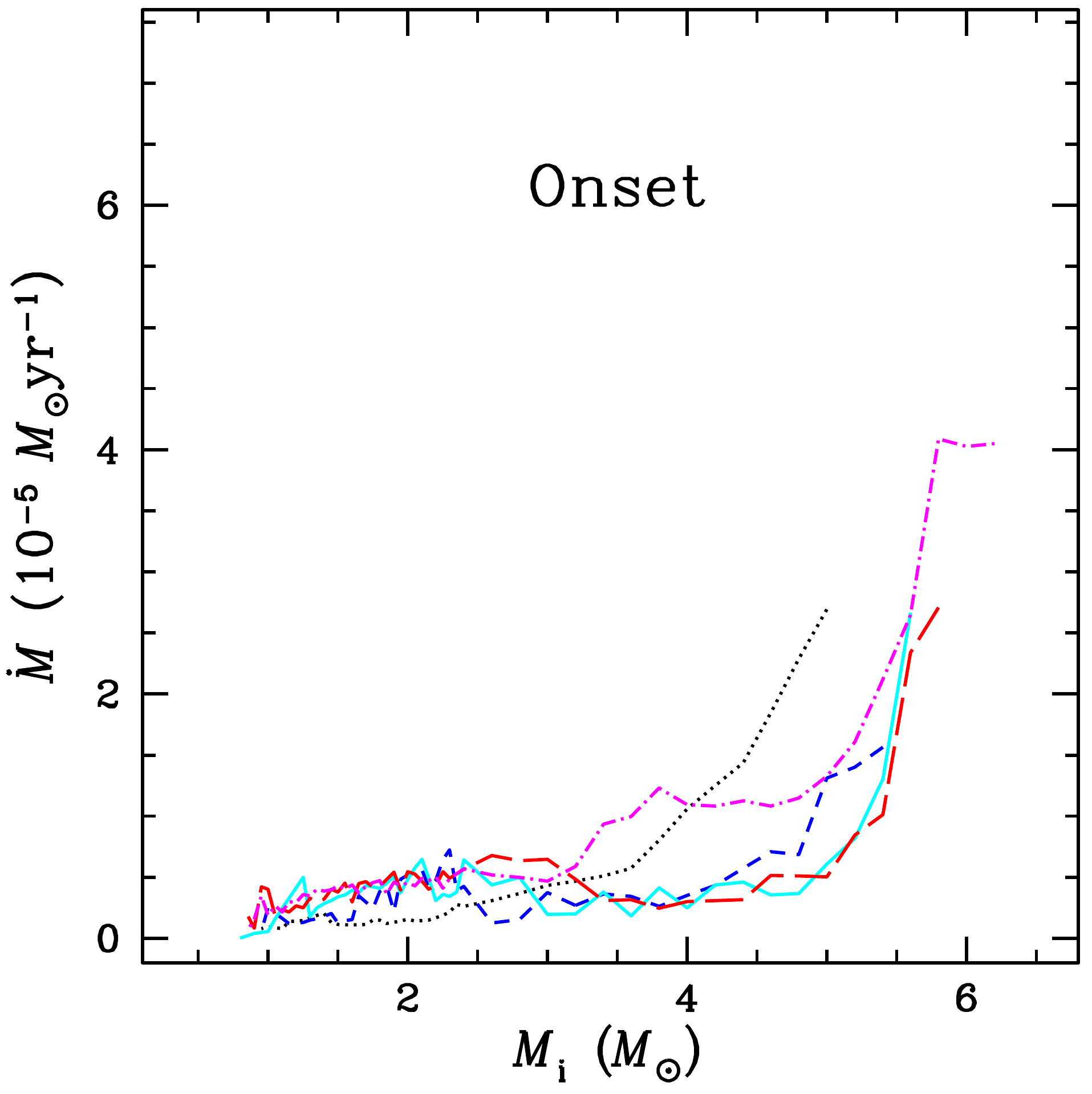}
&
\includegraphics[scale=0.30]{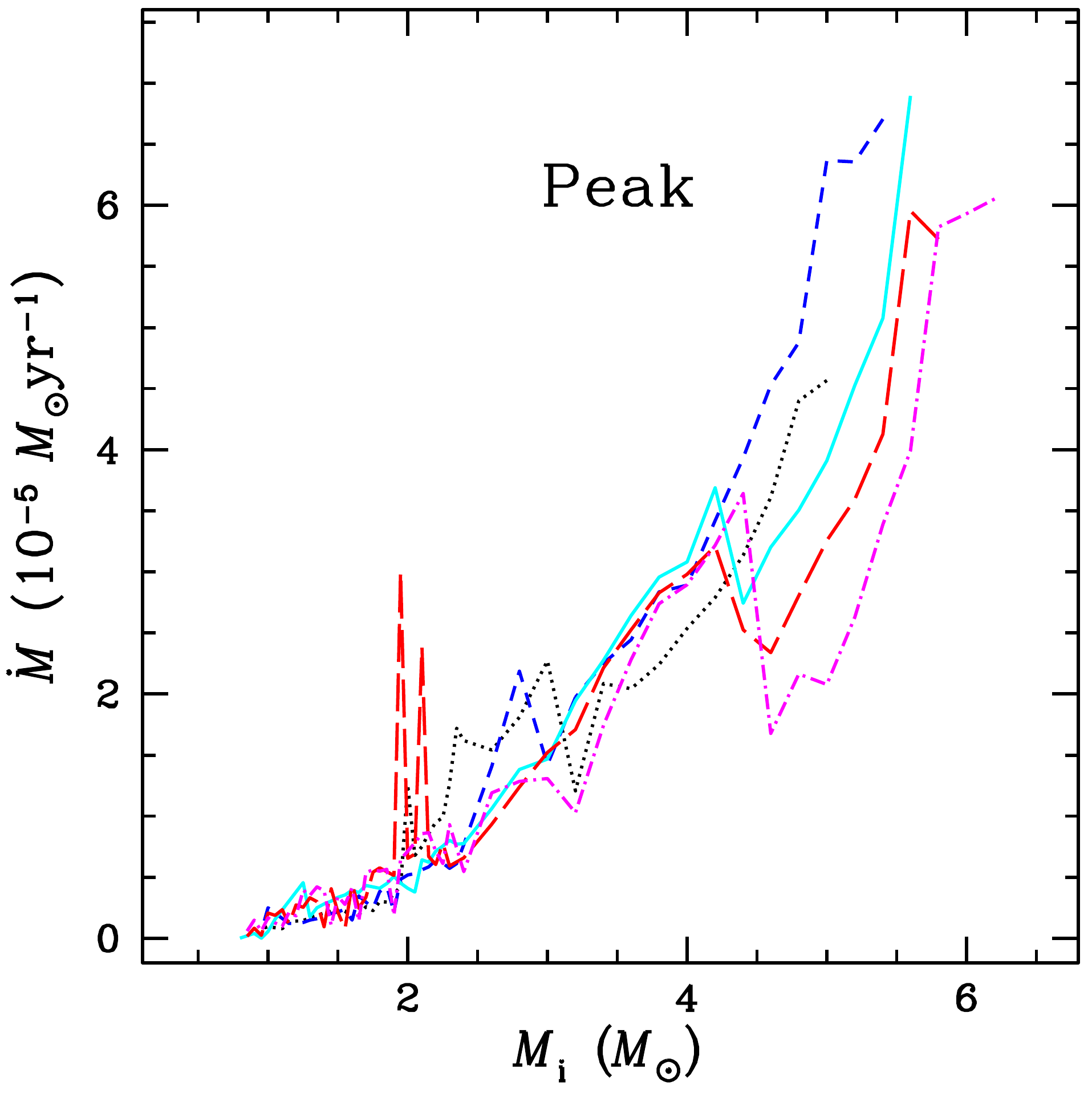}
&
\includegraphics[scale=0.25]{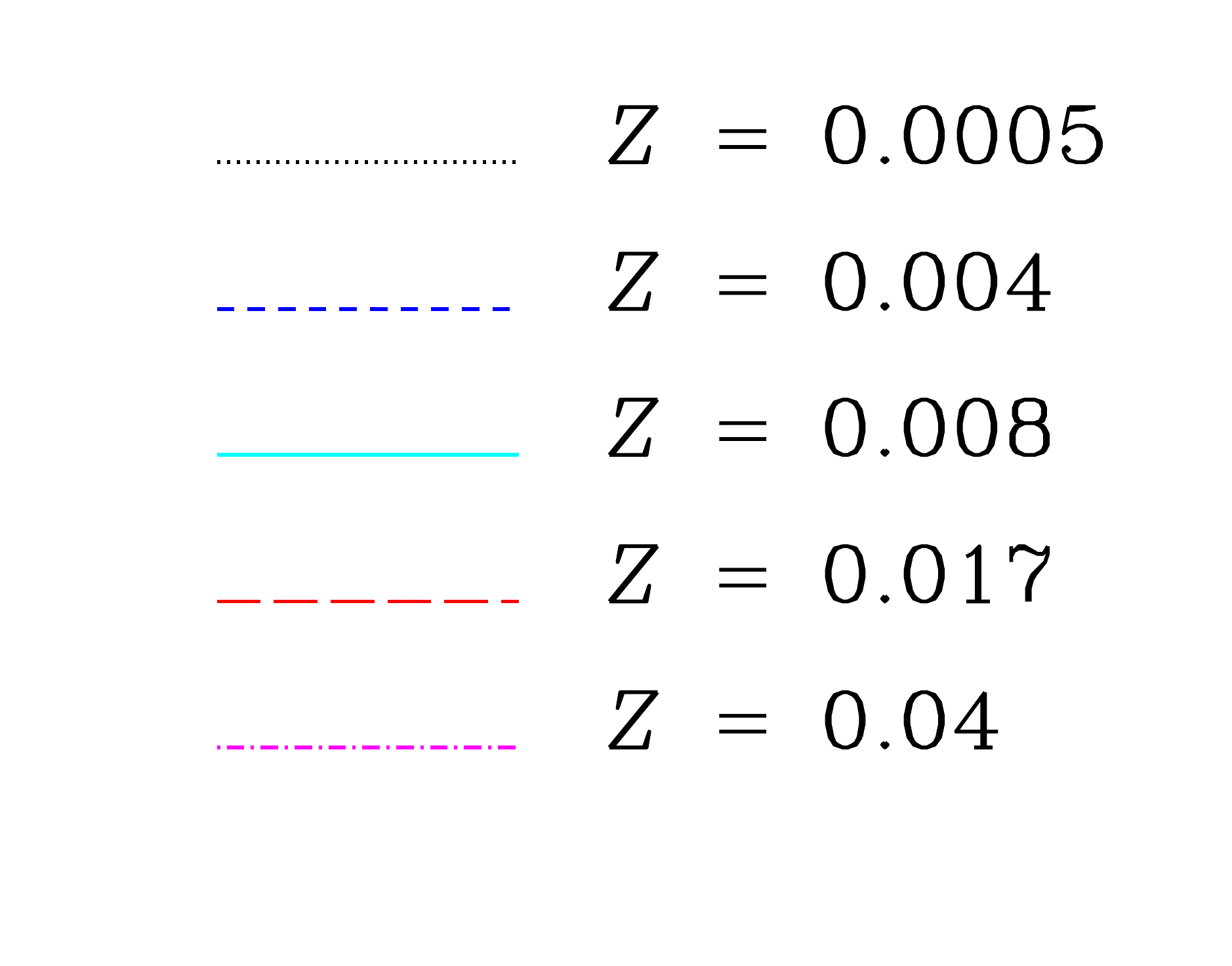}
\end{tabular}
\caption{Superwind phase fiducial mass-loss rate, $\dot M$, versus zero-age main-sequence stellar mass, $M_{\rm i}$.
Left: onset stage; right: peak stage. Black dotted line: $Z=$ 0.0005; blue short-dashed line:
$Z=$ 0.004; cyan solid line: $Z=$ 0.008; red long-dashed line: $Z=$ 0.017; magenta dotted-short-dashed line:
$Z=$ 0.04. 
}
\label{mlrs}
\end{figure*}

For this work, I adopt the view \citep[e.g.,][]{will00} that empirical relations between mass-loss and
stellar parameters are the result of very strong selection effects,
since stars with a low rate will not be detected as mass-losing,
whereas stars with a high rate will be obscured by dust and/or extremely short-lived.
In other words, regardless of the actual rate,
mass-loss will appear to follow a
Reimers' type relation \citep{reim75,reim77}, $\dot M = \eta LR/M$, where $M$ and $L$ are,
respectively, the stellar mass and luminosity, $R(L,M,Z)$ is
the stellar radius, and $\eta$ is a fitting parameter. Moreover, mass-loss
is not a smooth process: as $\dot M$ increases,
$L$ first grows, while $M$ stays constant, until the stellar configuration reaches 
a `cliff' in the log $M$ versus log $L$ plane; subsequently, mass-loss depends
on stellar parameters much more steeply than implied by empirical relations,
and the stellar envelope is shed in an extremely short time at roughly constant
$L$.  

Consequently, rather than, for example, varying $\eta$ while leaving
the stellar parameters unchanged, I will vary together the mass-loss rate and the
stellar parameters, in a consistent fashion. The whole procedure has been 
described in detail in Appendix A1 of \citet{gonz10}.

Briefly, $\dot M$ (computed from the 
mass differences along the track for a given star) is treated as the 
independent parameter. When $\dot M$ is changed with respect to its fiducial value,
a modified stellar luminosity $L$ and a modified stellar radius 
$R$ are obtained, respectively, from Fig.\ 2 in \citet{bowe91} and 
eq.\ 4 in \citet{iben84}. A new effective temperature is derived
using $L = R^2(T_{\rm eff}/5770)^4$.
The lifetimes $t$ within each of the superwind stages are then adjusted 
according to the fuel-consumption theorem \citep{renz86}, i.e., 
assuming that the product $Lt$ is constant (and equal to 
the value for fiducial $\dot M$) for each star and stage.
If the product $\dot Mt$ exceeds the stellar mass at the beginning of a 
stage, the star will not reach the following stage. Fundamental mode
pulsation periods, which change with stellar configuration, 
and C/O ratios of C-rich stars, which vary with $\dot M$, are modified
according to, respectively, eq.\ (12) and eq.\ (23) in \citet{mari07}.

The effects of dust on the stellar SEDs are included as explained in 
Appendix A2 of \citet{gonz10}, following a procedure outlined by \citet{piov03} and 
\citet{mari08}. Summing up, the optical depth at wavelength $\lambda$, $\tau_\lambda$, is proportional to
$\dot M \Psi \kappa_\lambda v_{\rm exp}^{-1}L^{-0.5}$, where $\Psi$ is the dust-to-gas ratio,
$\kappa_\lambda$ is the mass extinction coefficient, and 
$v_{\rm exp}$ is the wind expansion velocity. However, both $v_{\rm exp}$ 
and $\kappa_\lambda$ are functions of $\dot M$ (the latter through dust composition), 
and thus $\tau_\lambda$ must be found through an iterative process.
If $\Psi$ and hence $\tau$ diverge (Gonz\'alez-L\'opezlira et al.\ 2010, 
eqs.\ (A17) and (A3)), the star in question will be invisible during the
corresponding stage.

Finally, to produce the output SEDs of the TP-AGB stars, 
the radiative transfer in their 
dusty envelopes is calculated with the software DUSTY,
as reported in Appendix A3 of \citet{gonz10}. Dust mixtures for C-rich and O-rich
stars with varying envelope optical depths are adopted from \citet{suh99,suh00,suh02}. 

CB$^*$ models with fiducial mass-loss are reported
in Tables 2 and 3: mid-IR
colors as a function of age, for different metallicities,
are presented in Table 2;
fluctuation amplitudes are listed in Table 3.

\input table2.tex

\input table3.tex

\section{Broadband Colors.} \label{sec:brbndcols}

\subsection{Individual AGB Stars.} \label{subsec:ind_agb}

I show theoretical color-color diagrams
of individual TP-AGB stars along the 0.2 Gyr (dotted line), 0.5 Gyr (solid
line), and 9.5 Gyr (dashed line) isochrones, for populations with four metallicities
($Z =$ 0.004, black; 0.008, cyan; 0.017, red; and 0.04, magenta)
and five choices of spectra; these vary only due to the mass-loss rate adopted 
for stars in the TP-AGB. Figure~\ref{piov_mir} 
displays, from left to right, 
fiducial $\dot M/10$, fiducial $\dot M/2$, fiducial $\dot M$, 
fiducial $\dot M \times 5$, fiducial $\dot M \times 10$.
The top and bottom rows 
present, respectively, [5.8  - 8] versus [3.6  - 4.5] 
and [8 - 24] versus [3.6 - 8]. 

\begin{sidewaysfigure}[ht]
\begin{tabular}{lllll}
\hspace*{-1.2cm}\includegraphics[scale=0.20]{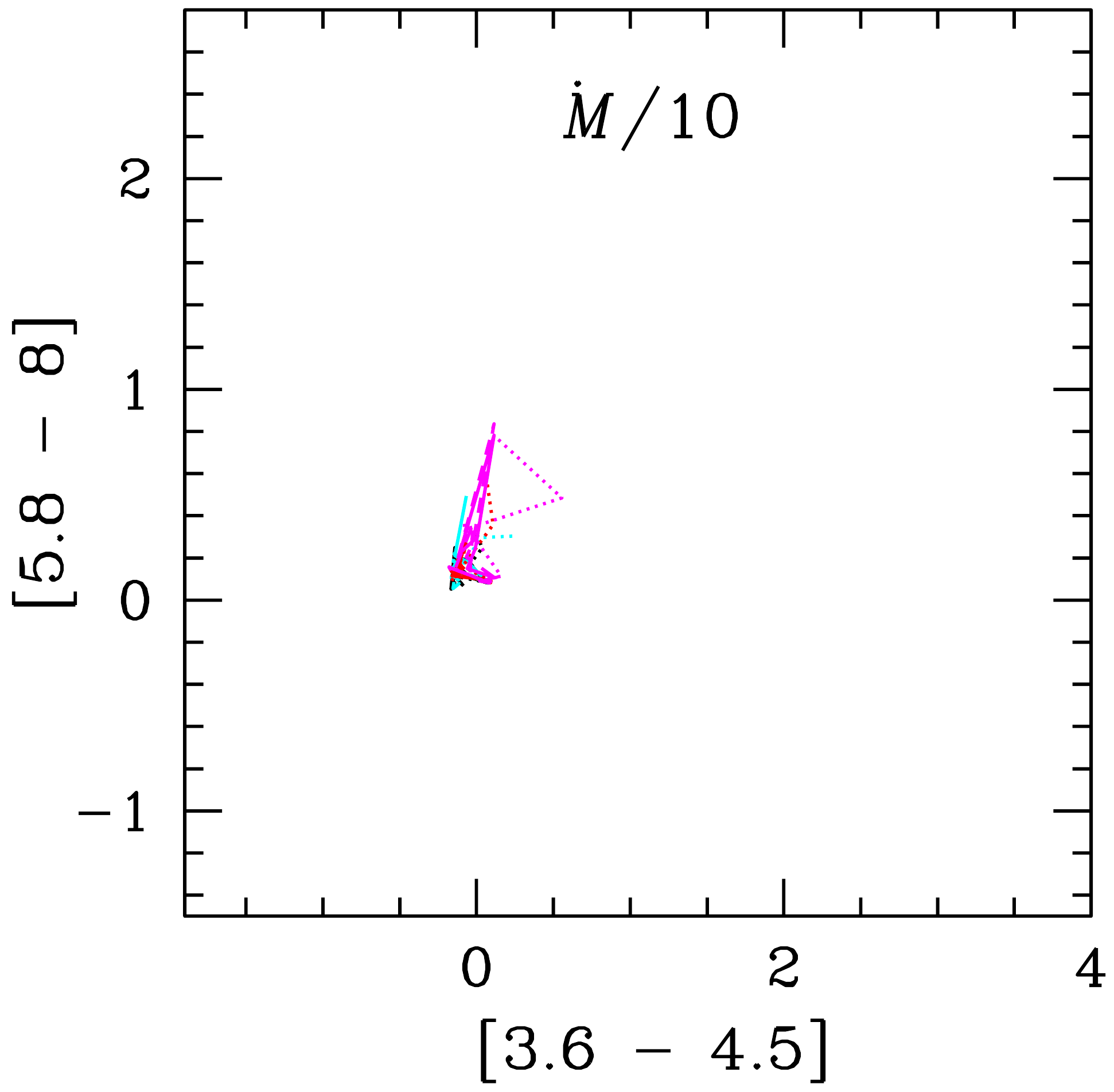}
&
\includegraphics[scale=0.20]{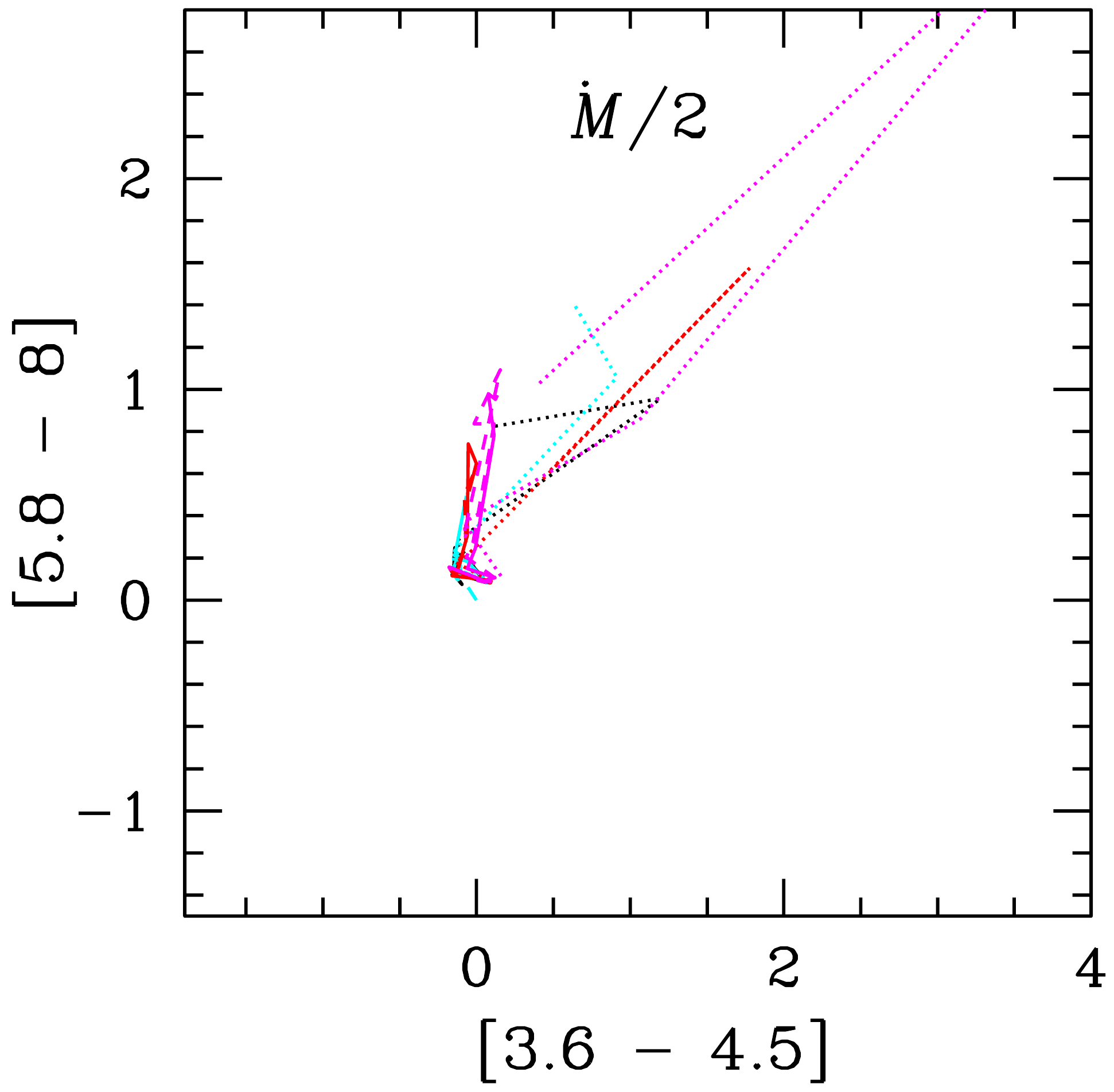}
&
\includegraphics[scale=0.20]{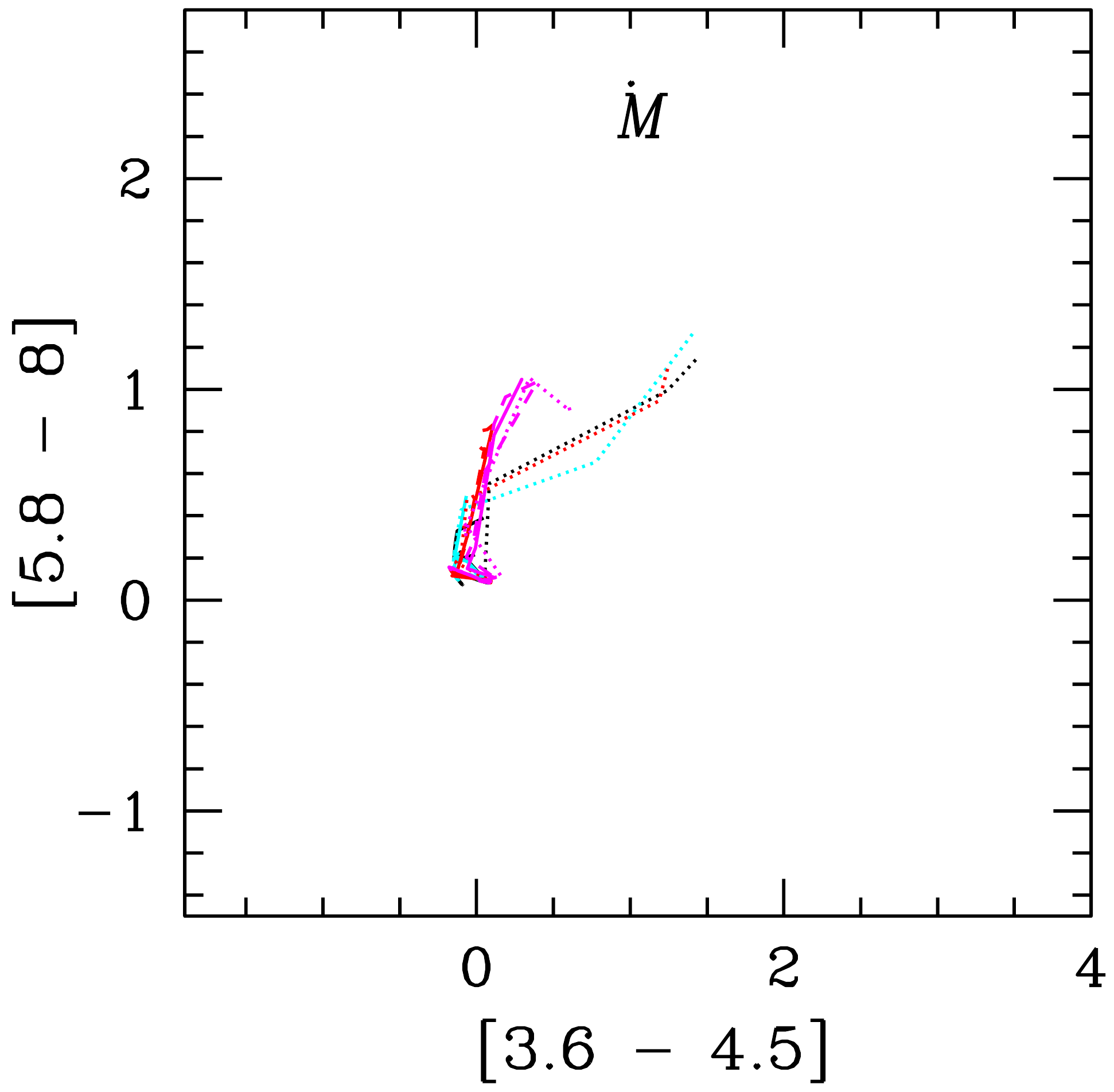}
&
\includegraphics[scale=0.20]{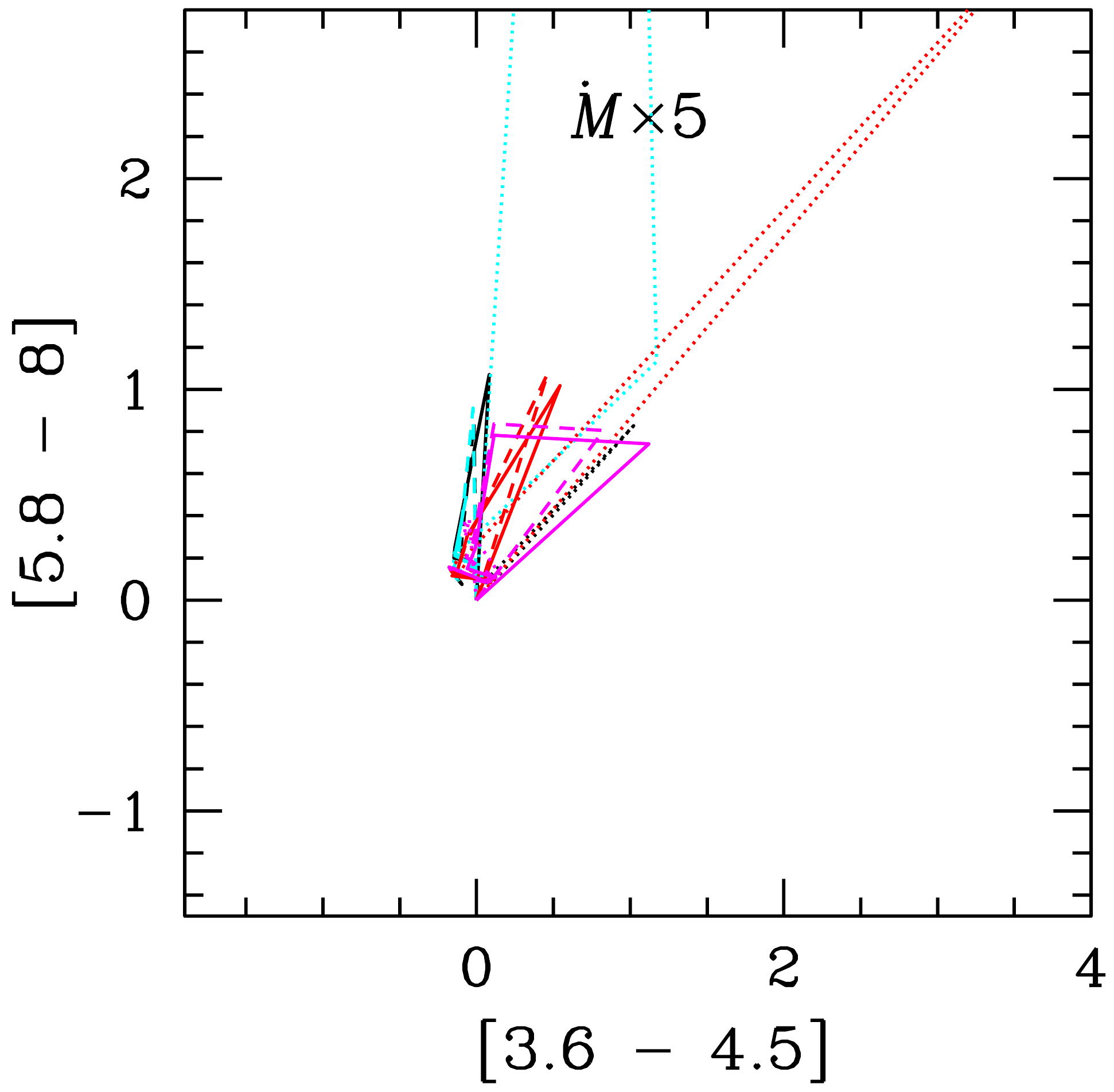}
&
\includegraphics[scale=0.20]{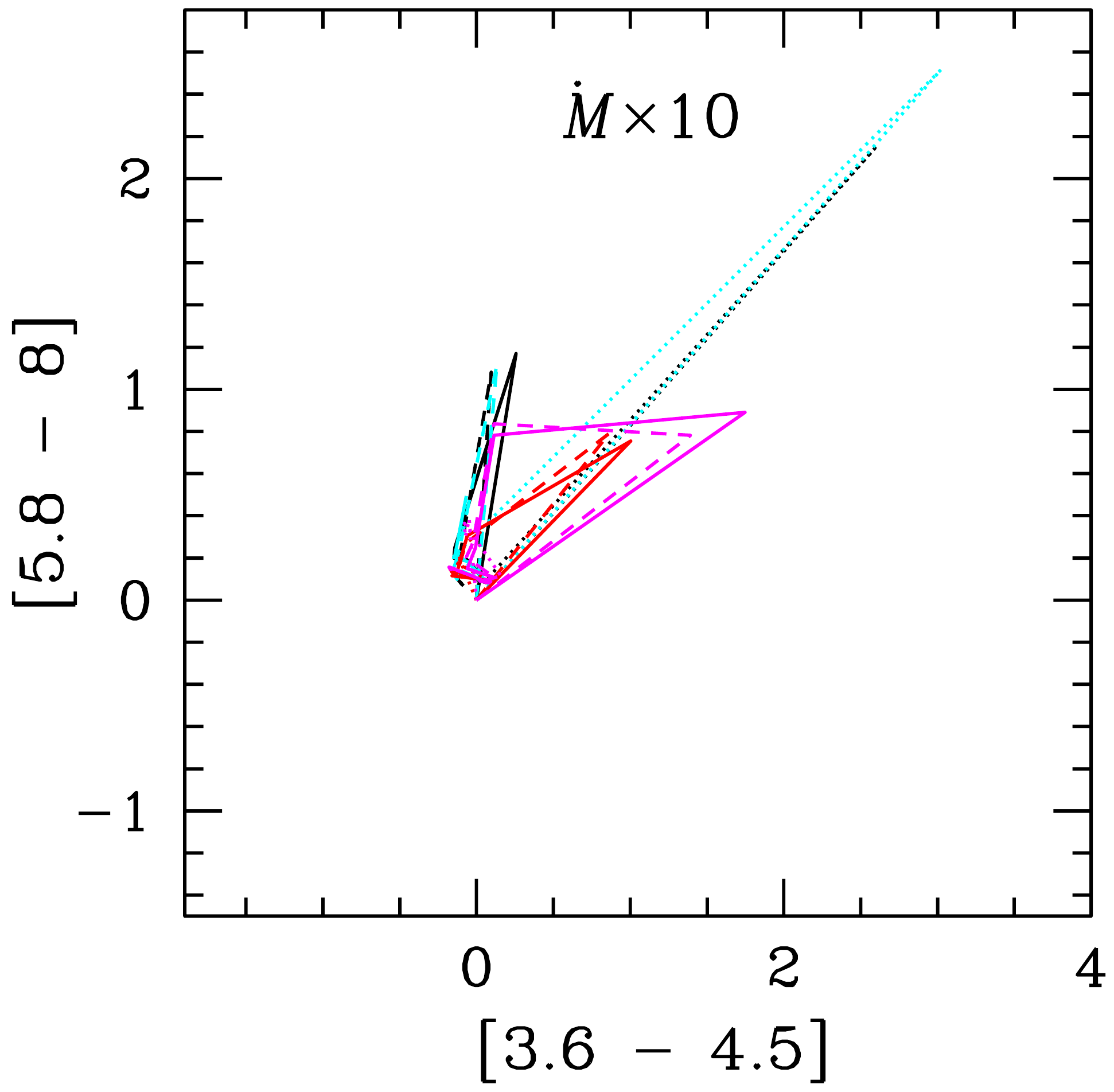}
\end{tabular}

\vspace*{0.3cm}

\begin{tabular}{llllll}
\hspace*{-1.2cm}\includegraphics[scale=0.20]{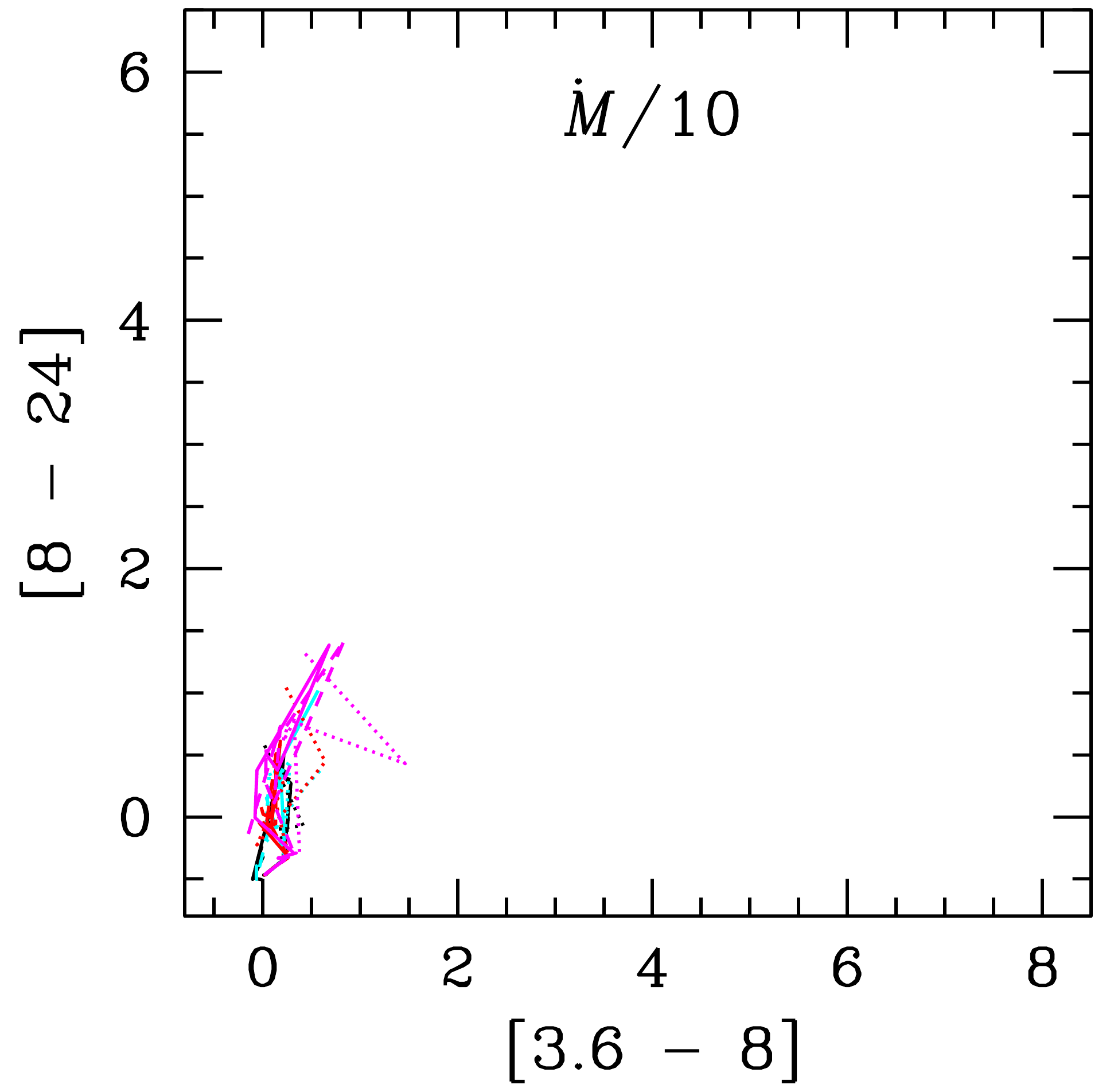}
&
\includegraphics[scale=0.20]{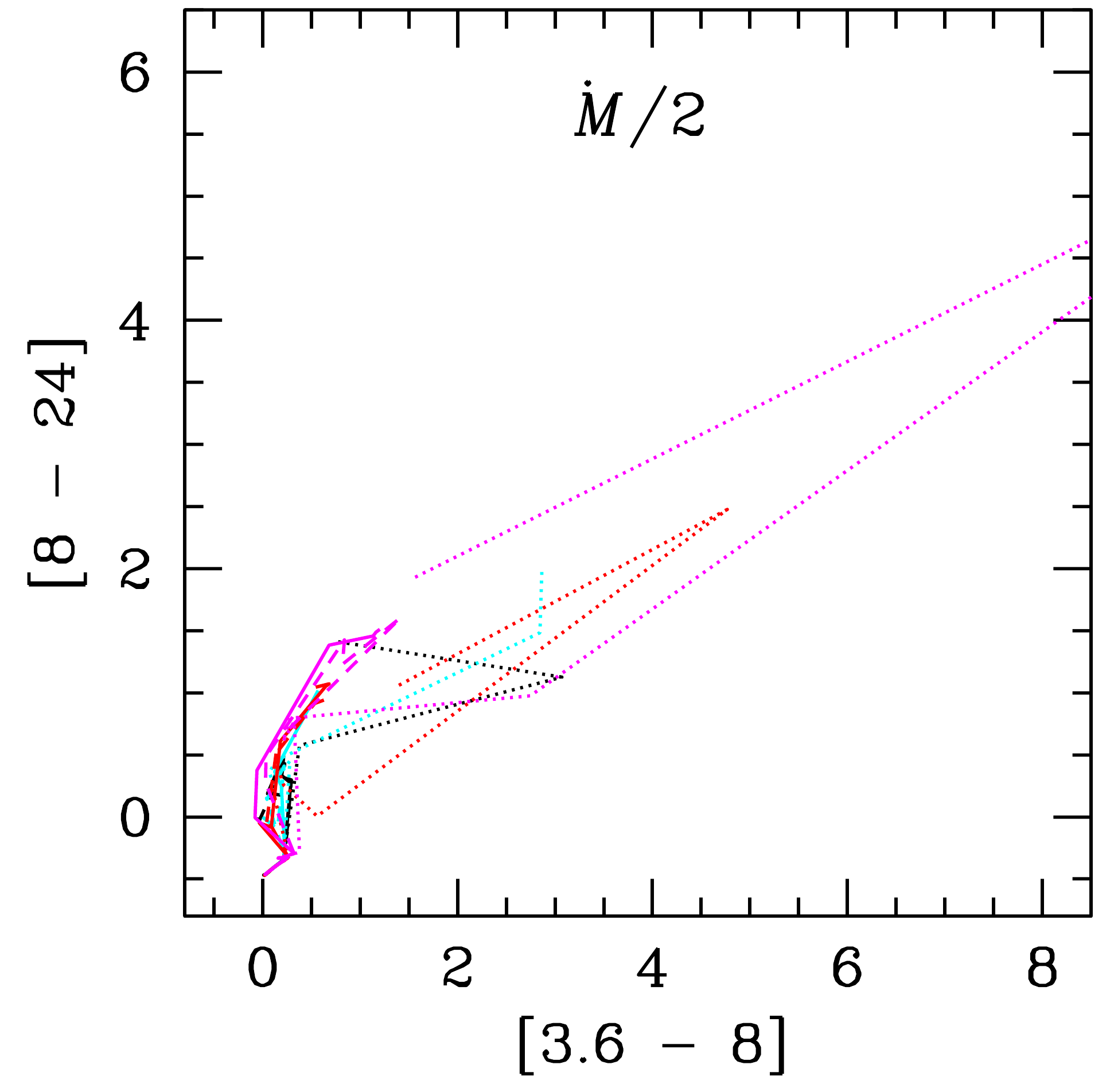}
&
\includegraphics[scale=0.20]{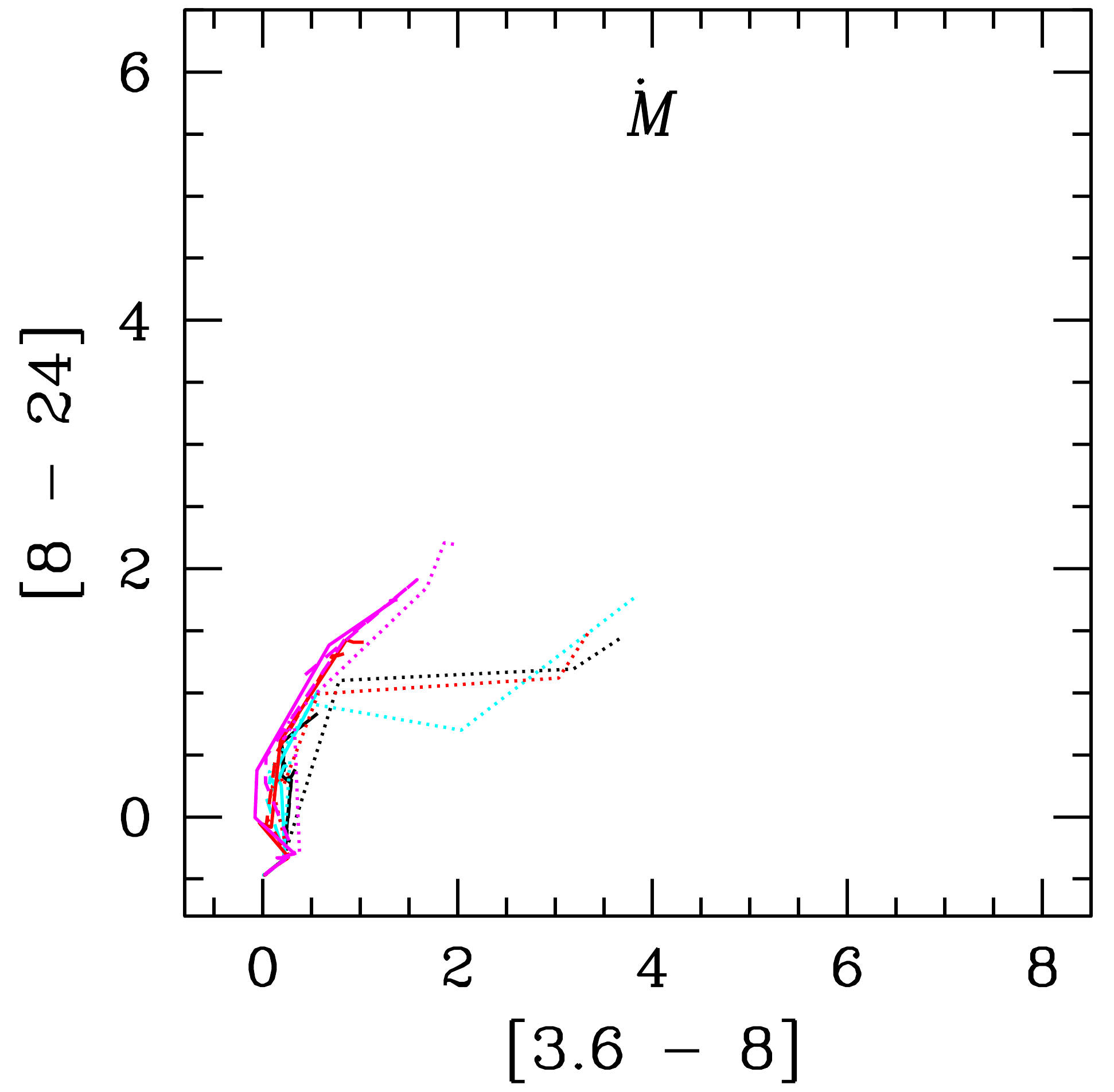}
&
\includegraphics[scale=0.20]{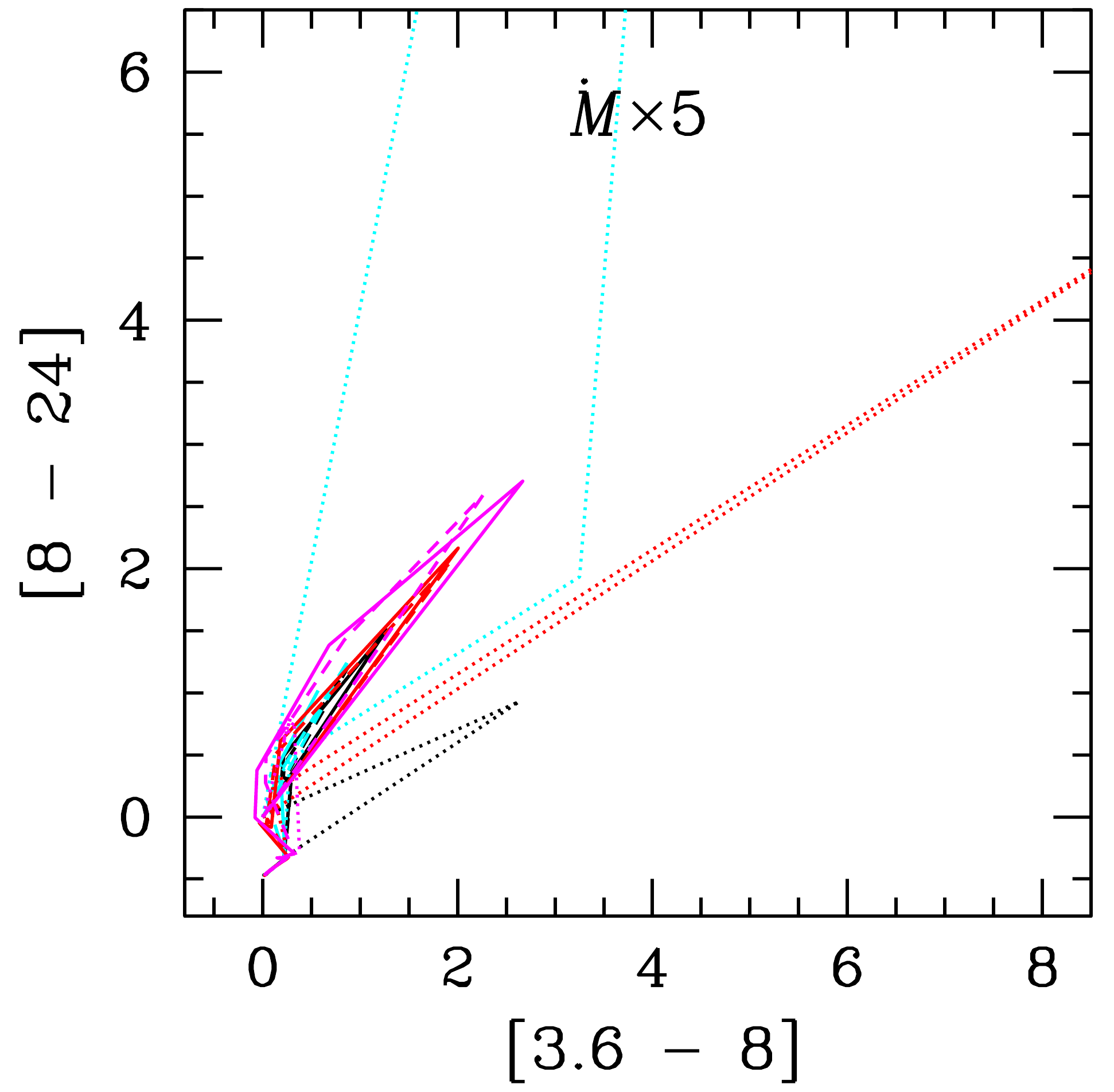}
&
\includegraphics[scale=0.20]{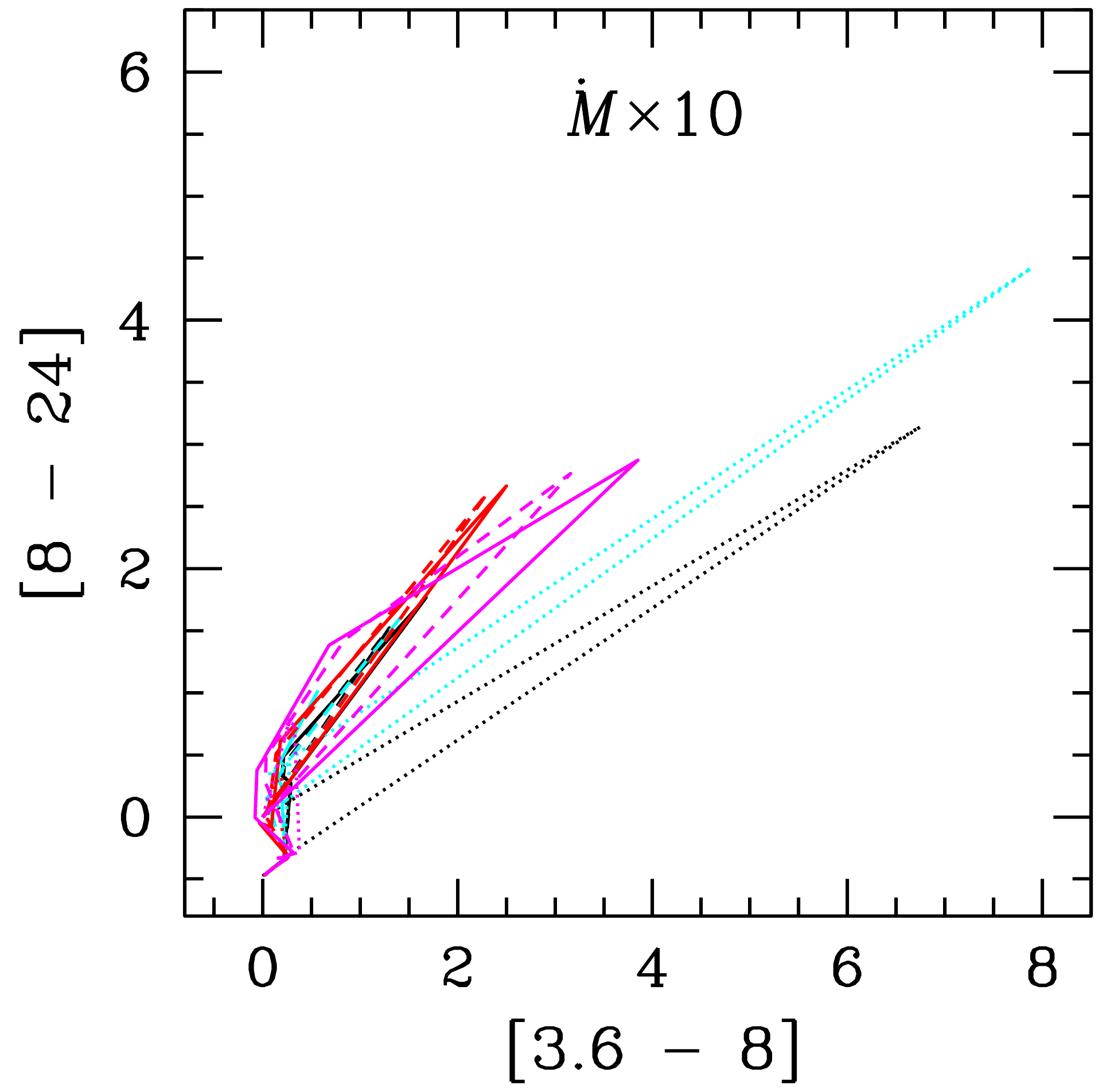}
\includegraphics[scale=0.20]{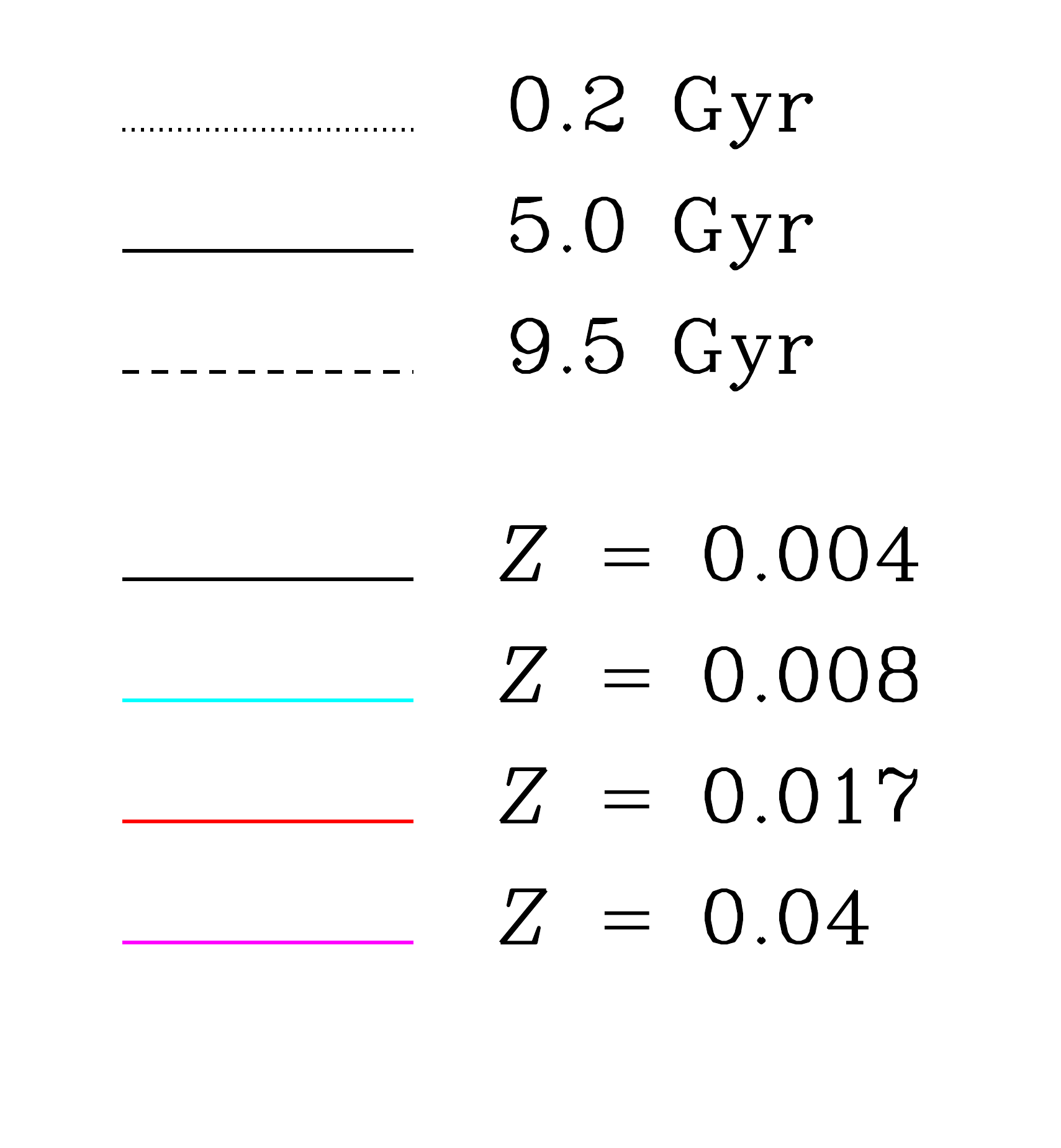}
\end{tabular}
\caption{
Theoretical two-color diagrams of
individual TP-AGB stars along the 0.2 Gyr (dotted line), 5.0 Gyr (solid
line), and 9.5 Gyr (dashed line) isochrones. 
Top row: [5.8  - 8] versus [3.6 - 4.5]; bottom row:
[8 - 24] versus [3.6 - 8]. From left to right: 
fiducial $\dot M/10$, fiducial $\dot M/2$, fiducial $\dot M$, 
fiducial $\dot M \times 5$, fiducial $\dot M \times 10$.
Different colors indicate diverse metallicities, i.e.,
black: $Z$ = 0.004; cyan: $Z$ = 0.008; red: $Z$ =
0.017; magenta: $Z$ = 0.04.}
\label{piov_mir}
\end{sidewaysfigure}

As a first test, I compare the model mid-IR broadband colors to the observed
[5.8 - 8] versus [3.6 - 4.5] and [8 - 24] versus [3.6 - 8]
color-color diagrams of individual AGB stars in the 
sample published by \citet{srin09}. In Figure~\ref{sundar_mir},
different colors are 
used for O-rich, C-rich, and ``extreme"
(based on their near- and mid-IR colors) AGB objects.
Next, in Figure~\ref{meixner_overlay}, 
models with both fiducial (thick lines) and 10 $\times$ fiducial 
(thin lines) mass-loss rates are superimposed on the stars, shown as a cloud of gray points. 
The ages and metallicities of the models are indicated as in 
Figure~\ref{piov_mir}.
The reddest extreme AGB stars are not fitted by models with 
fiducial $\dot M$.

\begin{figure}
\begin{tabular}{ll}
\hspace*{0.8cm}\includegraphics[scale=0.40]{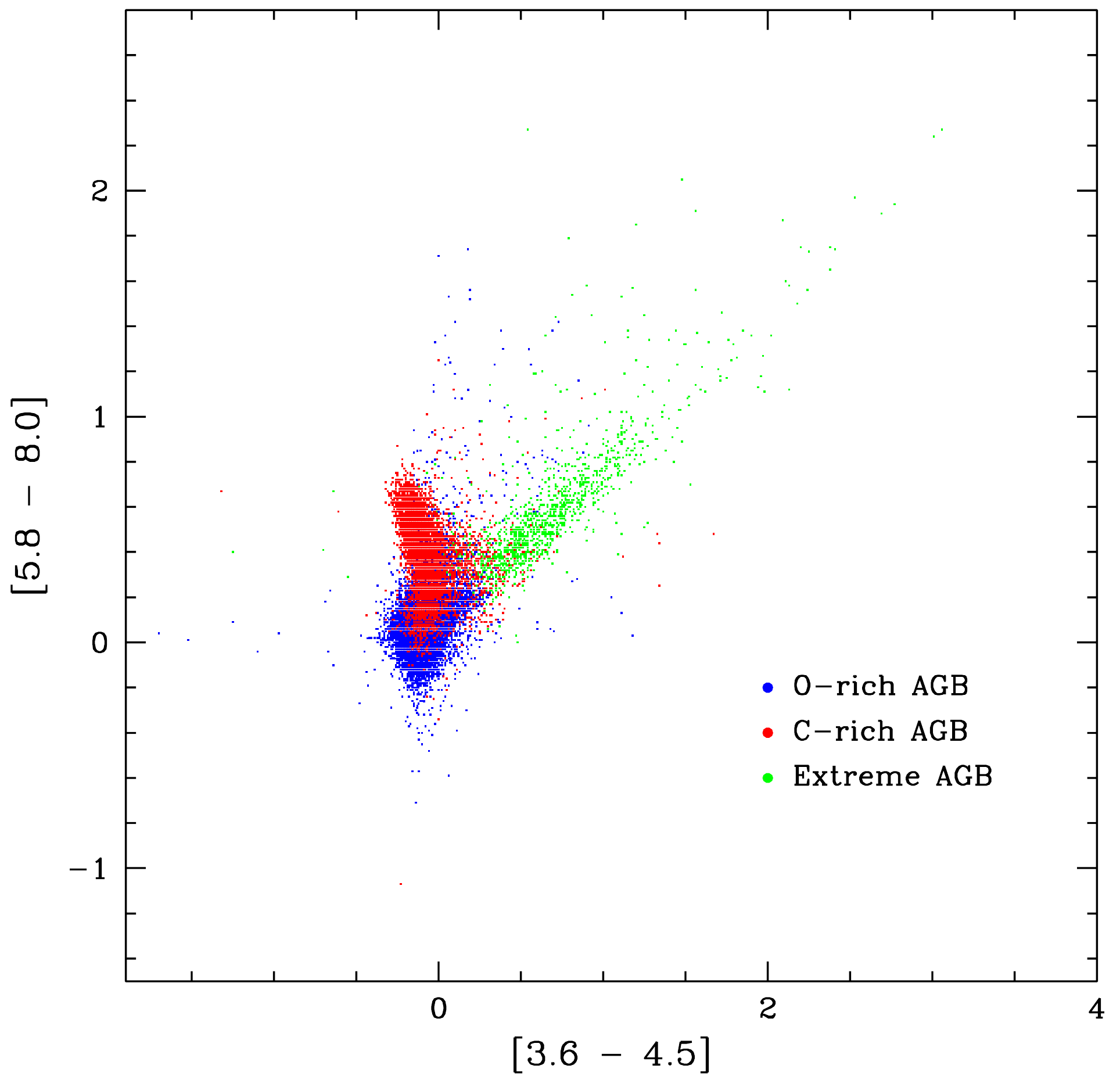}
&
\includegraphics[scale=0.40]{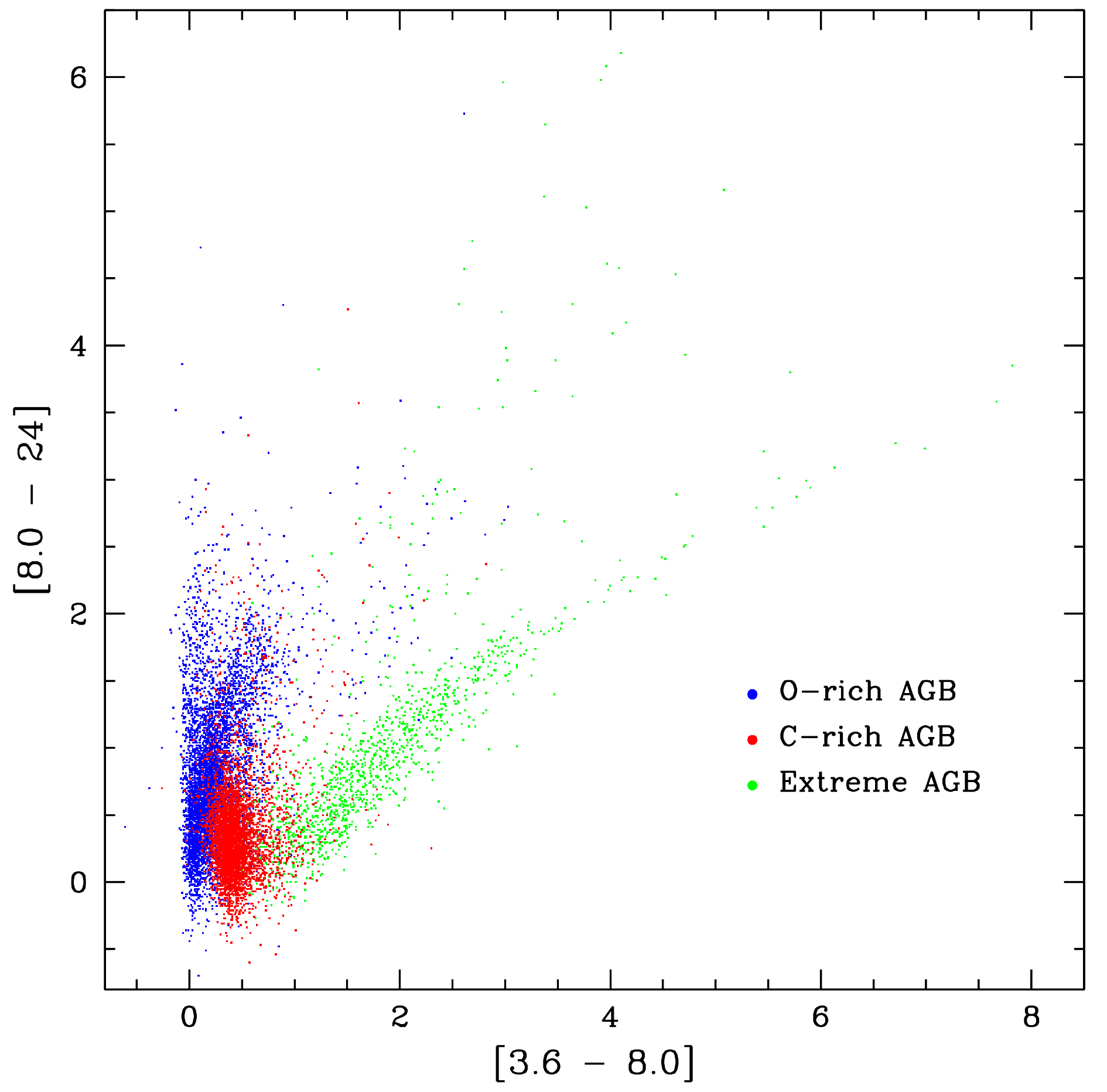}
\end{tabular}
\caption{
Integrated mid-IR colors of individual AGB candidates in the
sample of \citet{srin09}. Blue: O-rich stars. Red: C-rich stars.
Green: ``extreme" AGB stars; these are the most luminous AGB stars,
losing the most mass. Left: [5.8 - 8] versus [3.6 - 4.5]; right:
[8 - 24] versus [3.6 - 8]. For a 10th mag star at [8], typical photometric errors
are 0.15 mag at [3.6], 0.10 mag at [8], and 0.12 mag at [24].
Some extreme stars, however, can have errors of up to 0.15 mag in the IRAC bands, despite being brighter. 
}
\label{sundar_mir}
\end{figure}

\begin{figure}
\begin{tabular}{ll}
\hspace*{0.8cm}\includegraphics[scale=0.40]{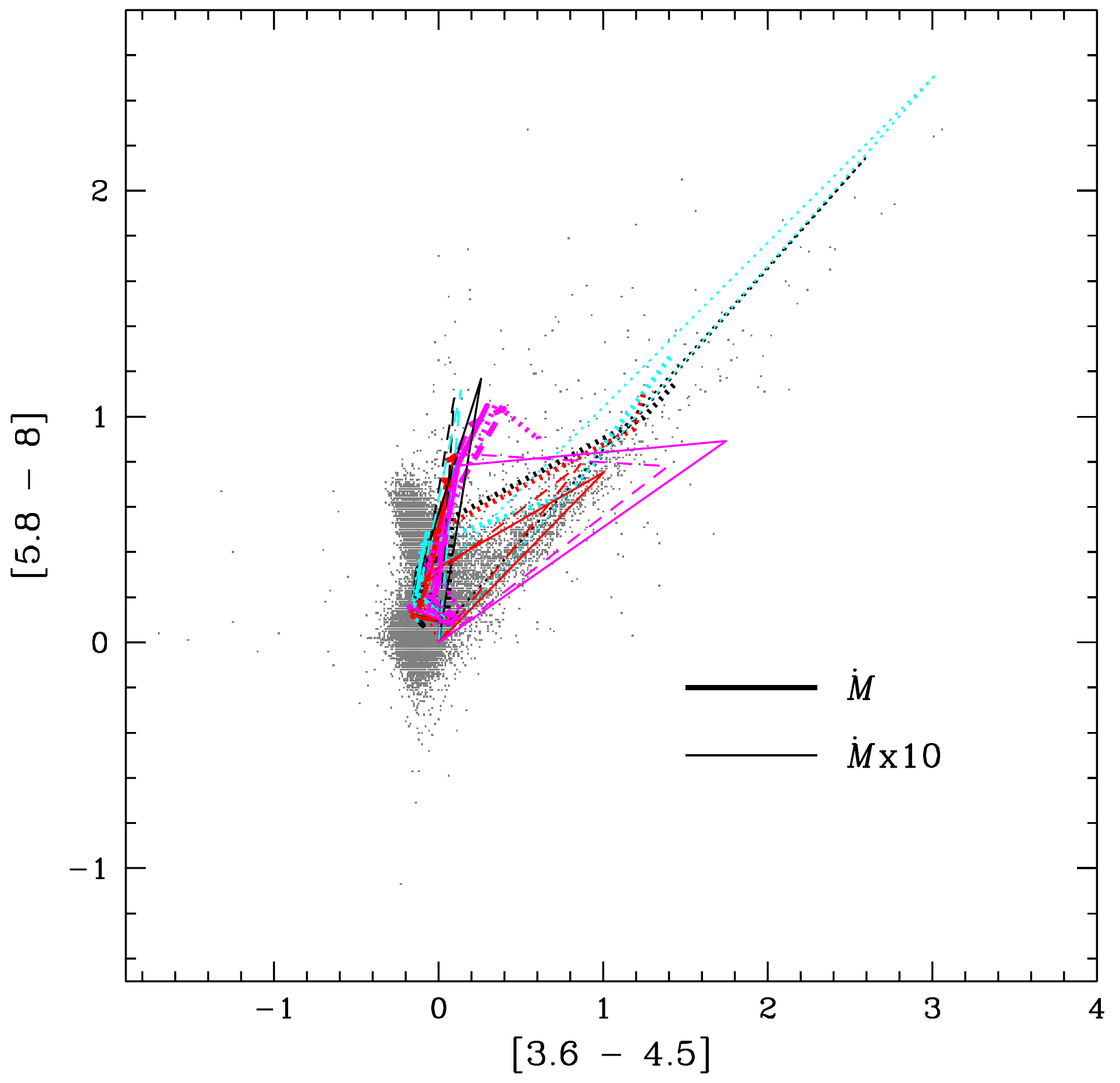}
&
\includegraphics[scale=0.40]{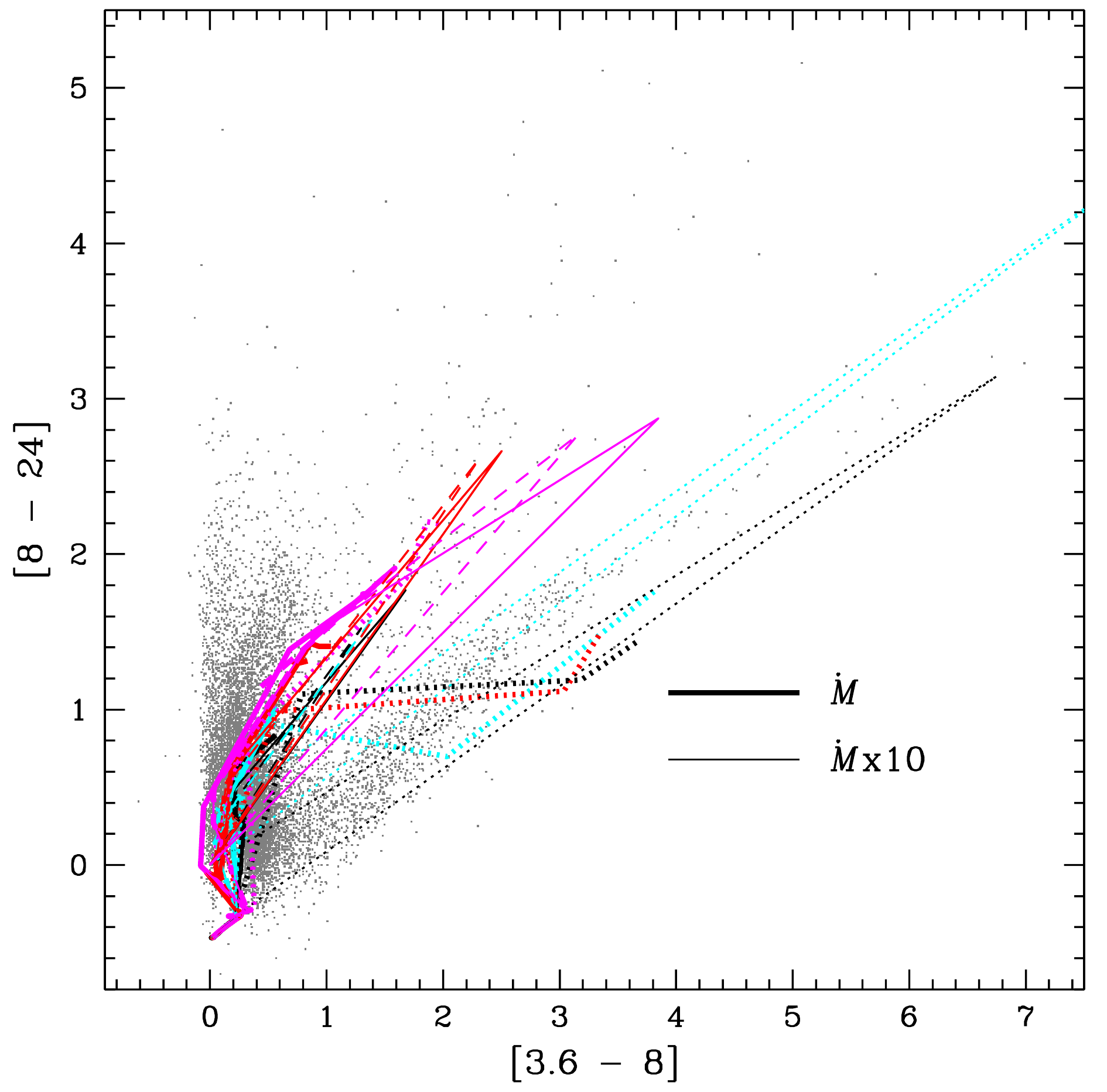}
\end{tabular}
\caption{
Superimposed on the sample of observed AGB stars, now shown as gray points, theoretical
mid-IR colors of individual TP-AGB stars along isochrones of populations
with different mass-loss rates and metallicities. Left: [5.8 - 8] versus [3.6 - 4.5]; right:
[8 - 24] versus [3.6 - 8].
Thick lines: fiducial
mass-loss rate $\dot M$; thin lines: high mass-loss rate
$\dot M \times$ 10. Metallicities and isochrone ages indicated as in Fig.~\ref{piov_mir}.
}
\label{meixner_overlay}
\end{figure}

\subsection{Star Clusters.} \label{subsec:starclust}

Figure~\ref{piov_ssp} presents theoretical two-color diagrams, 
[5.8 - 8] versus [3.6 - 4.5] and [8 - 24] versus [3.6 - 8],
for SSPs with different metallicities ($Z$ =
0.0005, blue; 0.008, cyan; 0.017, red) and, again, our five choices of
mass-loss and spectra for stars in the TP-AGB ($\dot M / 10$, $\dot M / 2$, $\dot M$, 
$\dot M \times 5$, and $\dot M \times 10$).
The model ages go from 10 Myr to 14 Gyr. 
Differing from models and observed colors of individual stars (see
Figures~\ref{piov_mir} and~\ref{meixner_overlay}),
the integrated colors of SSP models are confined to 
the  smaller parameter space -0.1 $\la$ [3.6 - 4.5] $\la$ 0.5, 
0. $\la$ [5.8 - 8] $\la$ 1.5, regardless of mass-loss rate.  
On the other hand, the expected [8 - 24] versus [3.6 - 8] colors
of SSPs occupy ranges similar to those of single stars.

\begin{sidewaysfigure}[ht]
\begin{tabular}{lllll}
\hspace*{-1.2cm}\includegraphics[scale=0.20]{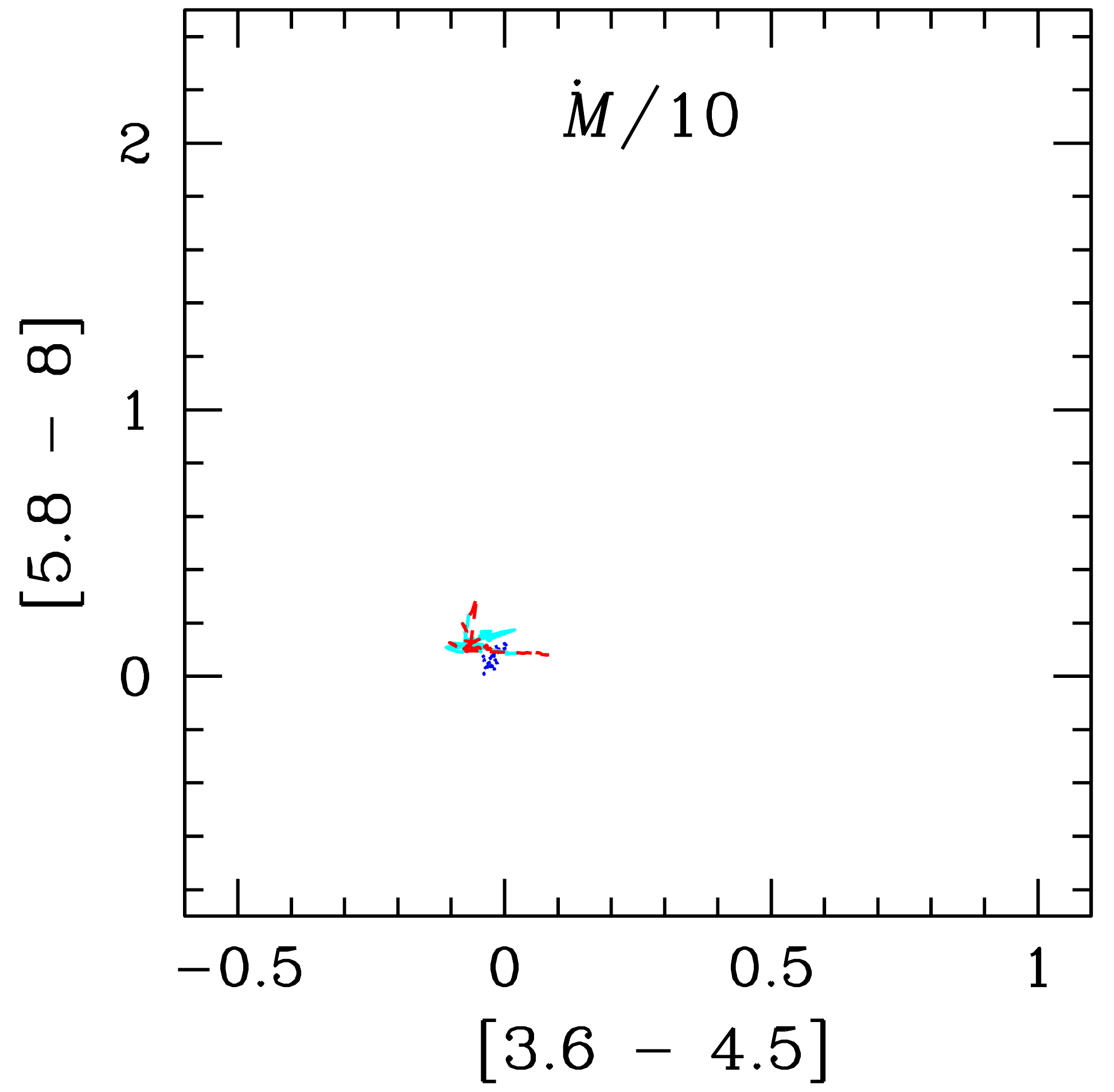}
&
\includegraphics[scale=0.20]{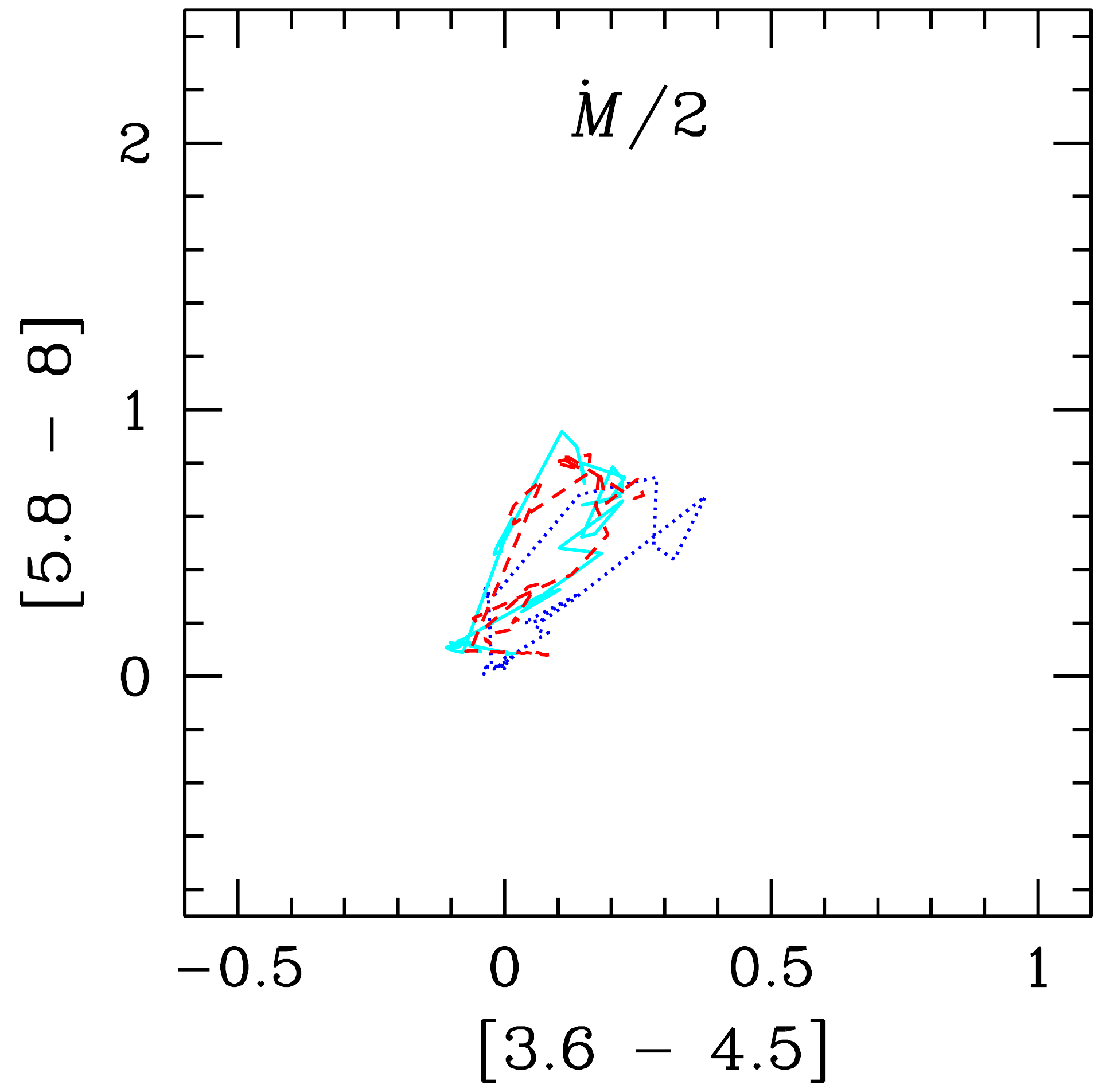}
&
\includegraphics[scale=0.20]{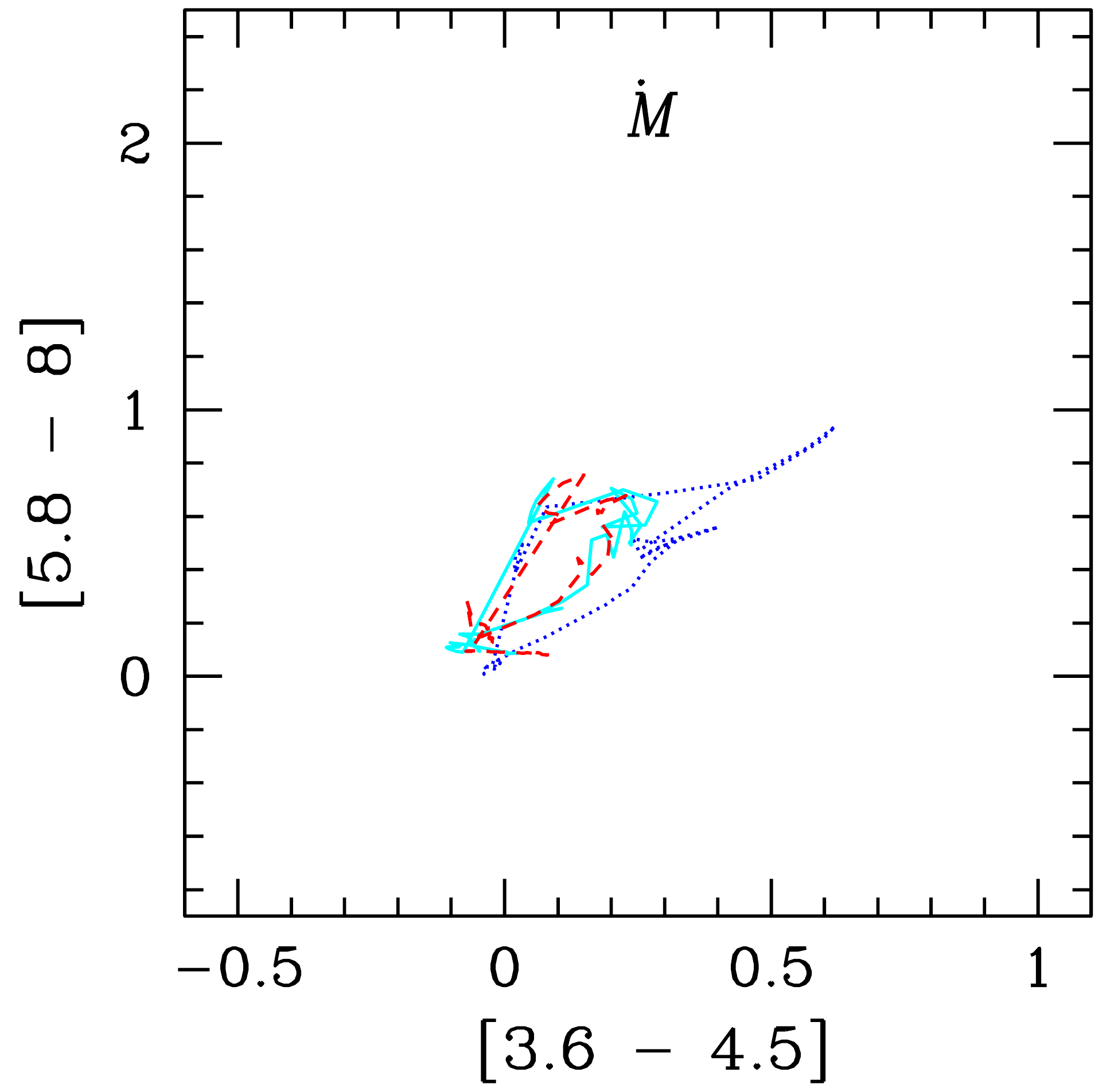}
&
\includegraphics[scale=0.20]{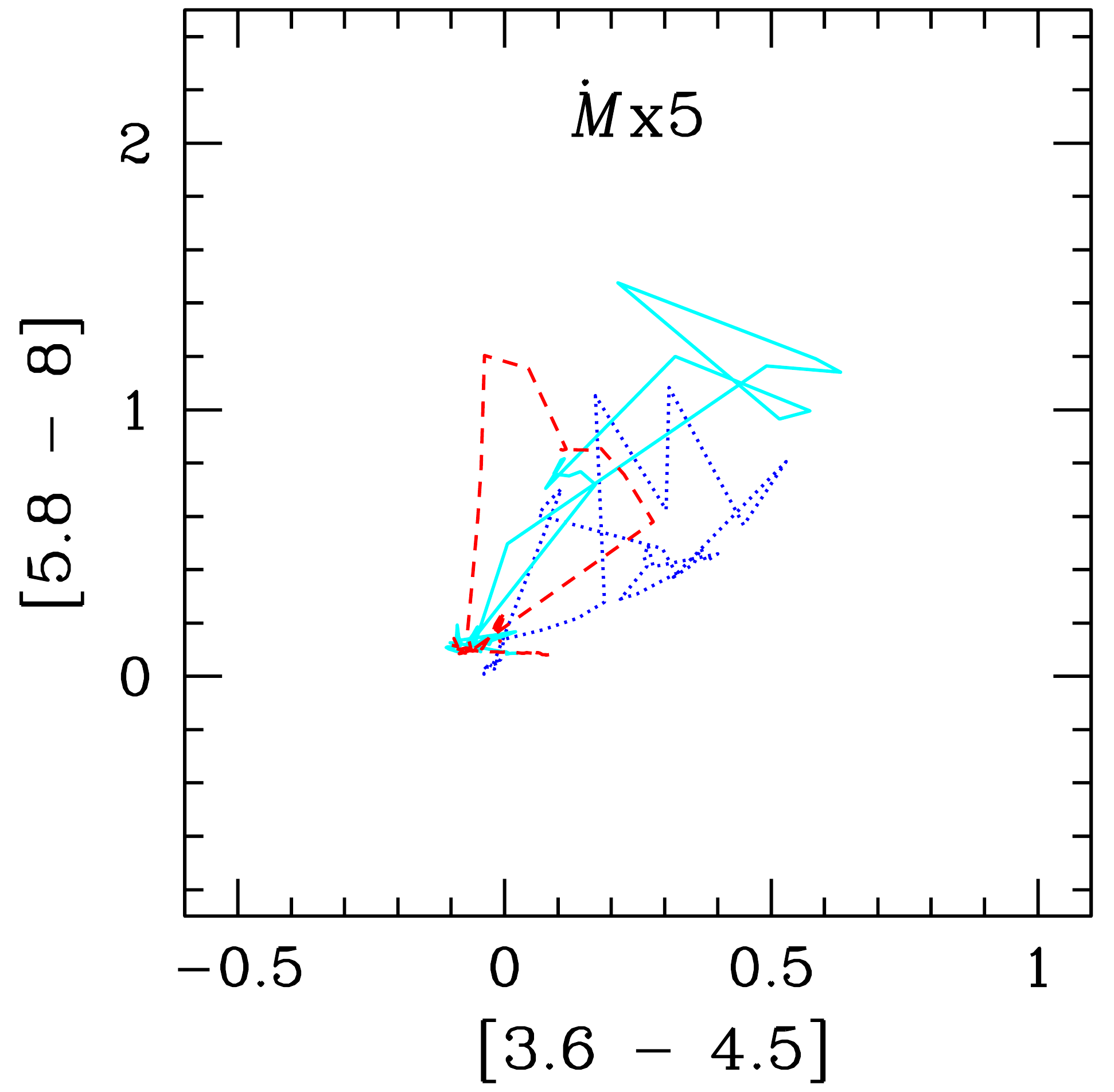}
&
\includegraphics[scale=0.20]{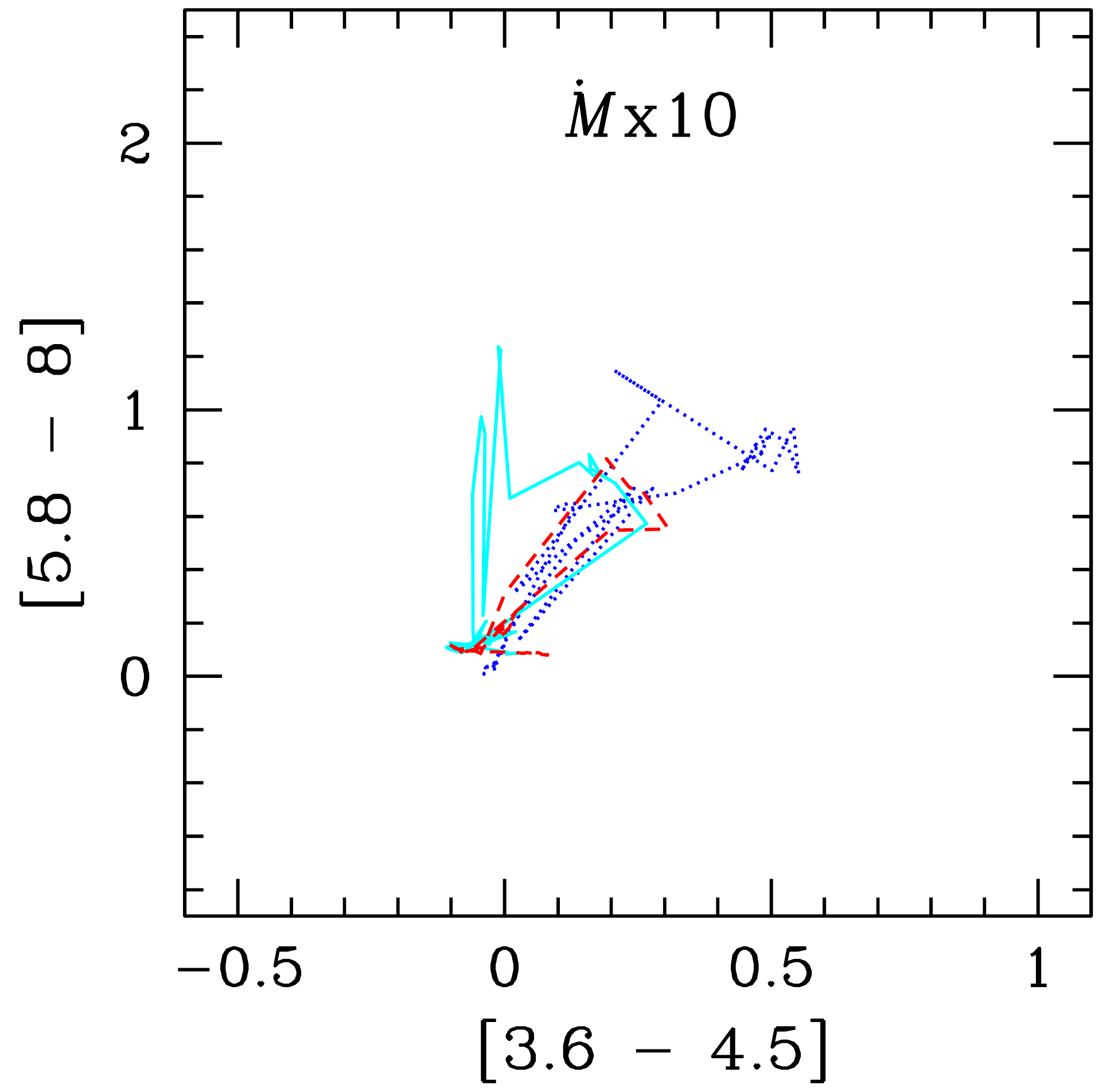}
\end{tabular}

\vspace*{0.3cm}

\begin{tabular}{llllll}
\hspace*{-1.2cm}\includegraphics[scale=0.20]{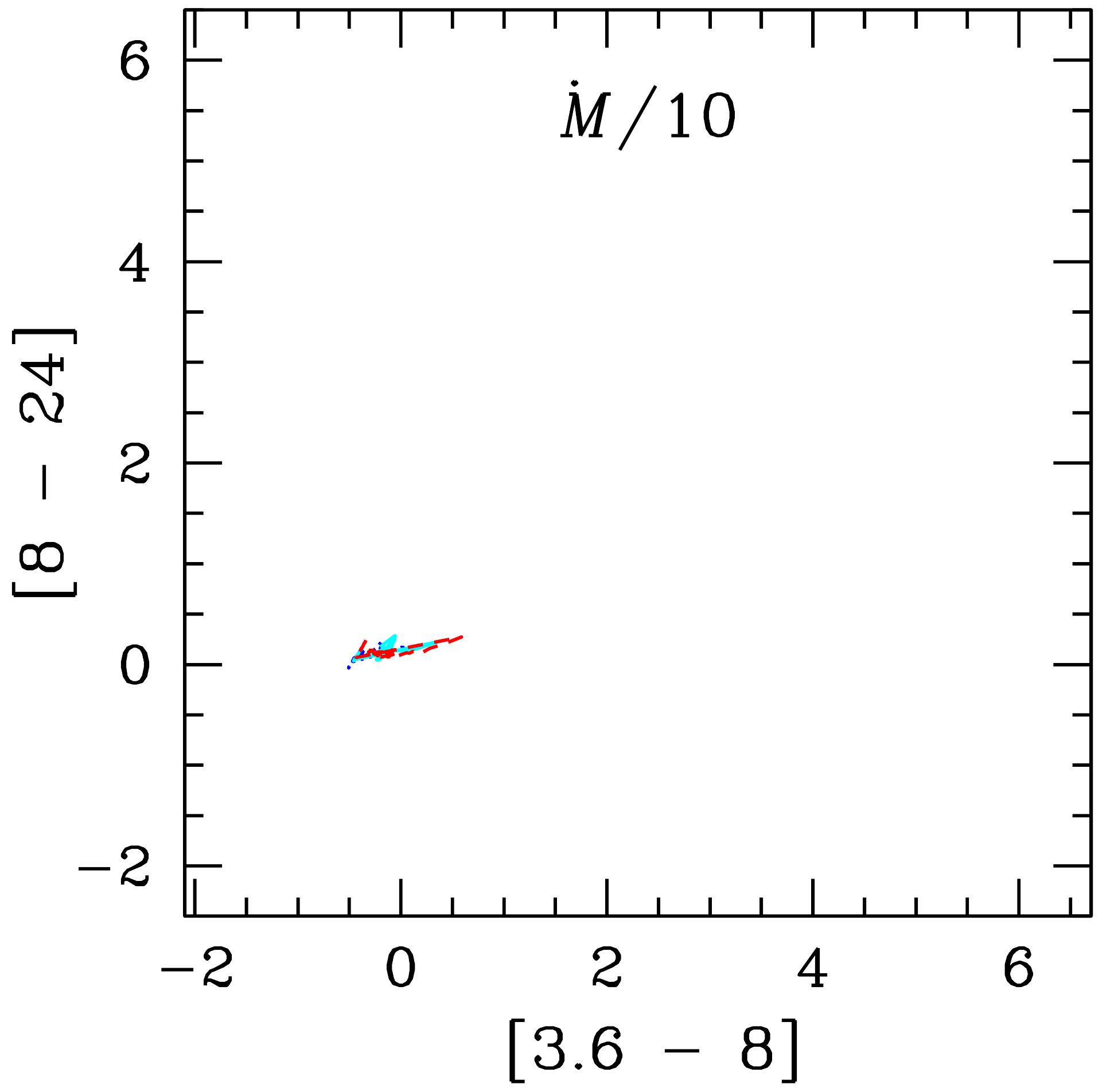}
&
\includegraphics[scale=0.20]{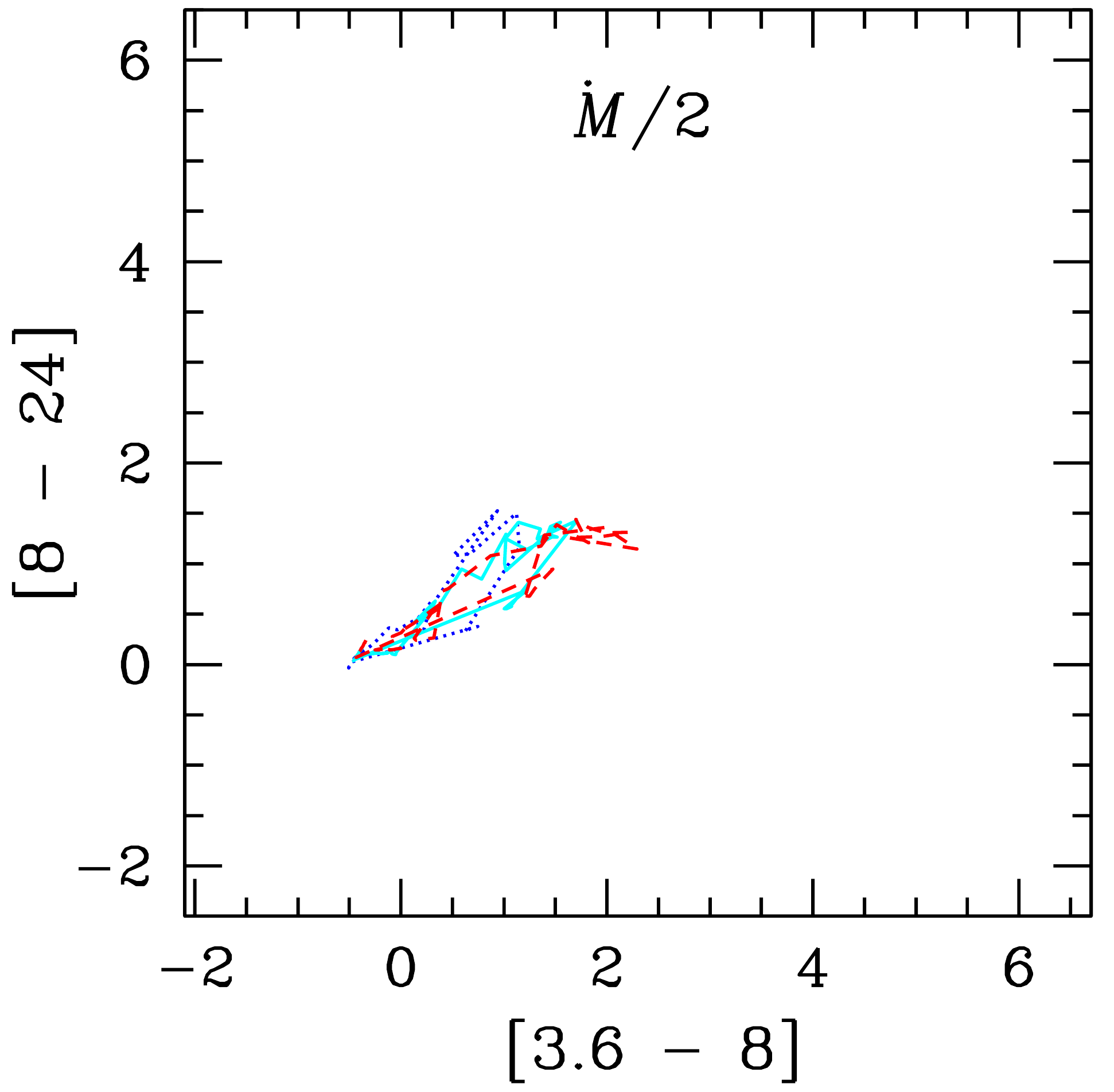}
&
\includegraphics[scale=0.20]{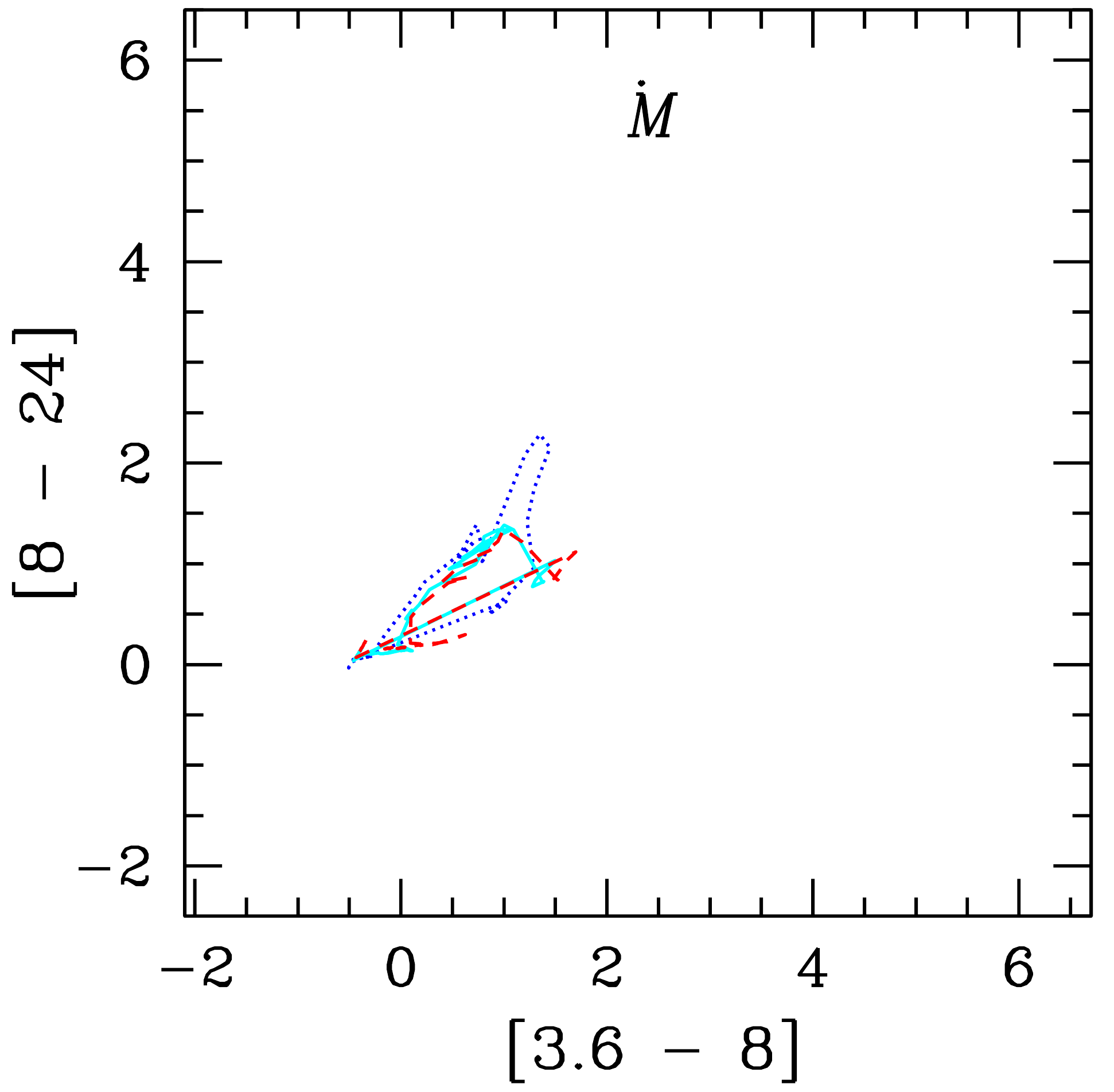}
&
\includegraphics[scale=0.20]{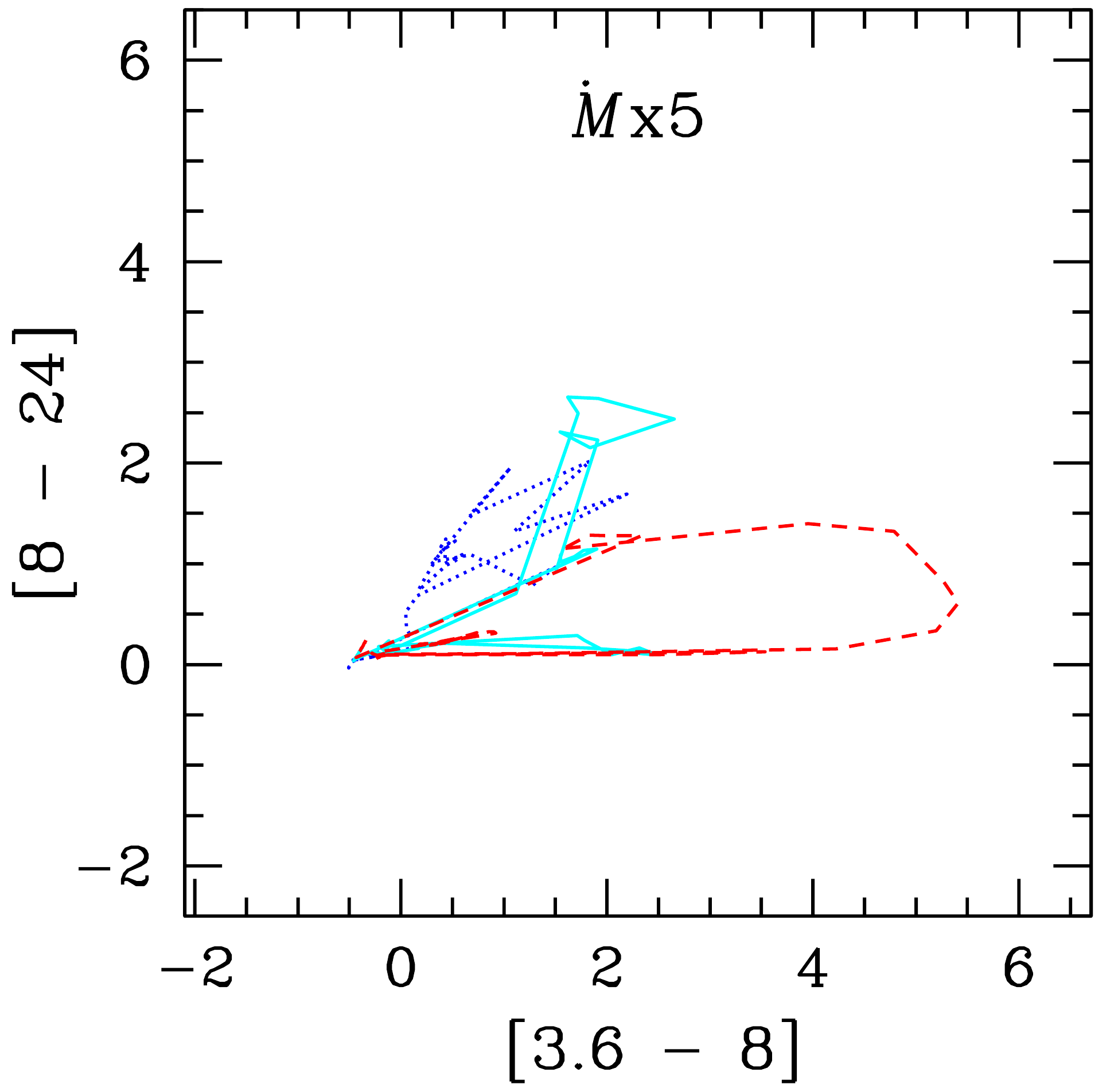}
&
\includegraphics[scale=0.20]{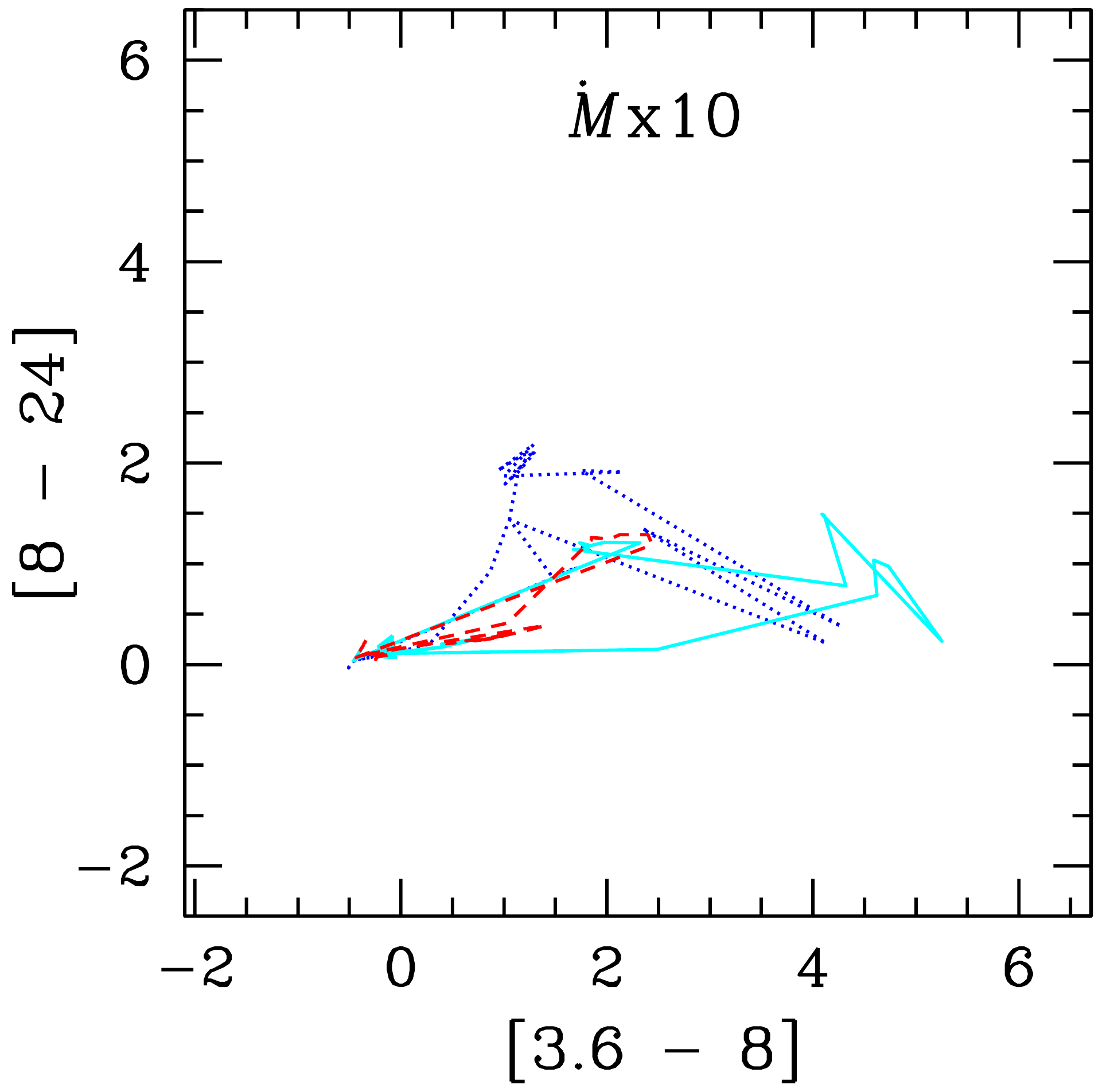}
\includegraphics[scale=0.20]{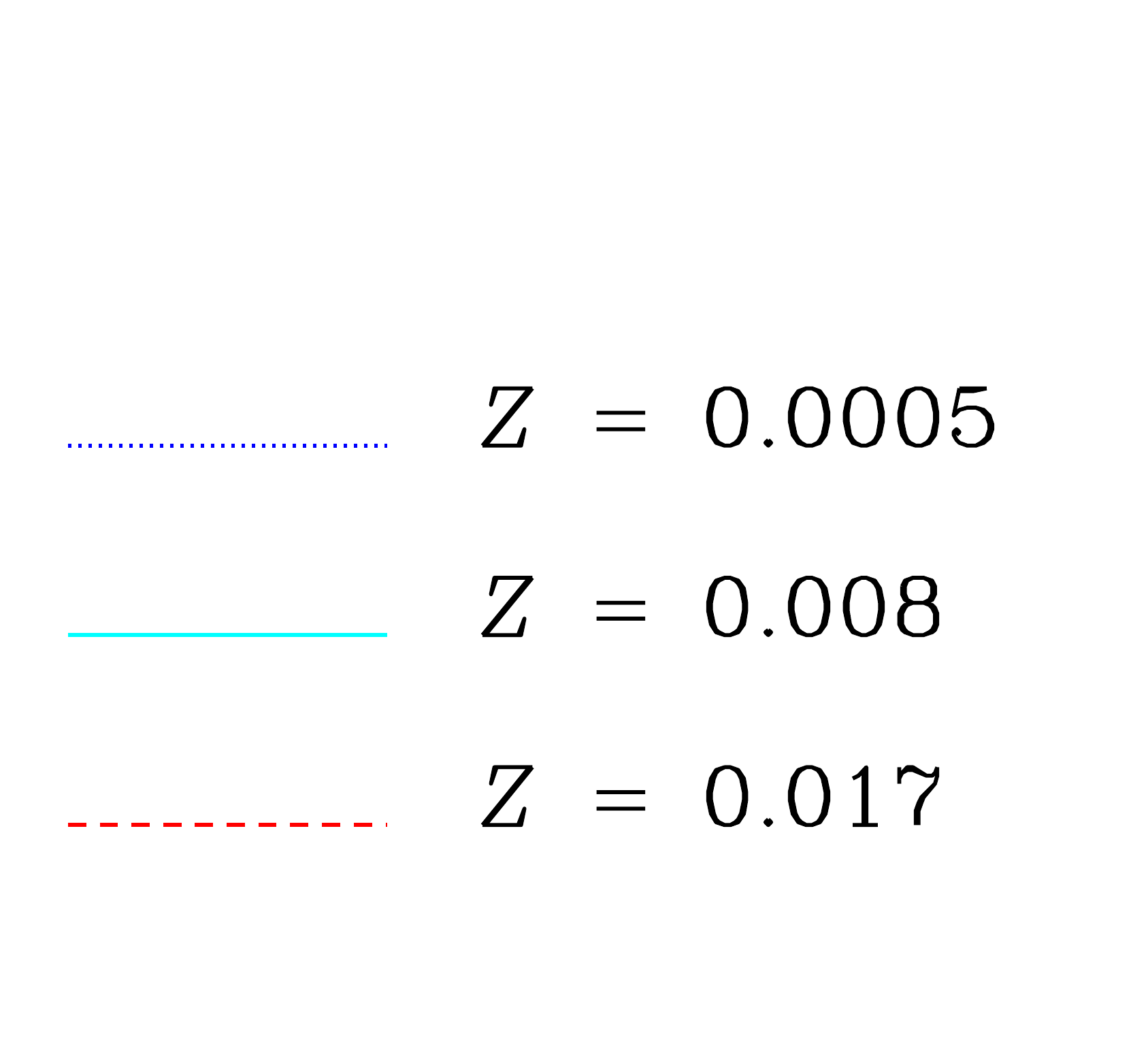}
\end{tabular}
\caption{Theoretical two-color diagrams
for SSPs with different metallicities and mass-loss rates in the TP-AGB.
Top row: [5.8 - 8] versus [3.6 - 4.5]; 
bottom row: [8 - 24] versus [3.6 - 8]. 
From left to right:
fiducial $\dot M/10$, fiducial $\dot M/2$, fiducial $\dot M$,
fiducial $\dot M \times 5$, fiducial $\dot M \times 10$.
Models span an age range between 10 Myr and 14 Gyr.
Blue dotted line: $Z$ = 0.0005; cyan solid line:
$Z$ = 0.008; red dashed line: $Z$ = 0.017.
}
\label{piov_ssp}
\end{sidewaysfigure}

For comparison with the models, I use data from the Spitzer Space Telescope surveys of the
LMC and SMC:
SAGE \citep{meix06,meix08} and SAGE-SMC \citep{gord11}, respectively.
SAGE consists in a uniform imaging survey of the LMC 
with the four IRAC channels (i.e., [3.6], [4.5], [5.8], and [8]), 
and the three Multiband Imaging Photometer (MIPS) bands ([24], [70], and [160]); I only use the
IRAC and MIPS [24] data in this work. The surveyed area was $7\fdg1 \times 7\fdg1$ with IRAC and 
$7\fdg8 \times 7\fdg8$ with MIPS. 

All four IRAC detectors are 256$^2$ pixels in size; the pixels are $1\farcs22 \times 1\farcs22$, for  
a $5\farcm2 \times 5\farcm2$ field of view (FOV). The two shorter wavelength channels employ InSb detectors,
whereas at [5.8] and [8] the camera works with Si:As impurity band conduction (IBC) arrays.
The actual angular resolution of the survey is $1\farcs7, 1\farcs7, 1\farcs9$ and $2\arcsec$, respectively, 
at [3.6], [4.5], [5.8], and [8]. The minimum effective exposure time per pixel in each channel was 43 s.
The 5$\sigma$ point-source sensitivity attained is 
17 mag at [3.6], 16 mag at [4.5], 14 mag at [5.8], and 13.5 mag at [8].   
For 24 $\mu$m imaging, the MIPS has a 128$^2$ pixel Si:As IBC array. 
The pixels are $2\farcs5 \times 2\farcs5$ in size, for a $5\arcmin \times 5\arcmin$ FOV. The
angular resolution of the 24 $\mu$m survey is $6\arcsec$. The minimum effective exposure time per pixel was 60 s, 
with a 5$\sigma$ point-source sensitivity of 10.4 mag.

SAGE-LMC covers an area of $\sim 30$ deg$^2$. Effective exposure times per pixel were 42 s for IRAC and 60 s for MIPS. The angular resolutions of the images are 2$\arcsec$ for IRAC and 6$\arcsec$ for MIPS at [24]. The 5$\sigma$ point-source sensitivity is 17 mag, 17 mag, 15 mag, 14.5 mag, and 10 mag, respectively, at 
[3.6], [4.5], [5.8], [8], and [24]. 

I have measured in the SAGE mosaics the integrated magnitudes of the LMC and SMC clusters listed in Table 4. 
Clusters that had visible nebulosity at [8] were not included in our sample.
The measurements were performed in apertures with $r =$ 1 arcmin, while the contributions 
of the sky and the field were estimated in an annulus
with 2\farcm0 $< r \leq$ 2\farcm5 and subtracted. Although not very important, a reddening correction was also applied. $E(B - V)$ values were taken from \citet{pers83} for individual clusters; when these were not available, we assumed $E(B - V) = 0.075$ and $E(B - V) =  0.037$, respectively, for the LMC and SMC \citep{schl98}. Table 4 shows the cluster names, cloud membership, and integrated magnitudes 
at [3.6], [4.5], [5.8], [8], and [24]. 
Integrated magnitudes are missing either because the clusters were not imaged by the survey
(mainly at [24]), or because they contained very prominent interstellar nebulosity. 
The clusters are grouped by SWB class \citep{sear80}, including a
very young pre-SWB category \citep{gonz04,gonzl05,gonz10}.

\clearpage

\input table4.tex

\vspace*{0.5cm}

A comparison between the individual clusters and the models is illustrated by the 
two-color diagrams shown in Figures~\ref{ind_5_8vs3_4} ([5.8 - 8] versus [3.6 - 4.5]) 
and~\ref{ind_8_24vs3_8} ([8 - 24] versus [3.6 - 8]).
The clusters are displayed as gray filled circles, and average photometric error bars are plotted for each figure, 
in its top left panel. 
The models shown have mass-loss rates that are either fiducial (top panels) or 5 $\times$ fiducial (bottom panels); 
metallicities $Z = 0.008$ (cyan bands; blue ticks) or $Z = 0.017$ (red bands; magenta ticks); and 
ages between 3.5 Myr and 14 Gyr. The run of ages is marked by the blue and magenta tick marks, located at 0.01, 0.3, 1, 5, and 13
Gyr, and whose sizes increase with age. 
The expected $\pm 1 \sigma$ error bars for the models, shown as colored bands,
have been calculated as in Gonz\'alez et al.\ (2004, Appendix). 
Roughly, if one assumes that the numbers of stars in different evolutionary
stages have a Poissonian distribution, then the 
errors of integrated colors scale as $M_{\rm tot}^{-1/2}$, where $M_{\rm tot}$
is the total mass of the stellar population \citep{cerv02}.
In these figures, I assume a mass for the model stellar populations of either 5$\times 10^4\ M_\odot$ (left panels) or 
5$\times 10^3\ M_\odot$ (right panels). The colors of the clusters are consistent with those of models with
$\dot M$ between fiducial and 5 $\times$ fiducial, and a total cluster mass $M_{\rm cl} = 5\times 10^4\ M_\odot$.

\begin{figure*}
\begin{tabular}{ll}
\includegraphics[scale=0.30]{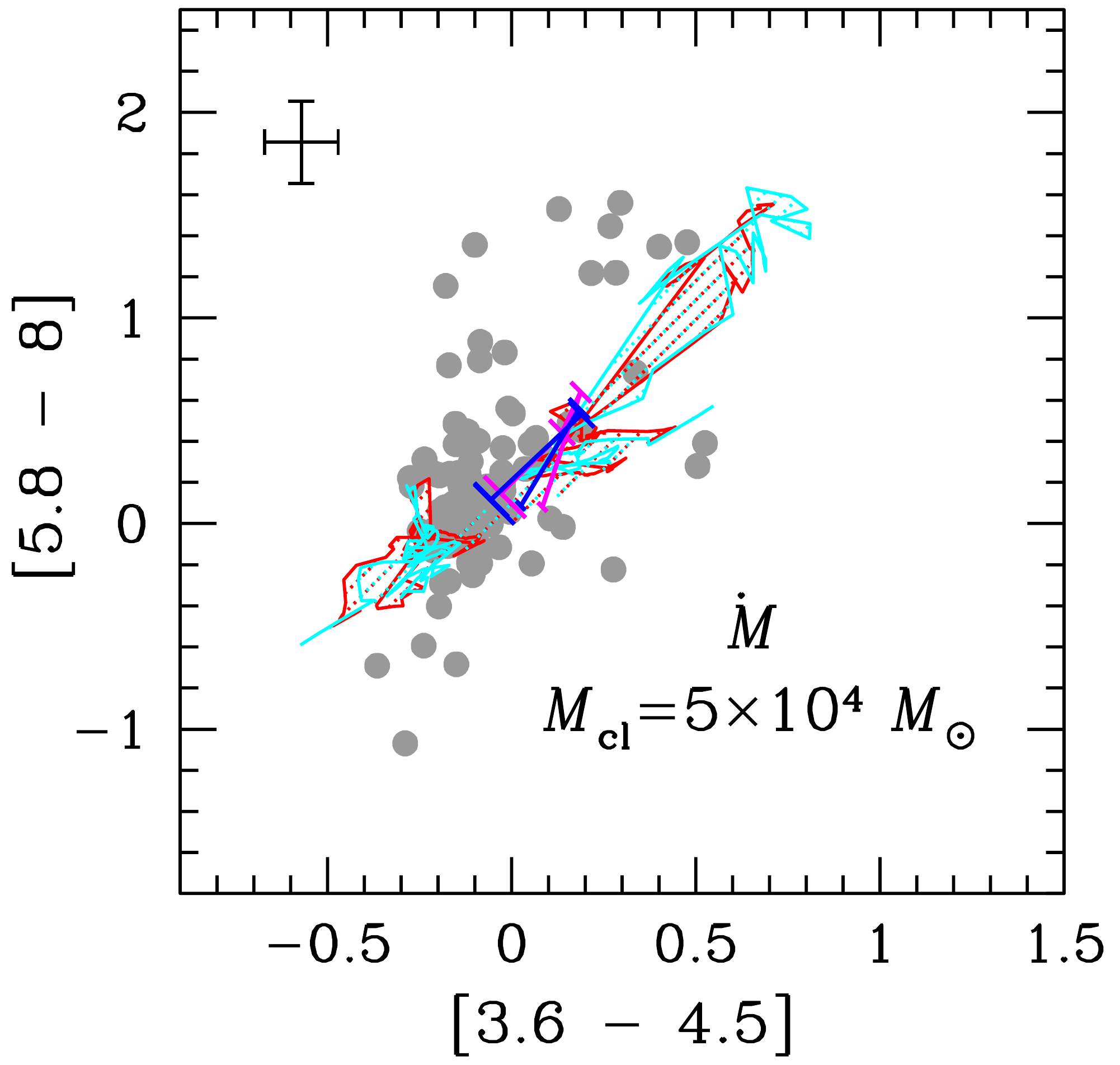}
&
\includegraphics[scale=0.30]{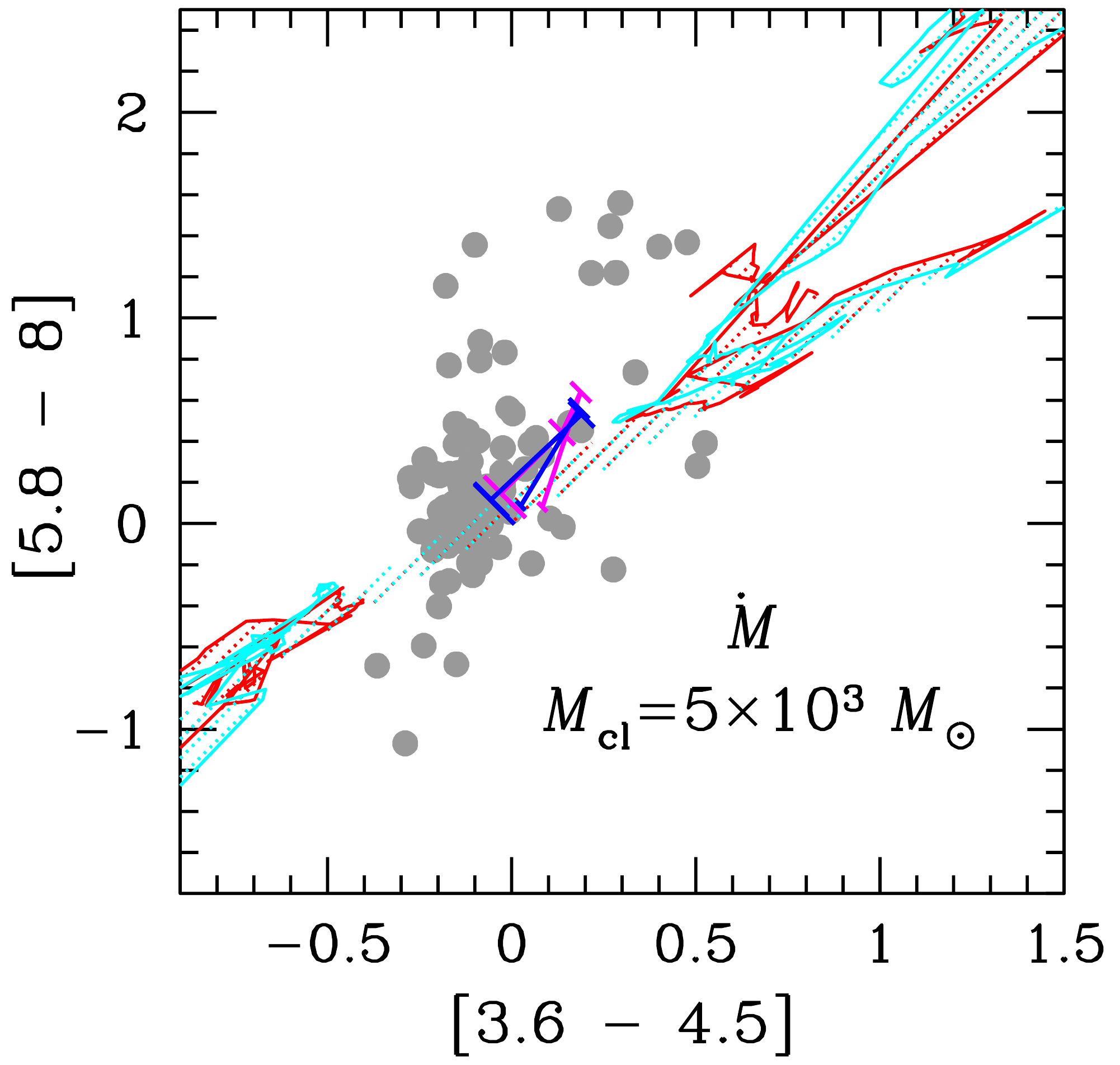}
\end{tabular}

\vspace*{0.3cm}

\begin{tabular}{ll}
\includegraphics[scale=0.30]{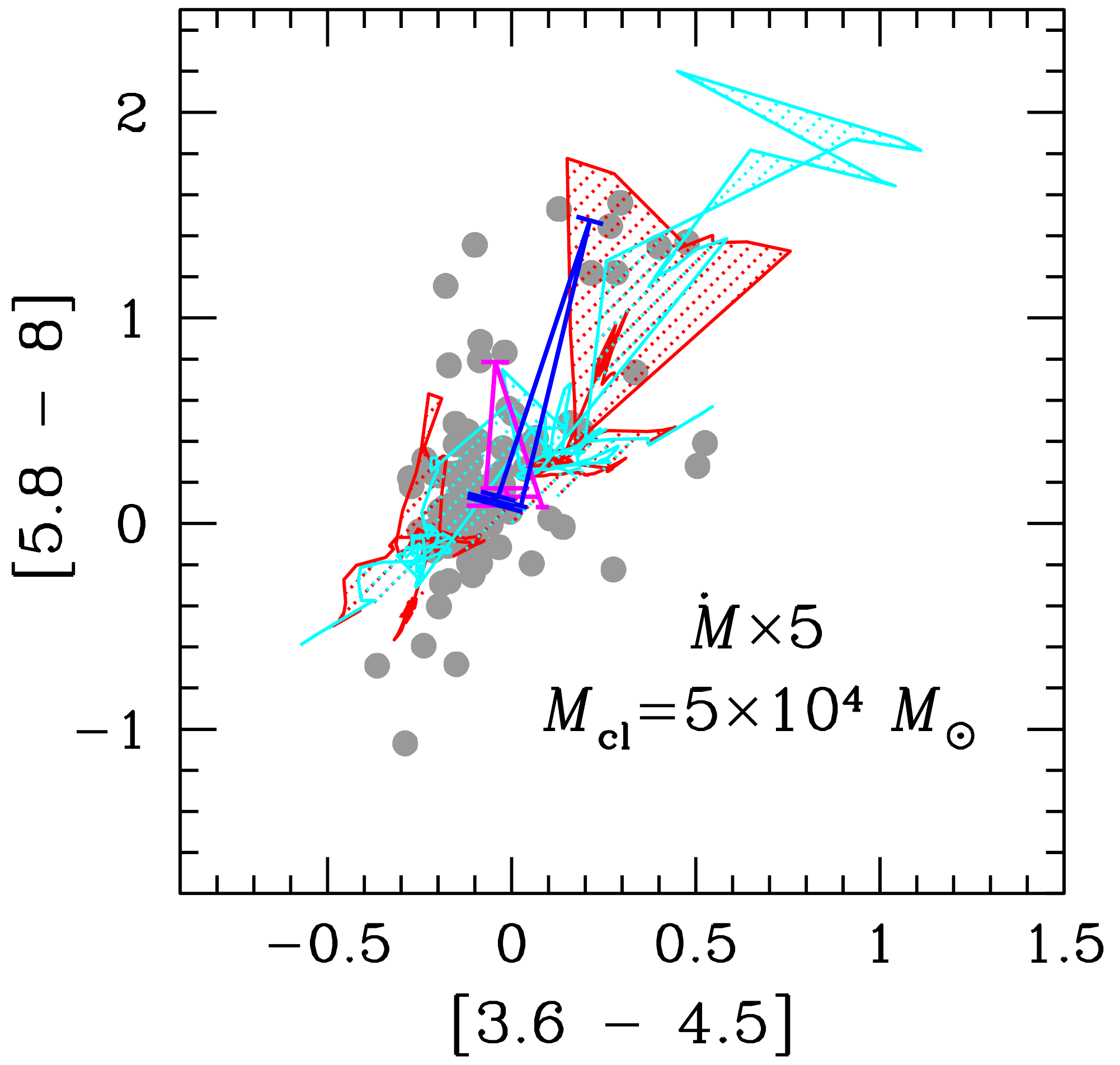}
&
\includegraphics[scale=0.30]{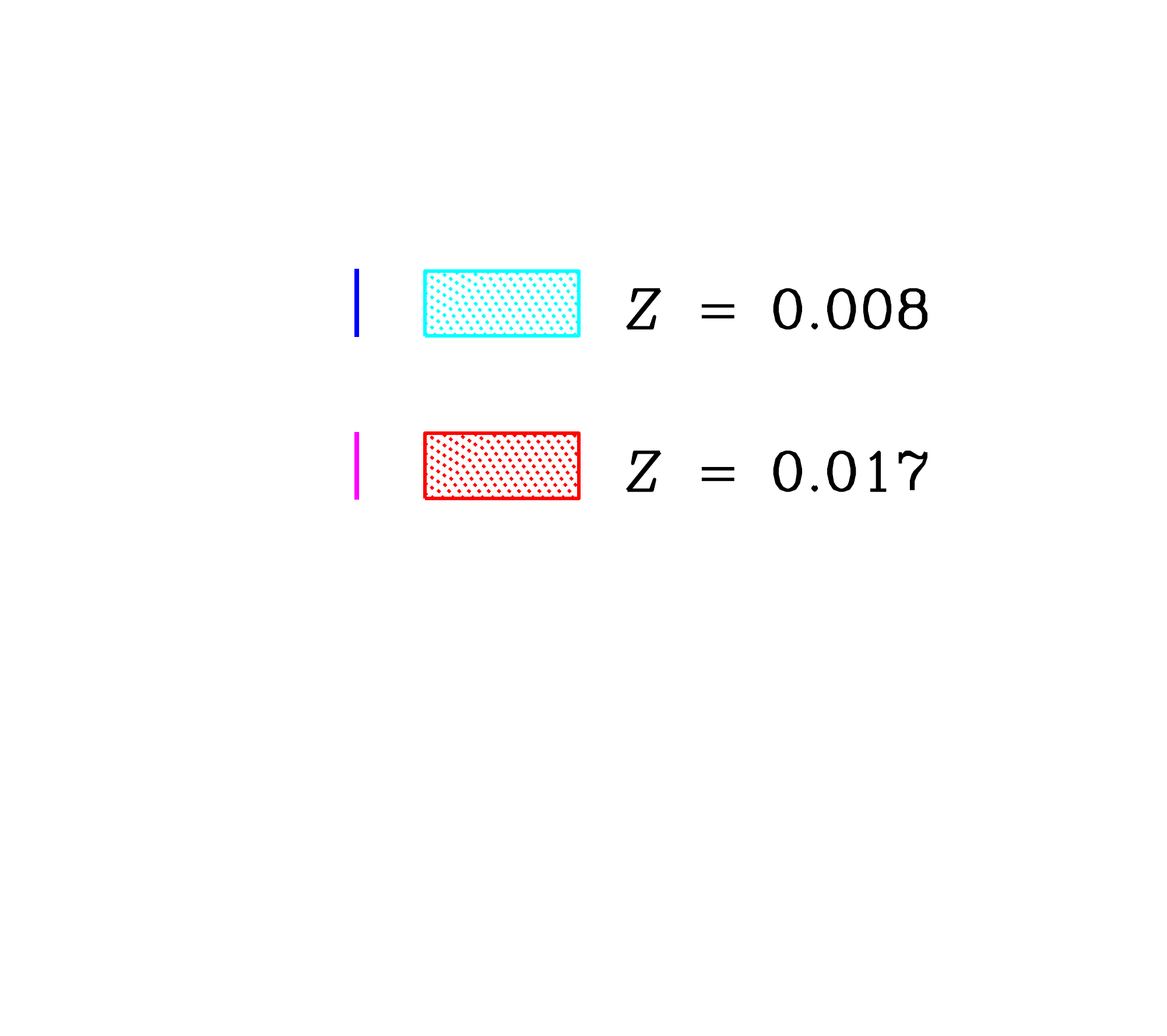}
\end{tabular}
\caption{Comparison between models and Magellanic individual star clusters: [5.8 - 8] versus [3.6 - 4.5].
Gray filled circles are reddening-corrected clusters from the sample 
in Table 4,
and black bars represent typical photometric errors. Colored regions
illustrate model SSPs and expected $\pm 1\ \sigma$ error bars. Top left: fiducial $\dot M$ and 5$\times 10^4 M_
\odot$;
top right: fiducial $\dot M$ and 5$\times 10^3 M_\odot$; bottom left: $5 \times$ fiducial $\dot M$ 
and
5$\times 10^4 M_\odot$. Cyan: $Z = 0.008$; red: $Z =  0.017$.  The models range in age between 3.5 
Myr and 14 Gyr.
Blue and magenta tick marks indicate 0.01, 0.3, 1, 5, 13 Gyr, respectively for $Z = 0.008$ 
and $Z = 0.017$; larger tick mark size
represents increasing age.}
\label{ind_5_8vs3_4}
\end{figure*}

\begin{figure*}
\begin{tabular}{ll}
\includegraphics[scale=0.30]{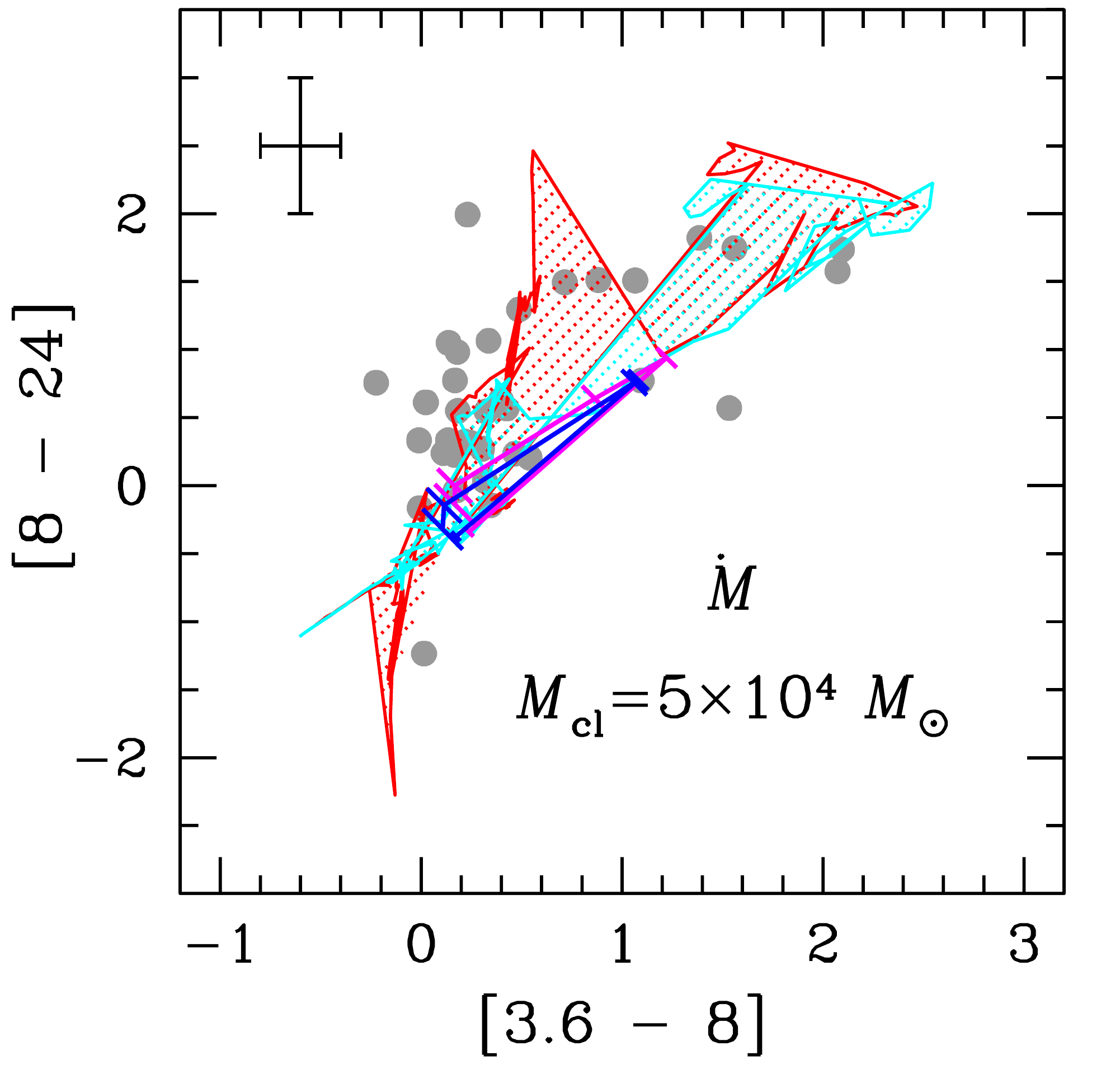}
&
\includegraphics[scale=0.30]{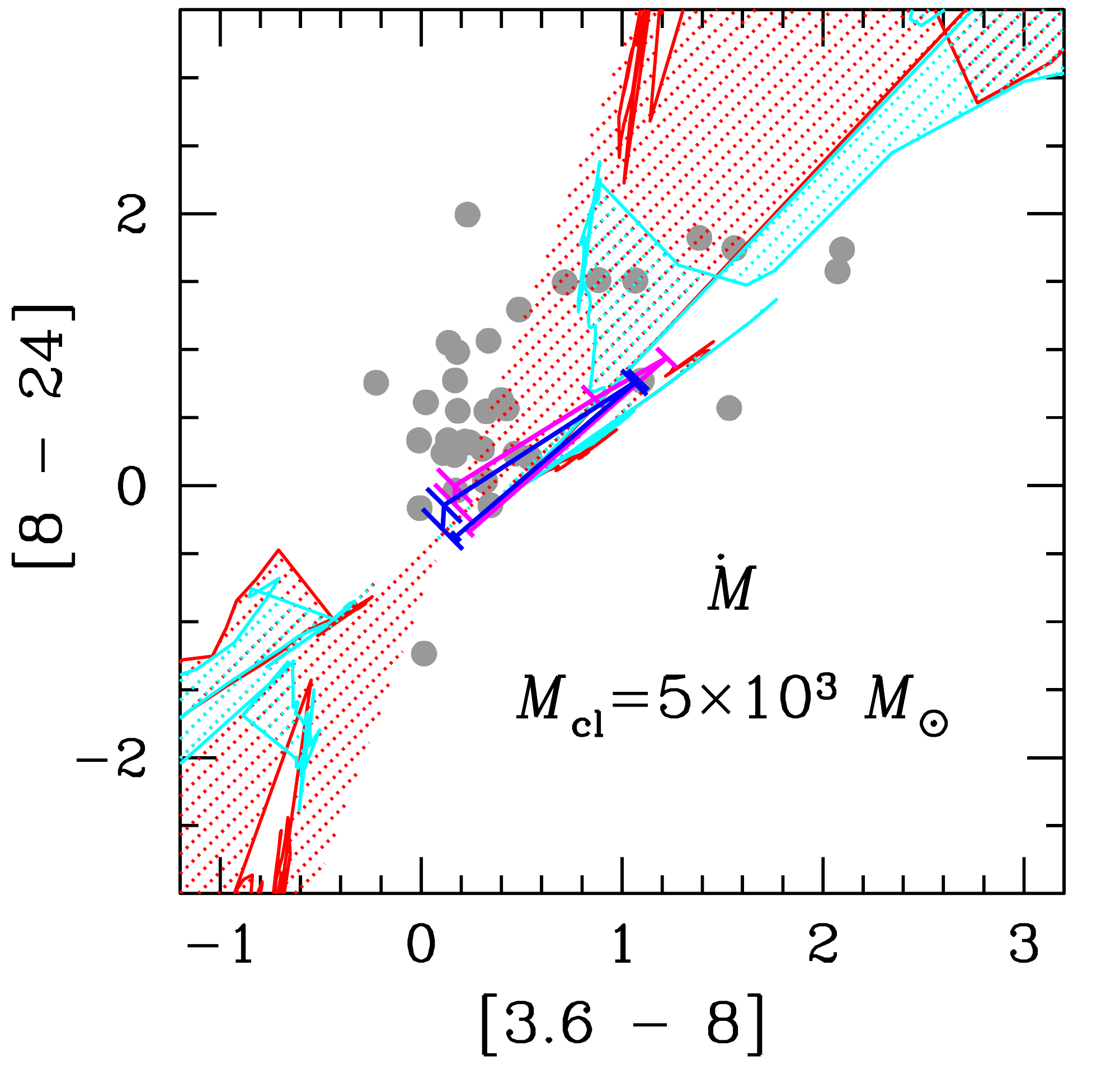}
\end{tabular}

\vspace*{0.3cm}

\begin{tabular}{ll}
\includegraphics[scale=0.30]{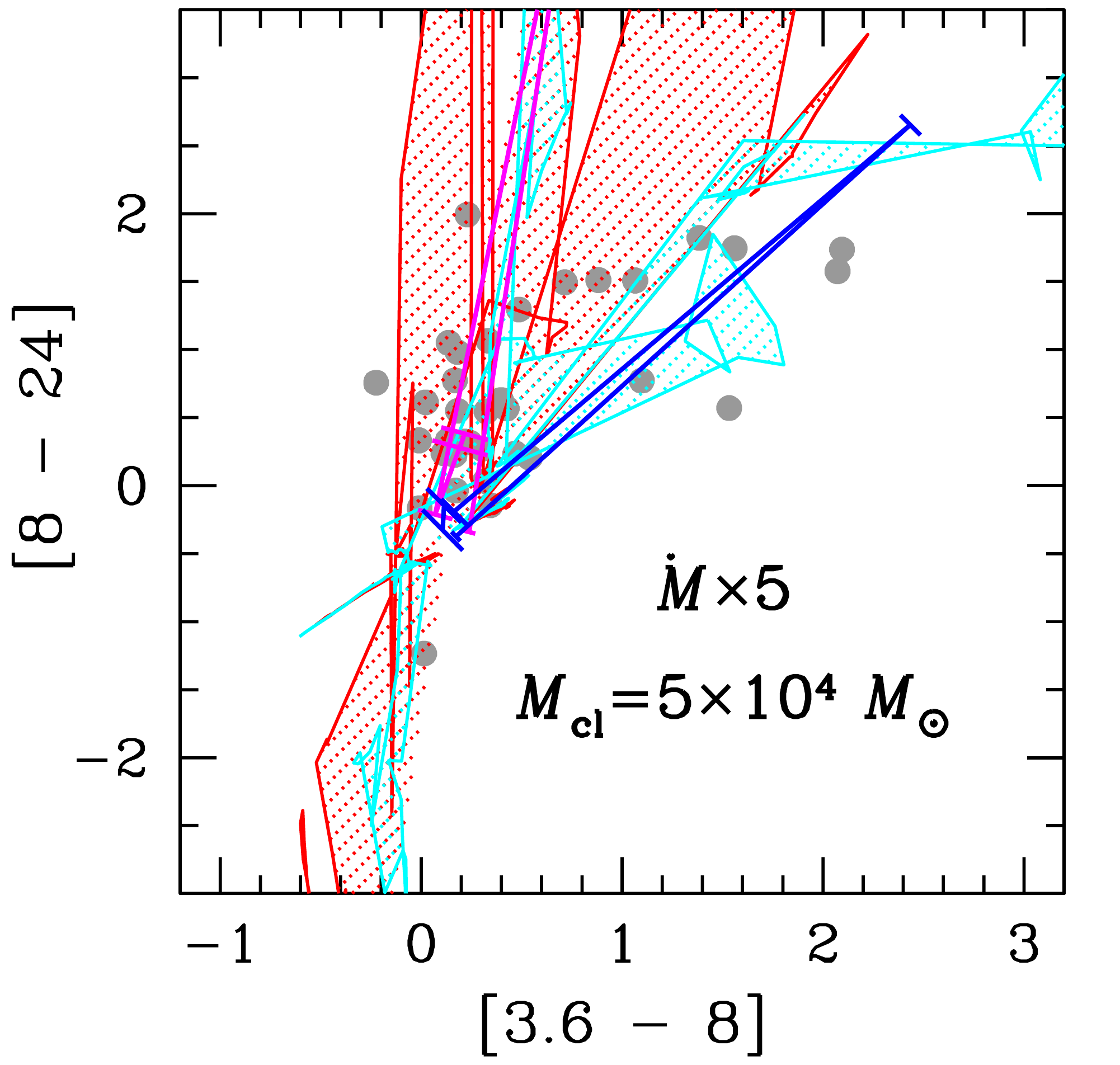}
&
\includegraphics[scale=0.30]{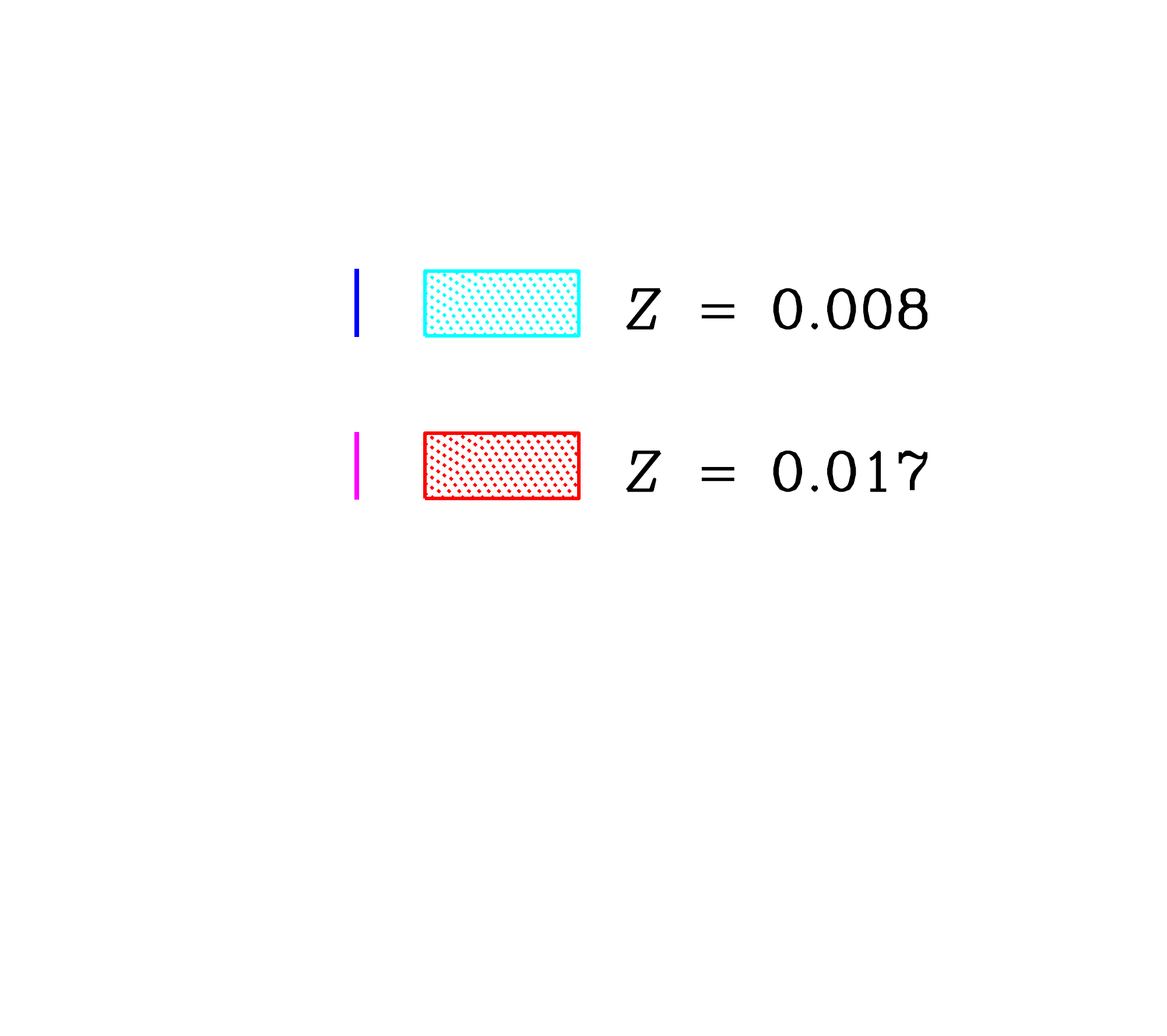}
\end{tabular}
\caption{
Comparison between models and Magellanic individual star clusters: [8  - 24] versus [3.6 - 8]. 
The symbols and models have the same meanings as in Fig.~\ref{ind_5_8vs3_4}.
}
\label{ind_8_24vs3_8}
\end{figure*}

\subsection{``Superclusters."} \label{subsec:superclust}

I have also obtained the integrated fluxes and the SBF measurements of artificial ``superclusters,"
assembled by adding together all the clusters in one SWB (or pre-SWB) class, as I have done before \citep{gonz04,gonzl05,gonz10}.
This procedure reduces the stochastic effects due to small numbers of stars in short-lived evolutionary phases, in particular
the TP-AGB. Before coaddition,  
the SMC clusters are geometrically magnified, conserving flux, to place them at the distance of the LMC,\footnote{
We assume $(m - M)_{\rm 0} =$ 18.50 $\pm$ 0.13 for the LMC, and 
$(m - M)_{\rm 0} =$ 18.99 $\pm$ 0.05 for the SMC, derived by \citet{ferr00} from
Cepheid measurements.}
and all (dereddened and sky-subtracted) clusters in a class are  
scaled to a common photometric zero-point and registered to a common center.
The integrated fluxes of the superclusters are derived in the same fashion as for individual clusters.
Measured colors for all superclusters are presented in Table~\ref{latabla},
together with their ages, metallicities, and photometric masses --derived from the comparison between 2MASS \citep{skru97} 
$J$, $H$, and $K_s$ supercluster mosaics and CB$^*$ models.

\begin{sidewaystable}
  \caption{Integrated Colors and Fluctuation Magnitudes of Magellanic Superclusters}
  \centering
  \begin{minipage}{480mm}
  \begin{ssmall}
   \begin{tabular}{@{}lcccccccccccc@{}}
   \hline
  Supercluster & Log age (year){\normalsize $^{\rm a}$} & $Z$  & Mass (10$^6 M_\odot$){\normalsize $^{\rm b}$} & [3.6 - 4.5] & [3.6 - 8] & [5.8 - 8] & [8 - 24] & $\barM_{3.6}$& $\barM_{4.5}$& $\barM_{5.8}$ & $\barM_{8}$ & $\barM_{24}$ \\
  \hline
pre\dotfill &6.78$\pm$0.62 & 0.010$\pm$0.005{\normalsize $^{\rm c}$} & 0.08 $\pm$ 0.02  &   0.19$\pm$0.29 &  2.50$\pm$0.52 & 1.23$\pm$0.23 &  3.56$\pm$0.36 & -11.35$\pm$0.19 &  -11.20$\pm$0.15 & -11.28$\pm$0.19 &    -13.37$\pm$0.36  &   -11.25$\pm$0.52 \\
I\dotfill   &7.51$\pm$0.32 & 0.010$\pm$0.005{\normalsize $^{\rm c}$} & 0.6 $\pm$ 0.1  &  -0.04$\pm$0.14 &  0.62$\pm$0.15 & 0.21$\pm$0.15 &  0.07$\pm$0.39 &   -10.69$\pm$0.10 &  -10.60$\pm$0.12 & -10.72$\pm$0.14 &    -11.5$\pm$0.17  & $\cdots$   \\
II\dotfill  &7.88$\pm$0.25 & 0.010$\pm$0.005{\normalsize $^{\rm c}$} & 0.5 $\pm$ 0.1  &  -0.17$\pm$0.11 & -0.09 $\pm$0.15 & -0.15$\pm$0.14 &  2.01$\pm$0.16 & -8.96 $\pm$0.20 &  -9.00$\pm$0.21 & -9.07$\pm$0.22 &    -9.75$\pm$0.23  &   -10.00$\pm$0.51 \\
III\dotfill &8.21$\pm$0.29 & 0.010$\pm$0.005{\normalsize $^{\rm d}$} & 0.4 $\pm$ 0.1   &  -0.14$\pm$0.08 &  1.00$\pm$0.10 & 0.61$\pm$0.07 &  2.23$\pm$0.21 &  -8.28$\pm$0.18 &  -8.20$\pm$0.14 & -8.33$\pm$0.19 &    -8.96$\pm$0.38  &    -13.26$\pm$0.48 \\
IV\dotfill  &8.65$\pm$0.36 & (3$\pm$2)e-3{\normalsize $^{\rm d}$} & 0.4 $\pm$ 0.0   &  -0.10$\pm$0.20 &  0.18$\pm$0.30 & 0.05$\pm$0.29 & -0.21$\pm$0.71 &  -9.30$\pm$0.20 &  -9.47$\pm$0.24 & -9.64$\pm$0.29 &    -10.57$\pm$0.22  &   -12.02$\pm$0.40 \\
V\dotfill   &9.09$\pm$0.29 & (4$\pm$2)e-3{\normalsize $^{\rm d}$} & 1.4 $\pm$ 0.1 &  -0.17$\pm$0.13 &  0.45$\pm$0.45 & 0.26$\pm$0.32 & -0.74$\pm$0.34 &  -8.69$\pm$0.12 &
-9.17$\pm$0.29 & -10.20$\pm$0.43 &    -11.64$\pm$0.40  &   -9.32$\pm$0.46 \\
VI\dotfill  &9.45$\pm$0.28 & (2$\pm$1)e-3{\normalsize $^{\rm d}$} & 2.4 $\pm$ 0.1 &  -0.07$\pm$0.09 &  2.27$\pm$0.07 & 1.74$\pm$0.10 &  0.00$\pm$1.26 & -8.49  $\pm$0.25
&  -8.73$\pm$0.40 & -8.79$\pm$0.47 &    -8.83$\pm$0.46  & $\cdots$  \\
VII\dotfill &9.82$\pm$0.29 & (7$\pm$4)e-4{\normalsize $^{\rm d}$} & 2.4 $\pm$ 0.3 & -0.08$\pm$0.08 &  0.23$\pm$0.07 & 0.07$\pm$0.07 &  0.31$\pm$0.11 & -6.79  $\pm$0.36
&  -6.54$\pm$0.50       &   -7.00  $\pm$  0.28  &      -7.88  $\pm$  0.60   & $\cdots$   \\

\hline
\end{tabular}
\\

\end{ssmall}
{$^{\rm a}$ From the calibration of the $S$-parameter by \citet{gira95}.} \\
{$^{\rm b}$ Masses from near-IR mass-to-light ratios (2MASS data and CB$^*$ models); errors are
equal to the dispersion of the results at $J$, $H$, and $K_s$.} \\
{$^{\rm c}$ \citet{cohe82}.} \\
{$^{\rm d}$ \citet{frog90}, assuming $Z_\odot = 0.017$.}

\end{minipage}
\label{latabla}
\end{sidewaystable}

Figures~\ref{piov_sclust_5.8_8vs3.6_4.5} and~\ref{piov_sclust_8_24vs3.6vs8} display the same two-color diagrams shown previously (respectively, [5.8 - 8] versus [3.6 - 4.5] and and [8 - 24] versus [3.6 - 8]),
now comparing the models to the Magellanic artificial superclusters.   
In all the panels, the data (solid black circles with error bars) are plotted together with models of 
different metallicities ($Z$ = 0.0005, blue; 0.008, cyan; 0.017, red), that
bracket those of the superclusters (0.0007 $ \leq Z \leq $ 0.01; Frogel et al.\ 1990, assuming
that $Z_\odot = 0.017$; Cohen 1982). Three different theoretical mass-loss rates are 
shown: fiducial $\dot M$ (top left), fiducial $\dot M/5$, and 5$\times$ fiducial $\dot M$.  
The expected $\pm 1 \sigma$ error bars for the models (colored bands)
have been calculated assuming a stellar population
of $5 \times 10^5~M_\odot$. Models with a mass-loss rate higher than fiducial seem more consistent
with the data, especially those of the pre-SWB and SWB VI superclusters. However, I note here that 
the pre-SWB supercluster may be subject to more stochastic fluctuations,
given its lower mass, and may be more affected by additional extinction than older objects.  
Also, as I have argued before \citep{gonz10}, 
the assumption that the addition of many small objects is statistically equal to a
large one will fail, if none of the small clusters are massive enough
to produce the most massive stars \citep[e.g.,][]{weid06}; 
this deficiency will be an issue during the first few 10$^7$ yr, when such stars   
contribute most of the cluster's light.

\begin{figure*}
\includegraphics[width=0.70\hsize,clip=]{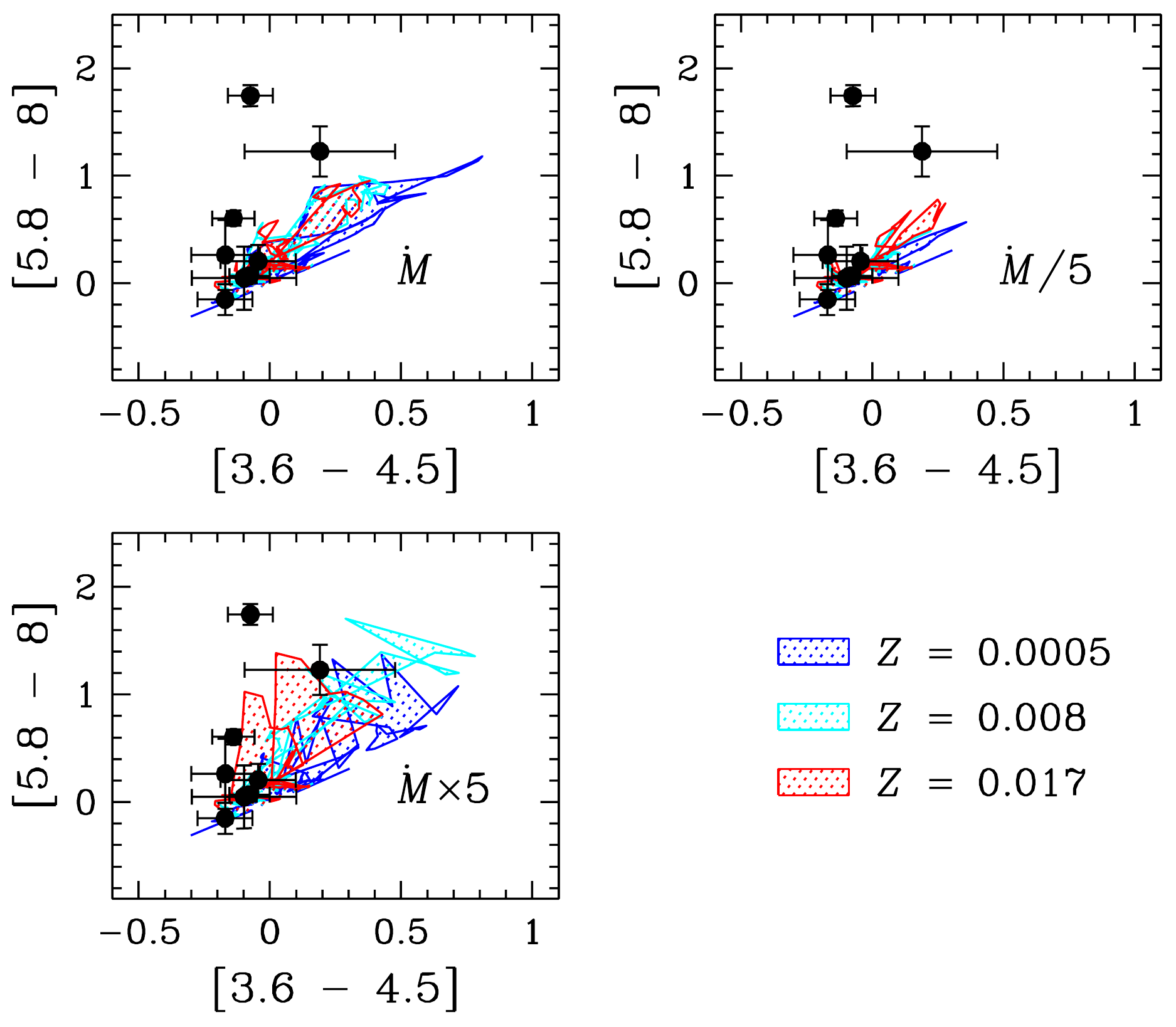}
\caption{Comparison between models and MC
``superclusters": two-color diagram, [5.8  - 8] versus [3.6 - 4.5].
Top left: fiducial $\dot M$; top right: fiducial $\dot M$/5;
bottom left: fiducial $\dot M \times 5$.
Filled circles are
artificial ``superclusters" built in this work, following 
\citet{gonz04}. Colored regions represent SSPs with
chosen mass-loss rate and expected $\pm 1 \sigma$ error bars for 5$\times
10^5~M_\odot$.  Blue: $Z =$ 0.0005; cyan: $Z =$ 0.008;
red: $Z =$ 0.017. Supercluster ages go from $\sim$ 6 Myr to $\sim$ 7 Gyr;
model ages span between 3 Myr and 14 Gyr.}
\label{piov_sclust_5.8_8vs3.6_4.5}
\end{figure*}

\begin{figure*}
\includegraphics[width=0.70\hsize,clip=]{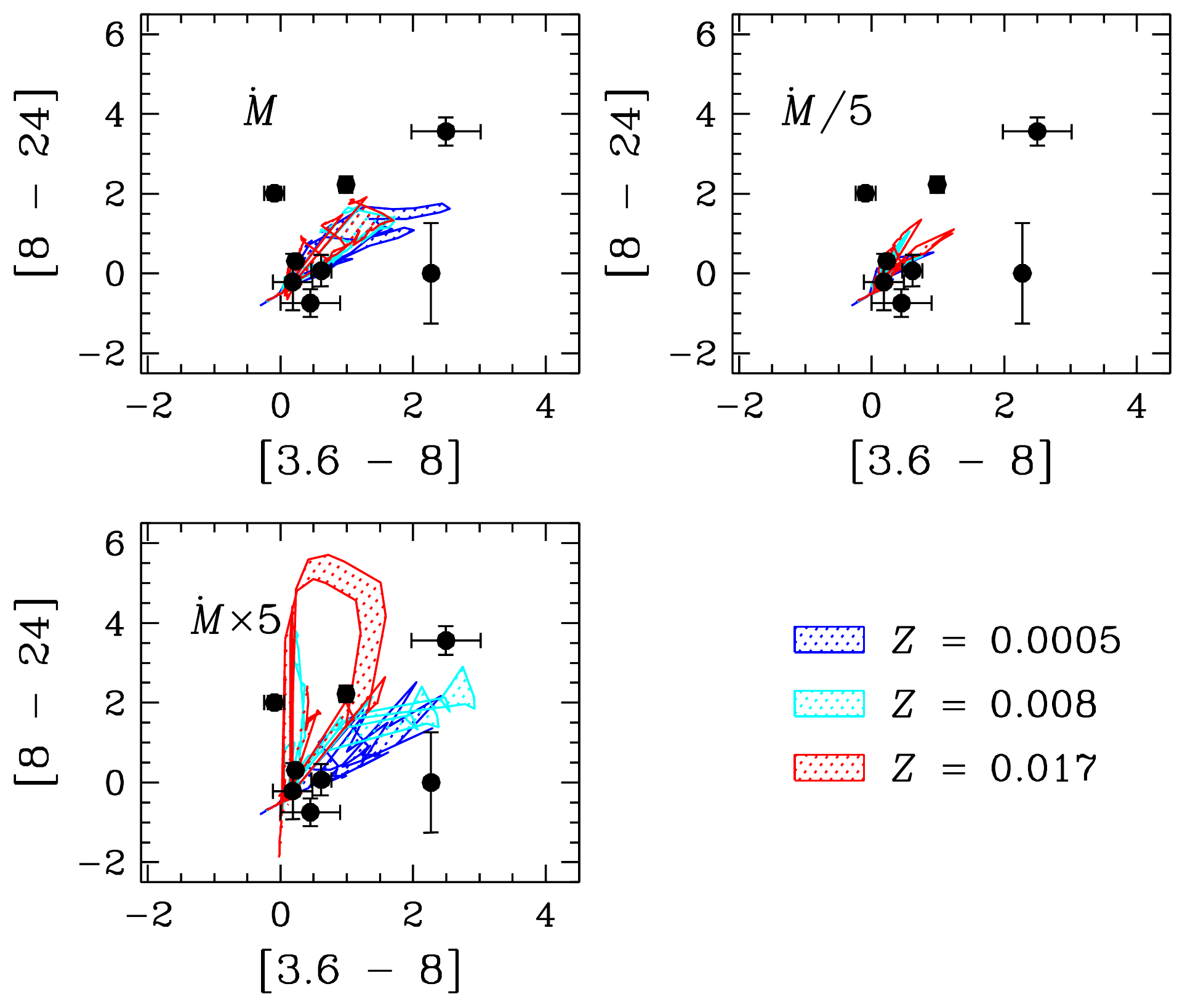}
\caption{Comparison between models and MC
``superclusters": two-color diagram, [8 - 24] versus [3.6 - 8].
The symbols and models have the same meanings as in Fig.\ \ref{piov_sclust_5.8_8vs3.6_4.5}.}
\label{piov_sclust_8_24vs3.6vs8}
\end{figure*}

It is also useful to compare data and models in the age-color plane.
I carry out this exercise in 
Figure~\ref{figage}, once again using models with
different metallicities and mass-loss rates. 
The [3.6 - 4.5] color data (top row) show a nearly flat behavior with age, with the
only possible exception being the pre-SWB cluster, which appears to have excess reddening.
The flat behavior of superclusters from SWB II to VI --the age range sensitive to 
changes in $\dot M$-- is best reproduced, for the low metallicities of the Magellanic clusters,
by fiducial $\dot M/5$, although $\dot M$ cannot be excluded, given measurement errors
and model uncertainties. Conversely, models with $\dot M$ and $5 \times \dot M$
appear to almost equally well reproduce [3.6 - 8], [5.8 - 8], and [8 - 24] color data. In particular,
at [3.6 - 8] and [5.8 - 8] (second and third rows),
the SWB VI cluster is slightly more consistent
with the lowest $Z$, 5 $\times \dot M$ models; the peak at 1--2 Gyr survives for models with $Z$ = 0.002, i.e., 
those closest to the metallicity of the SWB VI cluster.
In conclusion, although models are compatible with the observations, integrated colors cannot strongly constrain
the mass-loss rate, given the present data and theoretical uncertainties.
In the next section, I will discuss the relationship between surface brightness fluctuations 
and mass-loss rates in global stellar populations.

\begin{figure*}
\begin{tabular}{lll}
\hspace*{-0.3cm}\includegraphics[scale=0.23]{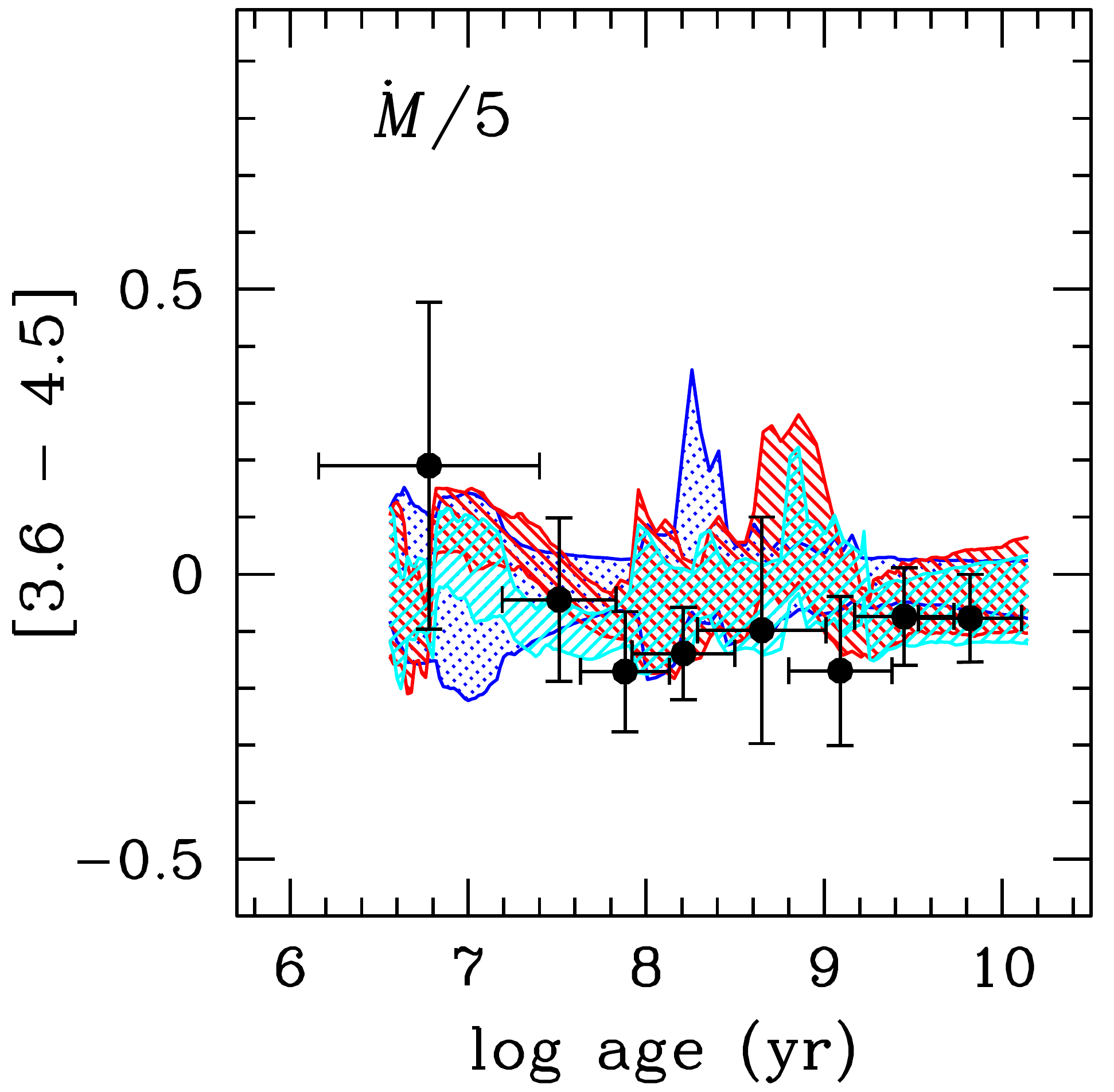}
&
\includegraphics[scale=0.23]{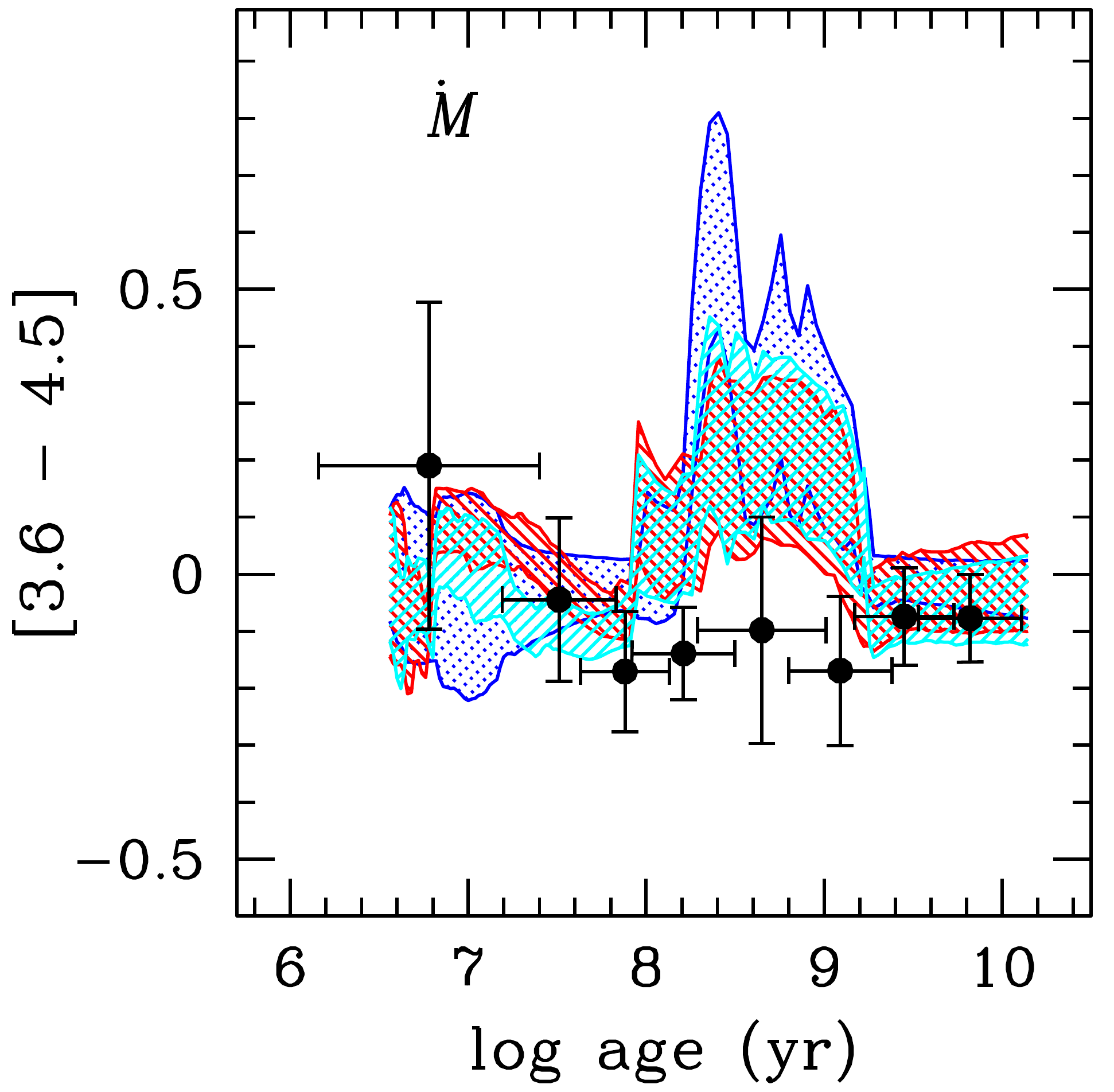}
&
\includegraphics[scale=0.23]{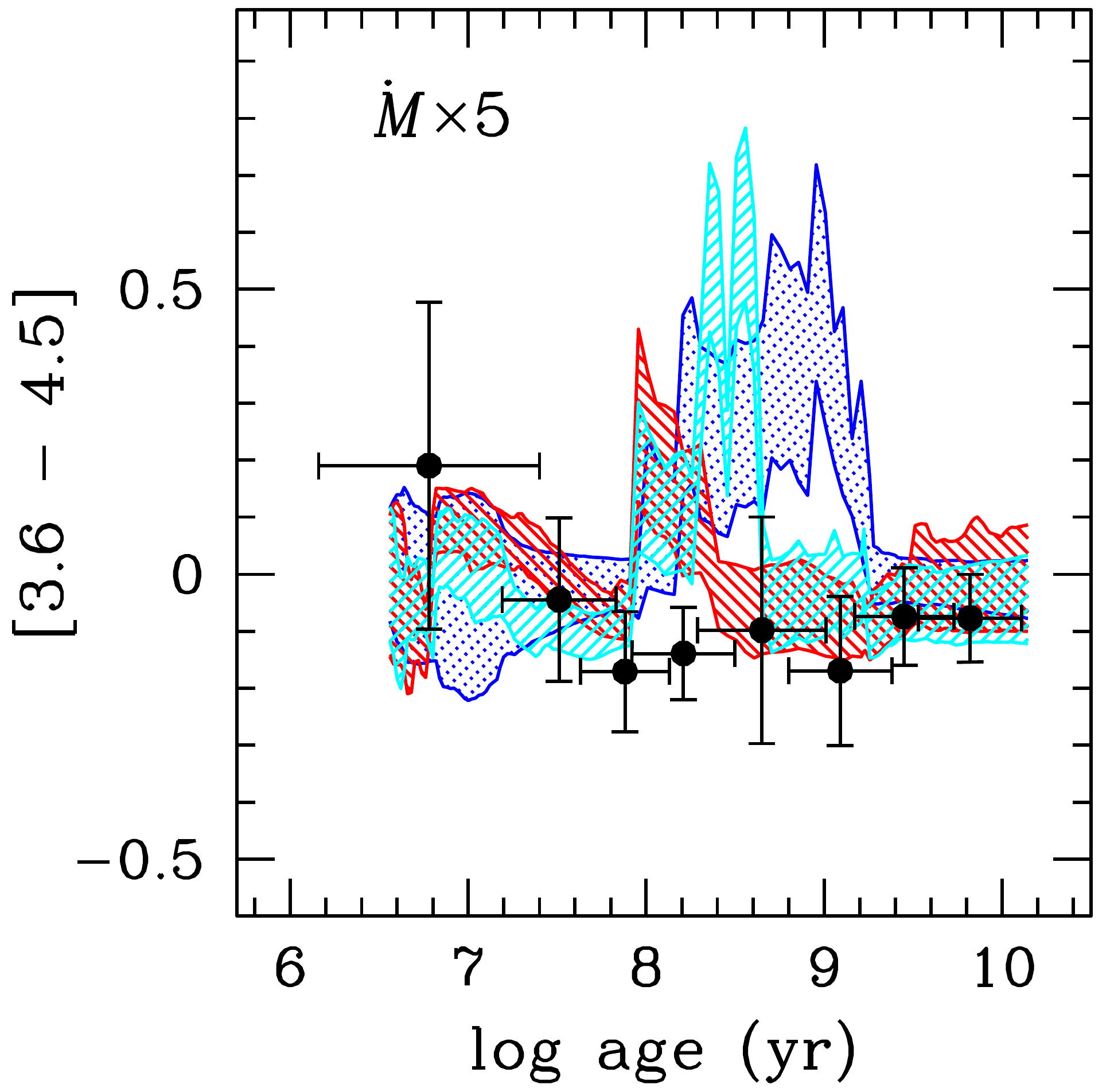}
\end{tabular}

\vspace*{0.3cm}

\begin{tabular}{lll}
\hspace*{-0.3cm}\includegraphics[scale=0.23]{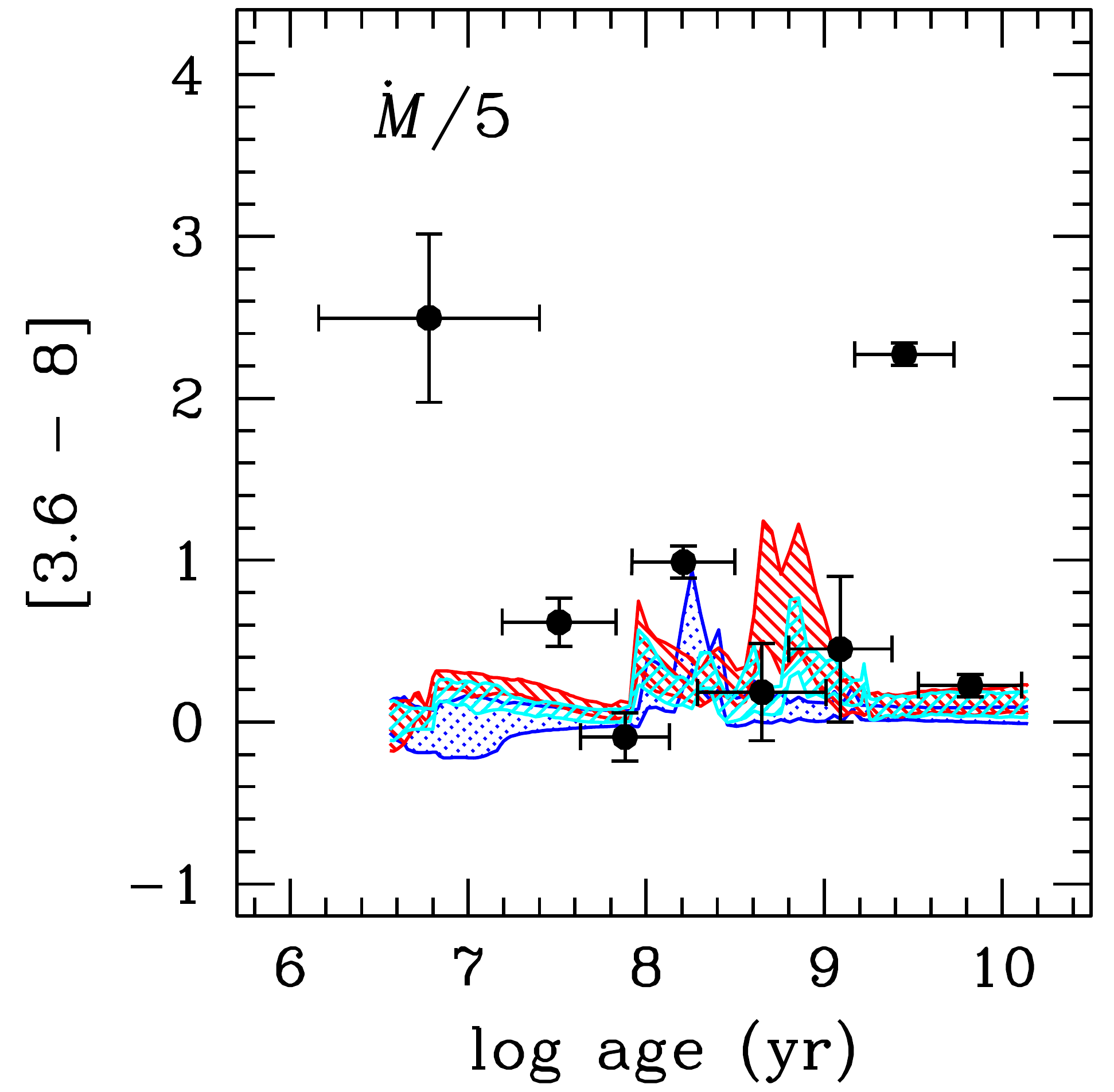}
&
\includegraphics[scale=0.23]{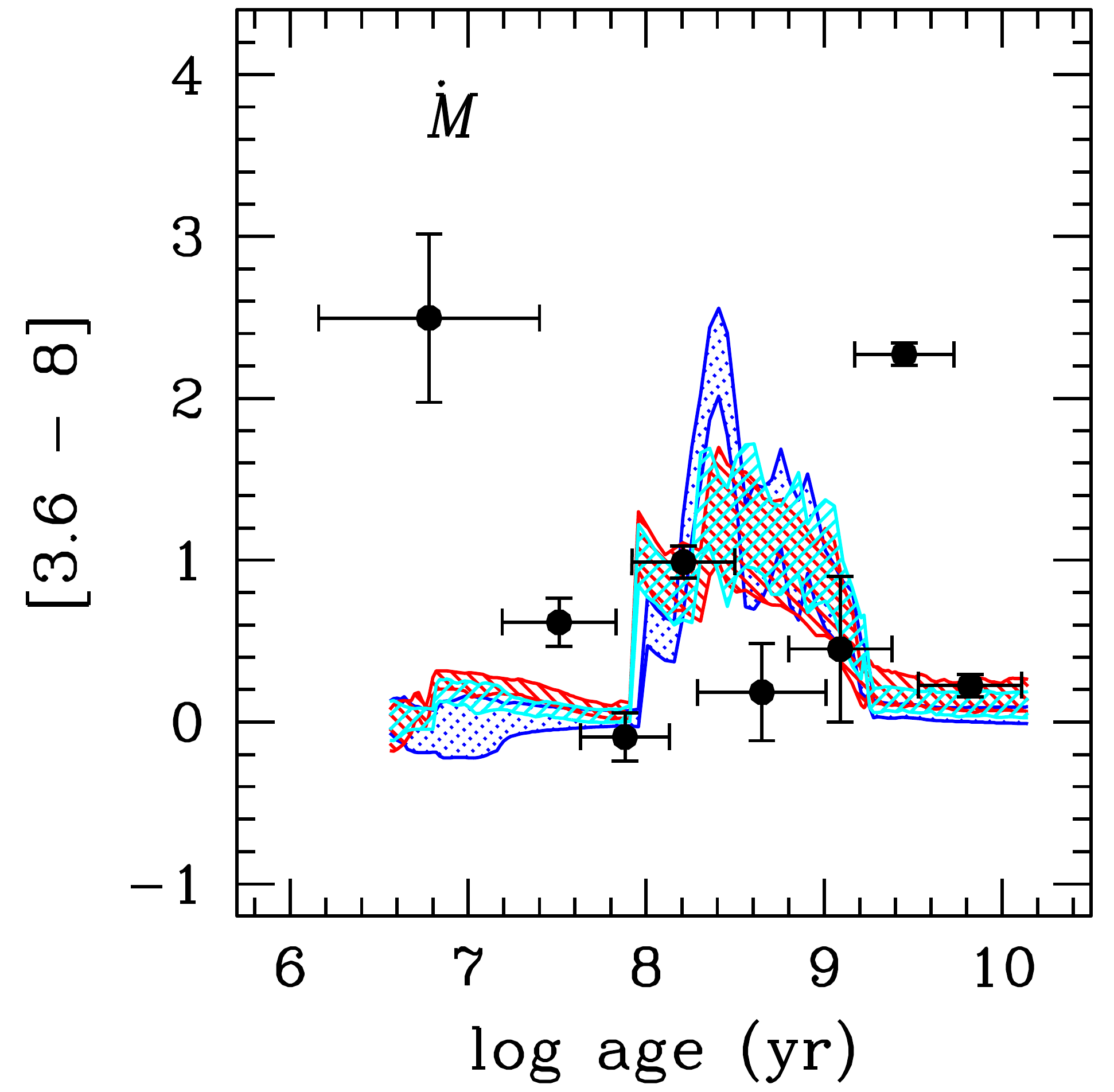}
&
\includegraphics[scale=0.23]{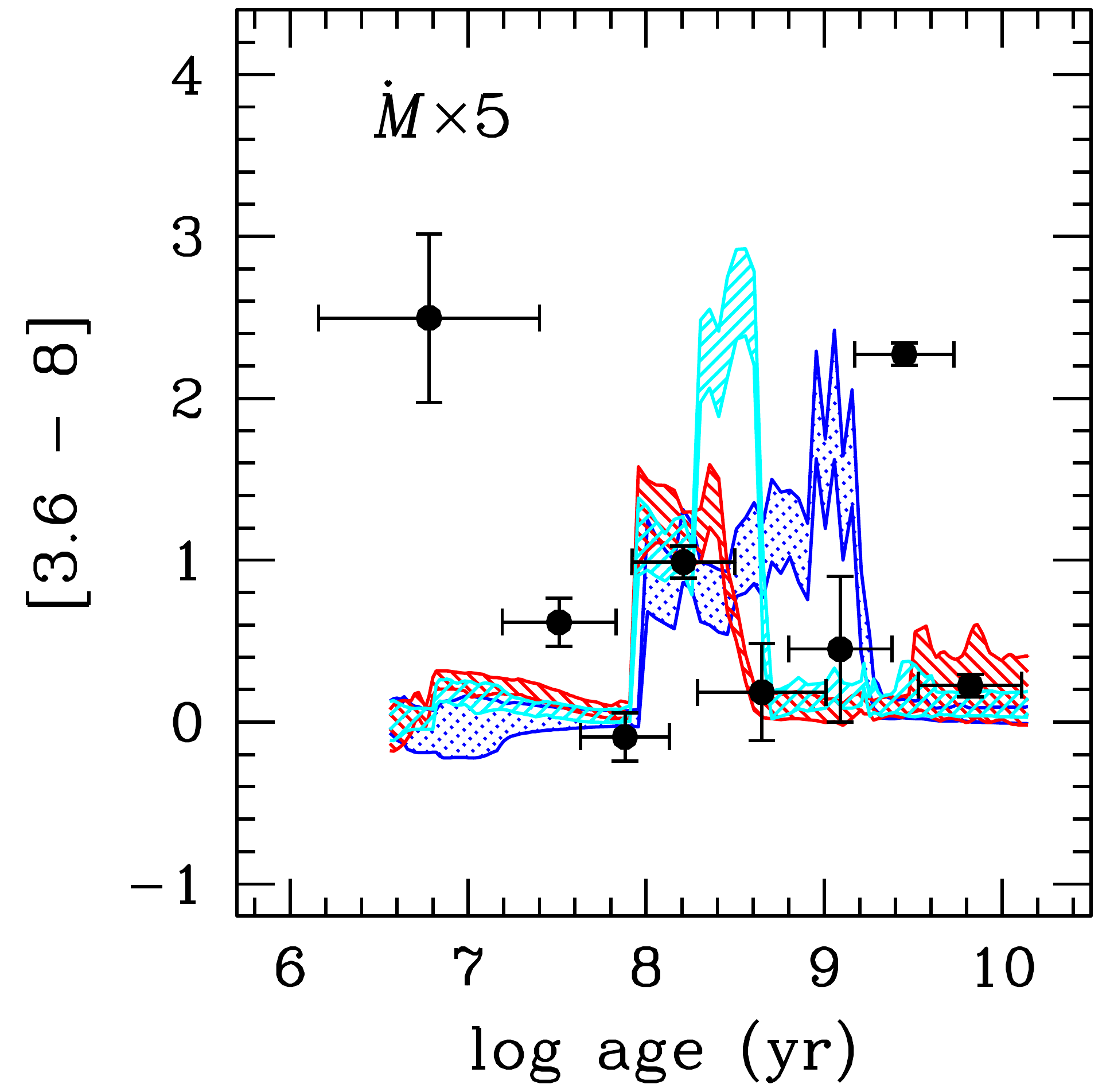}
\end{tabular}

\vspace*{0.3cm}

\begin{tabular}{lll}
\hspace*{-0.3cm}\includegraphics[scale=0.23]{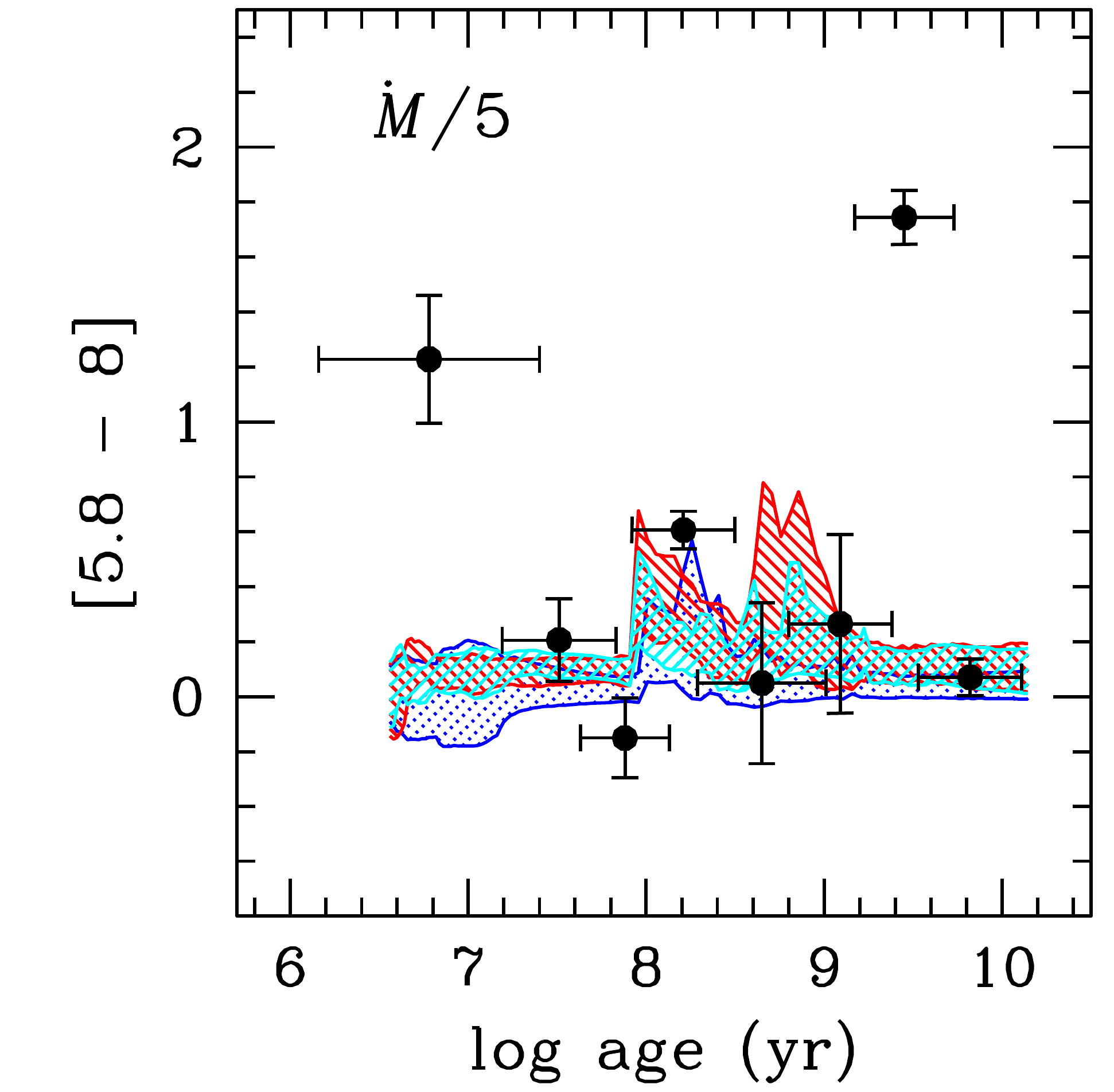}
&
\includegraphics[scale=0.23]{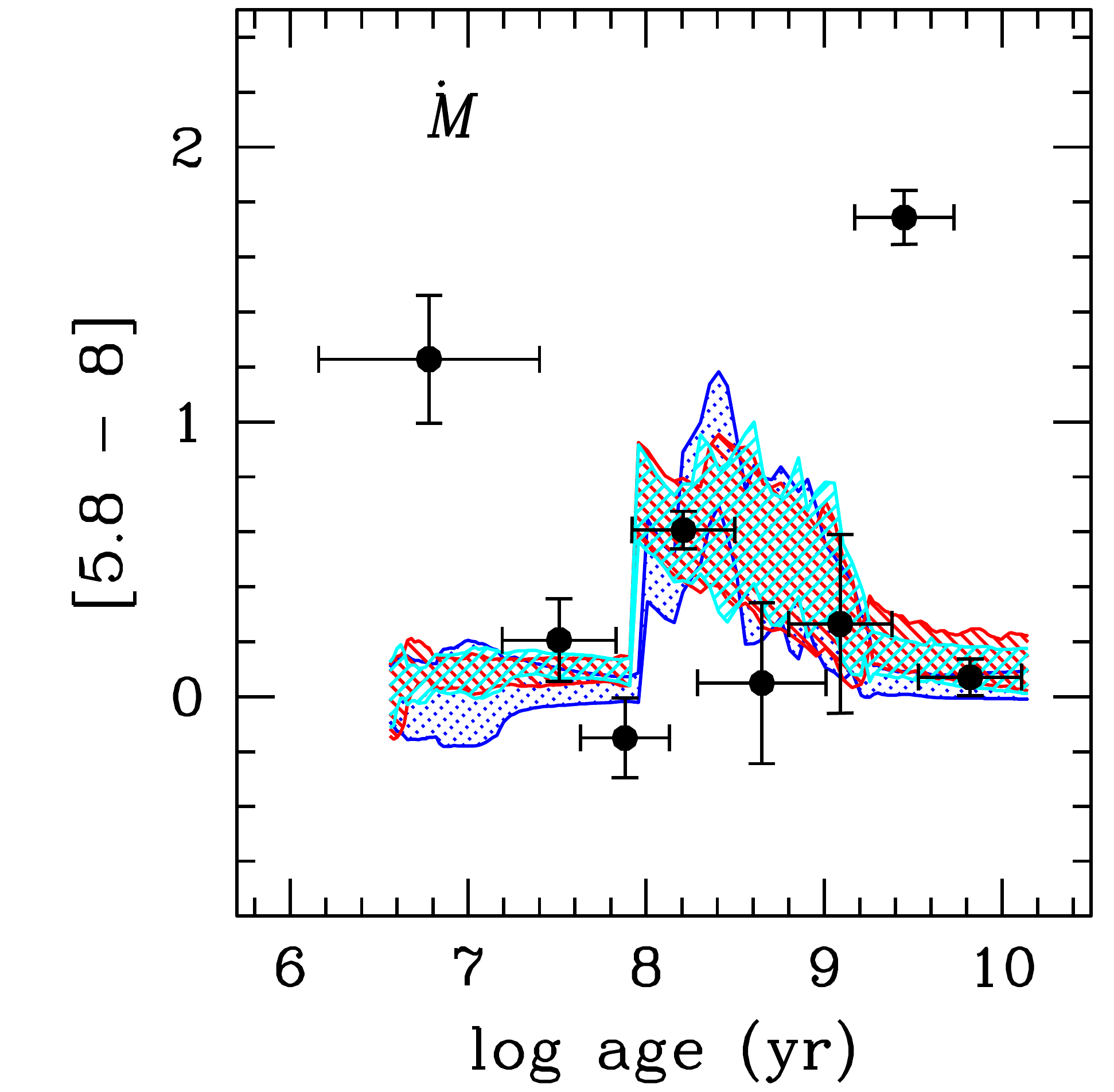}
&
\includegraphics[scale=0.23]{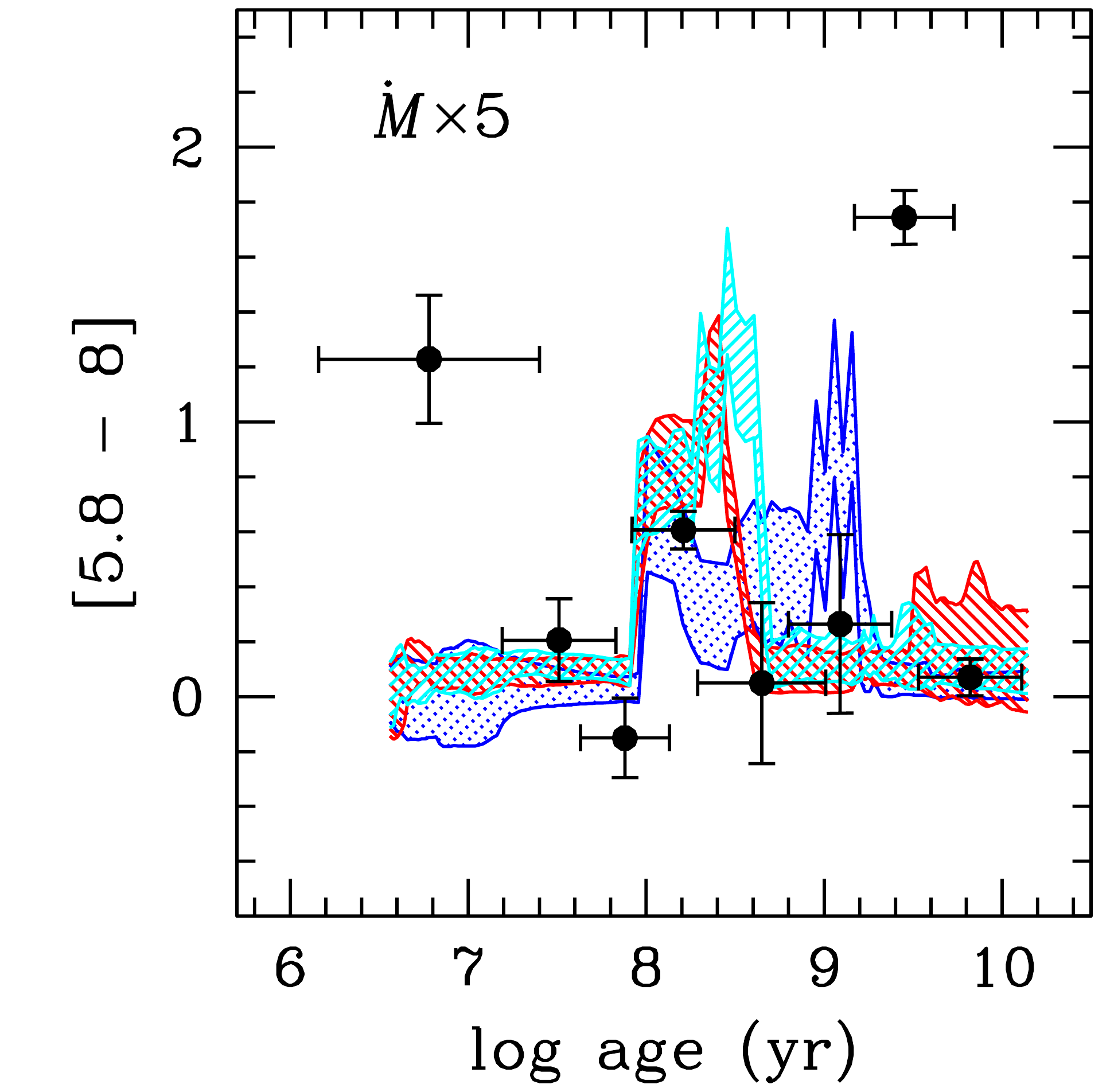}
\end{tabular}

\vspace*{0.3cm}

\begin{tabular}{llll}
\hspace*{-0.3cm}\includegraphics[scale=0.23]{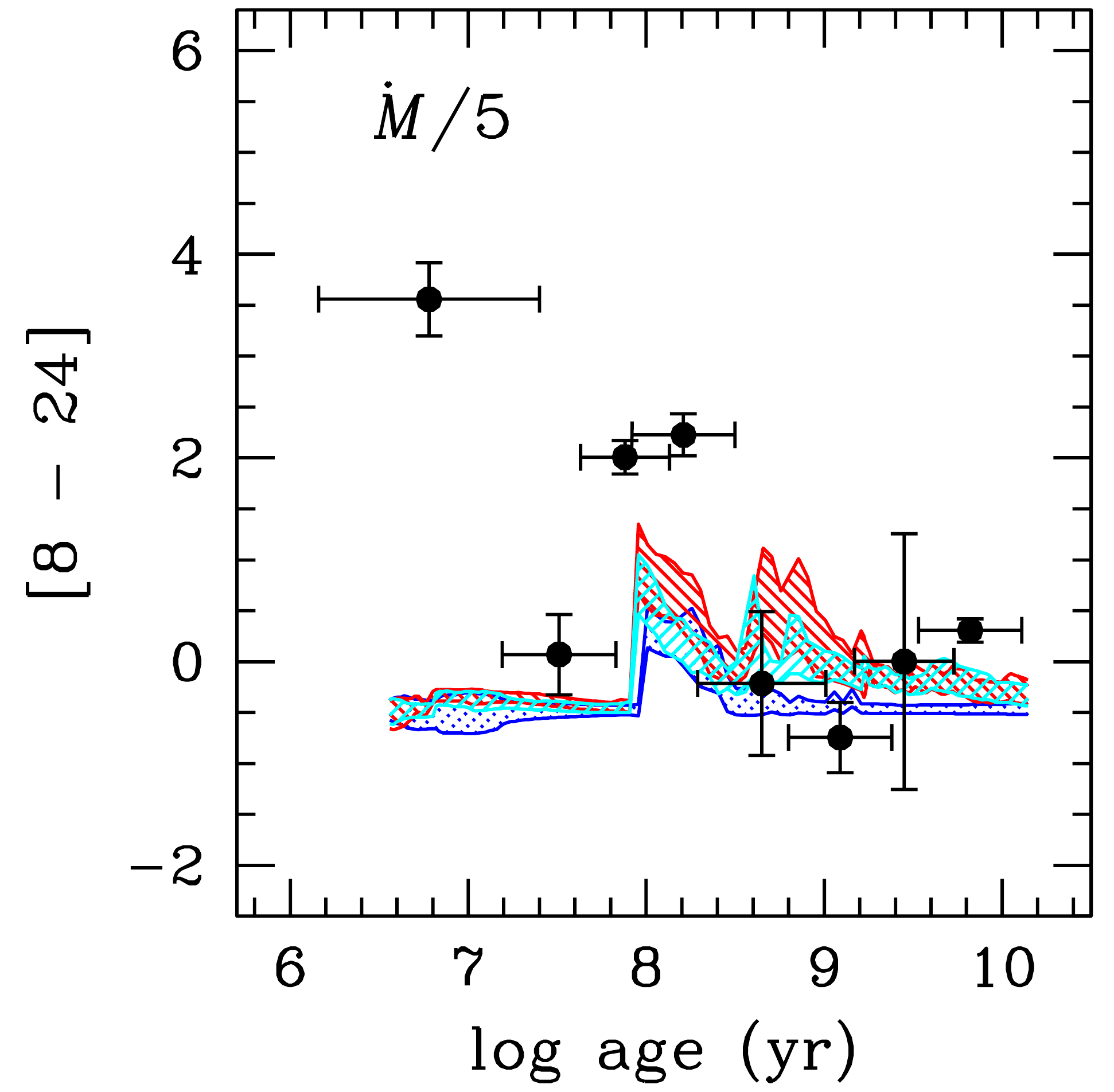}
&
\includegraphics[scale=0.23]{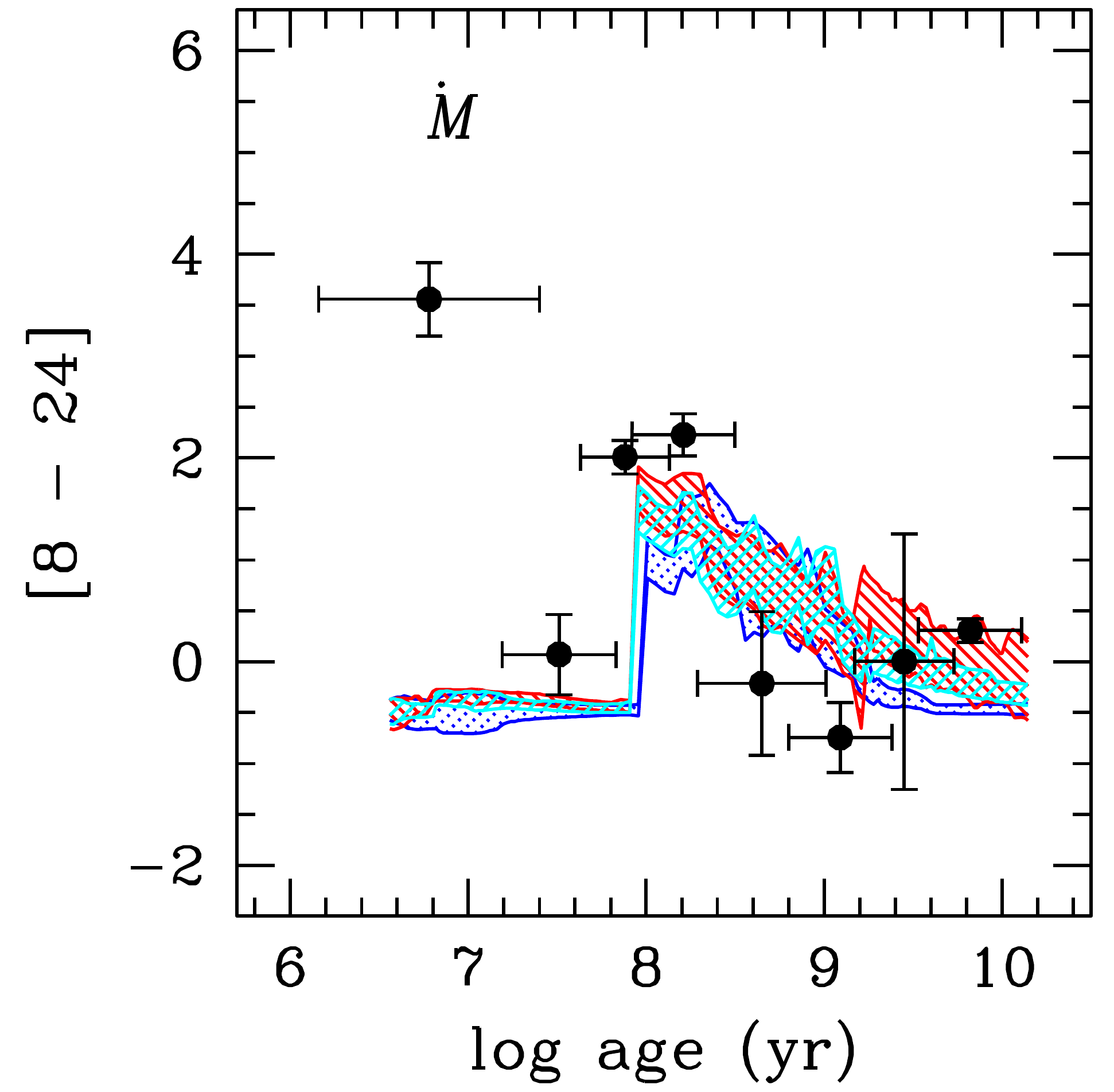}
&
\includegraphics[scale=0.23]{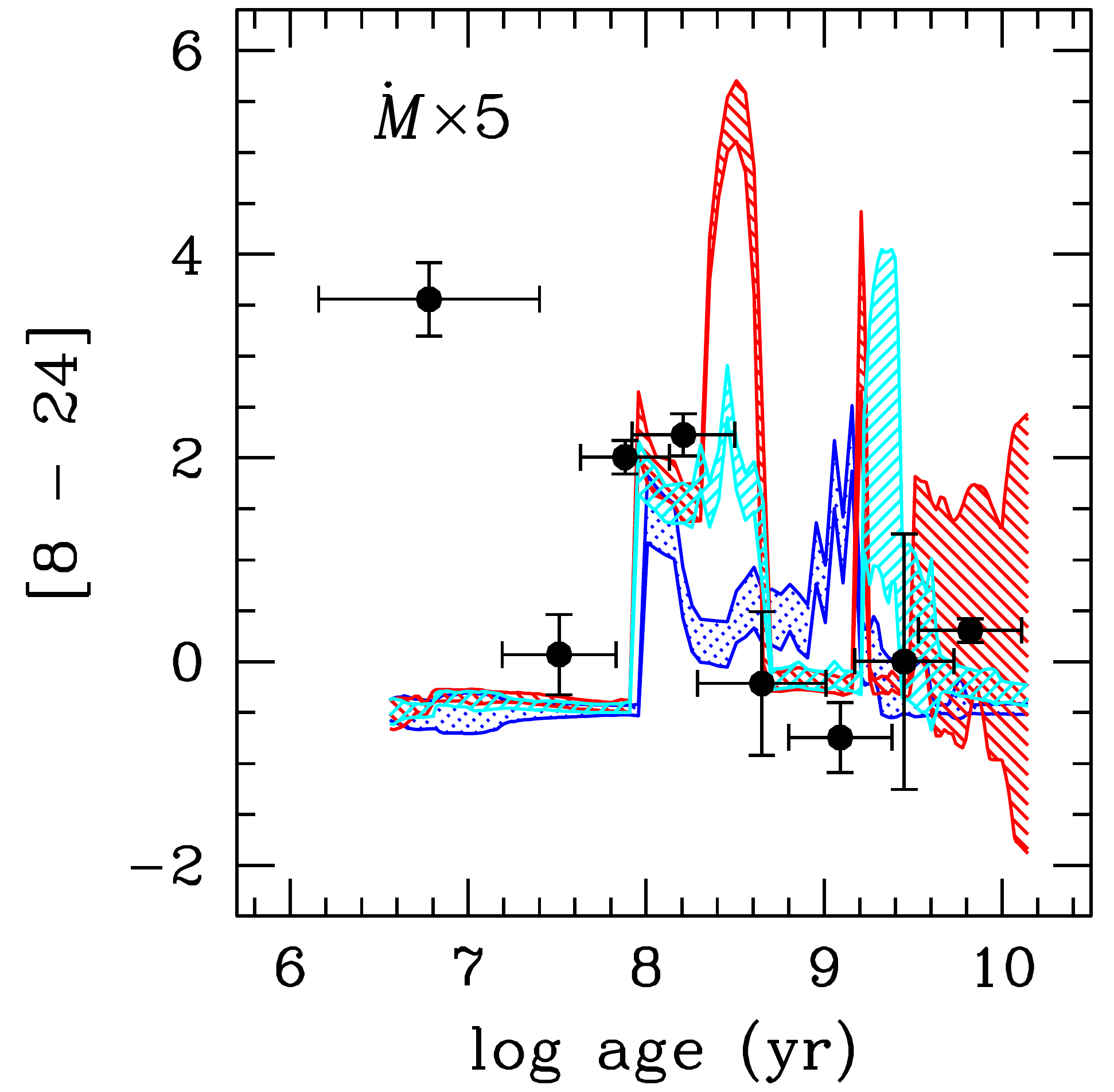}
&
\includegraphics[scale=0.23]{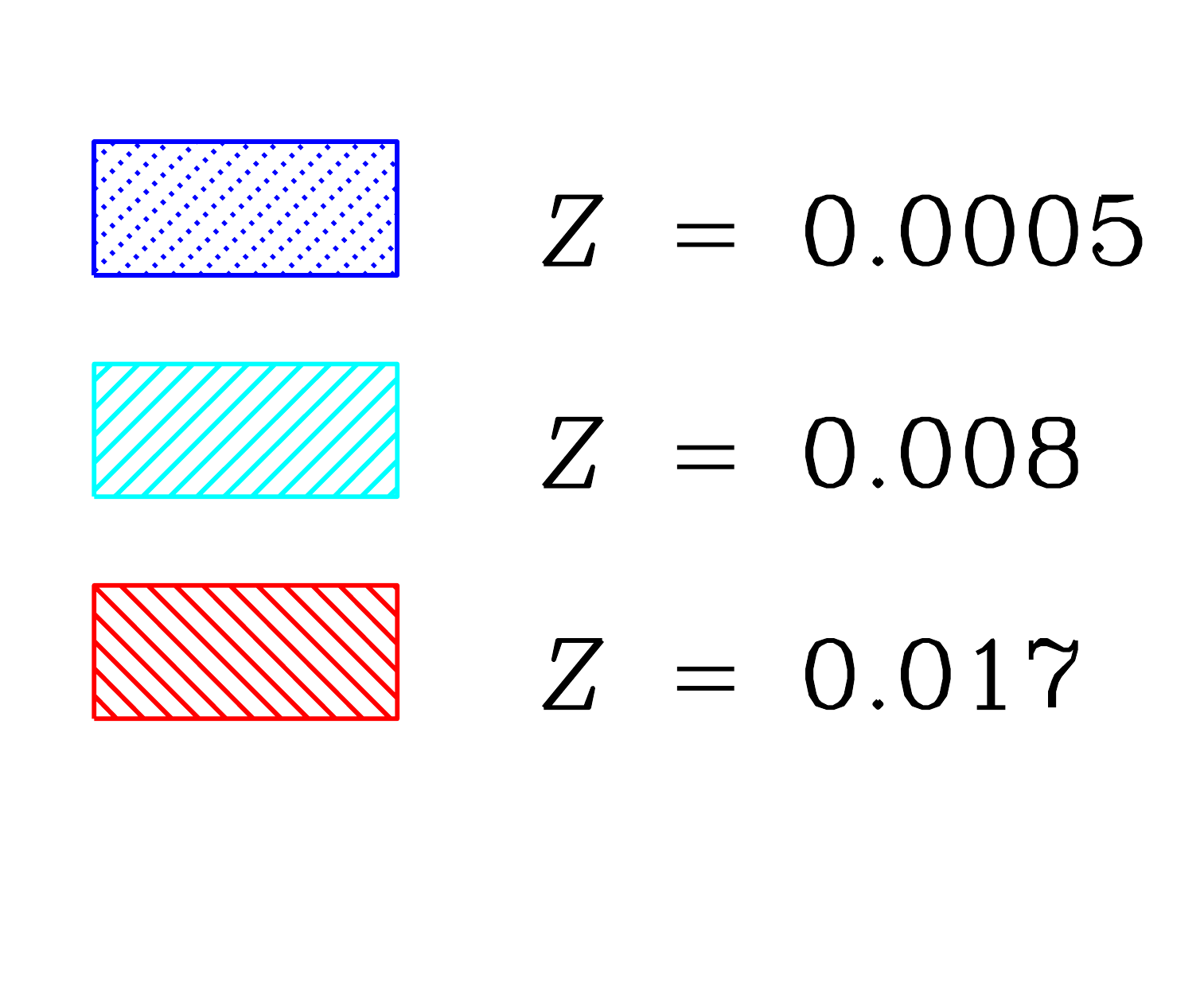}
\end{tabular}
\caption{Comparison between models and MC ``superclusters," colors versus log (age).
From top to bottom: [3.6 - 4.5], [3.6 - 8],
[5.8 - 8], [8 - 24]. 
The symbols and models have the same meanings as in Fig.\ \ref{piov_sclust_5.8_8vs3.6_4.5}. 
}
\label{figage}
\end{figure*}

\section{Surface Brightness Fluctuations.} \label{sec:sbfs}

The fluctuation magnitude (\barm) is the ratio between the variance and the mean
of the stellar luminosity function \citep{tonr88,tonr90}, normalized by
$(4 \pi d^2)$, where $d$ is the distance. This can be 
expressed by the following equation:

\begin{equation}
\barm = -2.5\ {\rm log}\ \frac {1}{4 \pi d^2} \frac{\Sigma  n_i l_i^2}{\Sigma n_i l_i} + {\rm zero point}, 
\label{theequation}
\end{equation}

\noindent where $n_i$ and $l_i$ are
the number of stars of type $i$ and their luminosity, respectively. 

Because of the dependency on the square of the stellar luminosity, the sum in the numerator is
dominated by the brightest stars. Hence,
\barm\ is especially sensitive to and informative about the most luminous stars in a given band and  
at a particular evolutionary phase of a population. TP-AGB stars in the mid-IR are clearly a case in point. 
In what follows, I will calculate the fluctuation magnitudes of the MC superclusters in the {\it Spitzer} IRAC bands 
and MIPS [24] filter. 

The integrated fluxes of the superclusters, calculated as described in Section\ \ref{subsec:superclust}, 
were used for the denominator of eq.~\ref{theequation}
--the sum of $n_i l_i$ converges slowly, and low-mass stars fainter than the detection limit
cannot be added individually.
The sum in the numerator, on the other hand, converges quickly, and was found by summing the flux, squared,
of resolved bright stars in the SAGE Winter '08 IRAC Epoch 1 and Epoch 2 Archive, 
and the SAGE Winter '08 MIPS 24 $\mu$m Epoch 1 and Epoch 2 Catalog (IPAC 2009).\footnote{
\url{http://irsa.ipac.caltech.edu/cgi-bin/Gator/nph-scan?mission=irsa&submit=Select&projshort=SPITZER}.
}
Before adding the squared fluxes in the numerator, though, the field stars were statistically removed following the procedure
presented in \citet{migh96}, and described also in \citet{gonz10} for 2MASS near-IR data. 
Basically, the [3.6  - 8] versus [8] CMD of the stars within 1$\arcmin$ of the
supercluster center (the ``cluster region," which presumably includes both cluster and
field stars) is compared to the CMD of the stars in the annulus with 2\farcm0 $< r \leq$
2\farcm5 (i.e., the ``field"). For each star in the cluster region with 
mag [8] $\pm \sigma_{[8]}$ and color [3.6  - 8] $\pm \sigma_{[3.6  - 8]}$,
I count the number of stars in the same CMD with [3.6  - 8] colors within
$\pm$MAX(2$\sigma_{[3.6  - 8]}$,0.100) mag and [8] mag within
$\pm$MAX(2$\sigma_{[8]}$,0.200) mag. I call this number $N_{\rm scl}$.
I also count the number of stars in the field CMD within the same
$\Delta$ [8] by $\Delta$ [3.6  - 8] bin determined from the cluster star.
This is  $N_{\rm fld}$. The probability $p$ that the star
in the cluster region CMD actually belongs to the supercluster
can be expressed as

\begin{equation}
p \approx 1 - {\rm MIN}\left( \frac{\alpha(N_{\rm
fld}+1)}{N_{\rm scl}+1},1.0\right),
\label{probdec}
\end{equation}

\noindent
where $\alpha$, in this case 0.44, is the ratio of the area of the cluster region ($\pi$ arcmin$^2$)
to the area of the field region (2.25 $\pi$ arcmin$^2$).
Once $p$ is calculated for a given star, it is compared
to a randomly drawn number $0 \leq p^\prime \leq 1$. If $p \geq p^\prime$,
the star is
accepted as a supercluster member; otherwise, it is 
considered as a field object and rejected.

Figures~\ref{cmdsI} and~\ref{cmdsII} display the [3.6 - 8] versus [8] CMDs for our 8 
``superclusters." The decontaminated cluster sources are represented with black solid dots, while the 
contaminating field stars are shown with gray dots.   
Theoretical isochrones from CB$^*$ models have been
overplotted on the decontaminated sources as solid brown lines.
The mean ages and metallicities of the model isochrones are approximately the same as 
those reported in Table\ \ref{latabla}, 
except for the pre-SWB supercluster, for which
the isochrone is a few Myr older (10 Myr versus 6 Myr).
For comparison, analogous [$J - K_s$] versus $K_s$ CMDs, 
from 2MASS data,
are shown to the right of each mid-IR graph. 
Photometric errors as a function of magnitude for individual stars 
are shown in Figure~\ref{photerrind}; from top to bottom: $\Delta$ [3.6], $\Delta$ [8], 
$\Delta\ J$, and $\Delta\ K_s$.

There are not enough 24 $\mu$m sources to perform a meaningful decontamination of the MIPS data
following the above procedure. Instead, I determine the cluster star list in this band by matching the 
sources in the 
``cluster region" with the decontaminated IRAC sample. 

\begin{figure}
     \centering
     \subfigure[ Pre-SWB. Isochrone $Z =$ 0.008; age = 10 Myr.  
     ]{
          \label{cmdpre}
          \includegraphics[width=.45\textwidth]{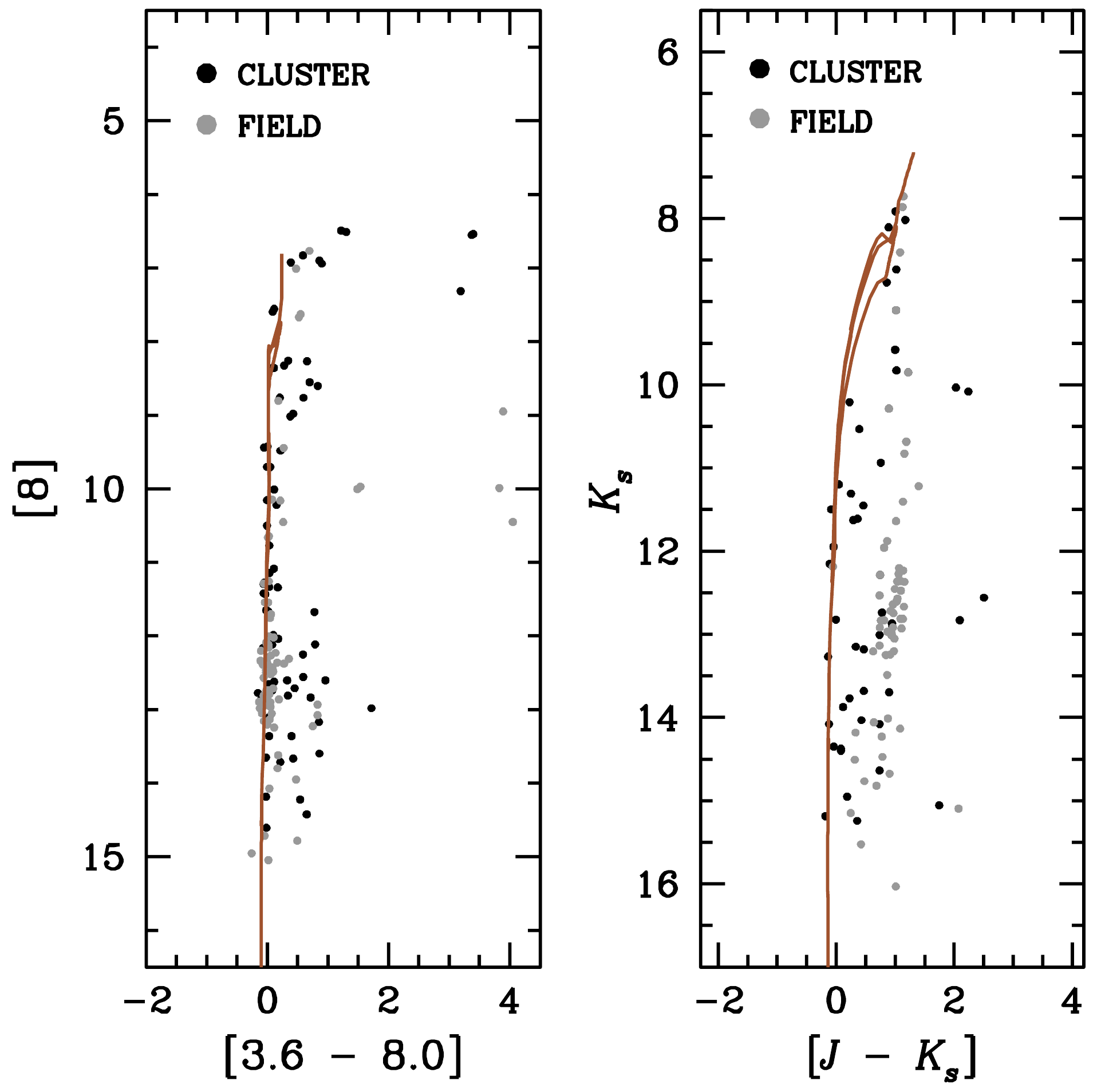}
           }
     \subfigure[  SWB I. Isochrone $Z =$ 0.008; age =  32 Myr. 
         ]{
          \label{cmdone}
          \includegraphics[width=.45\textwidth]{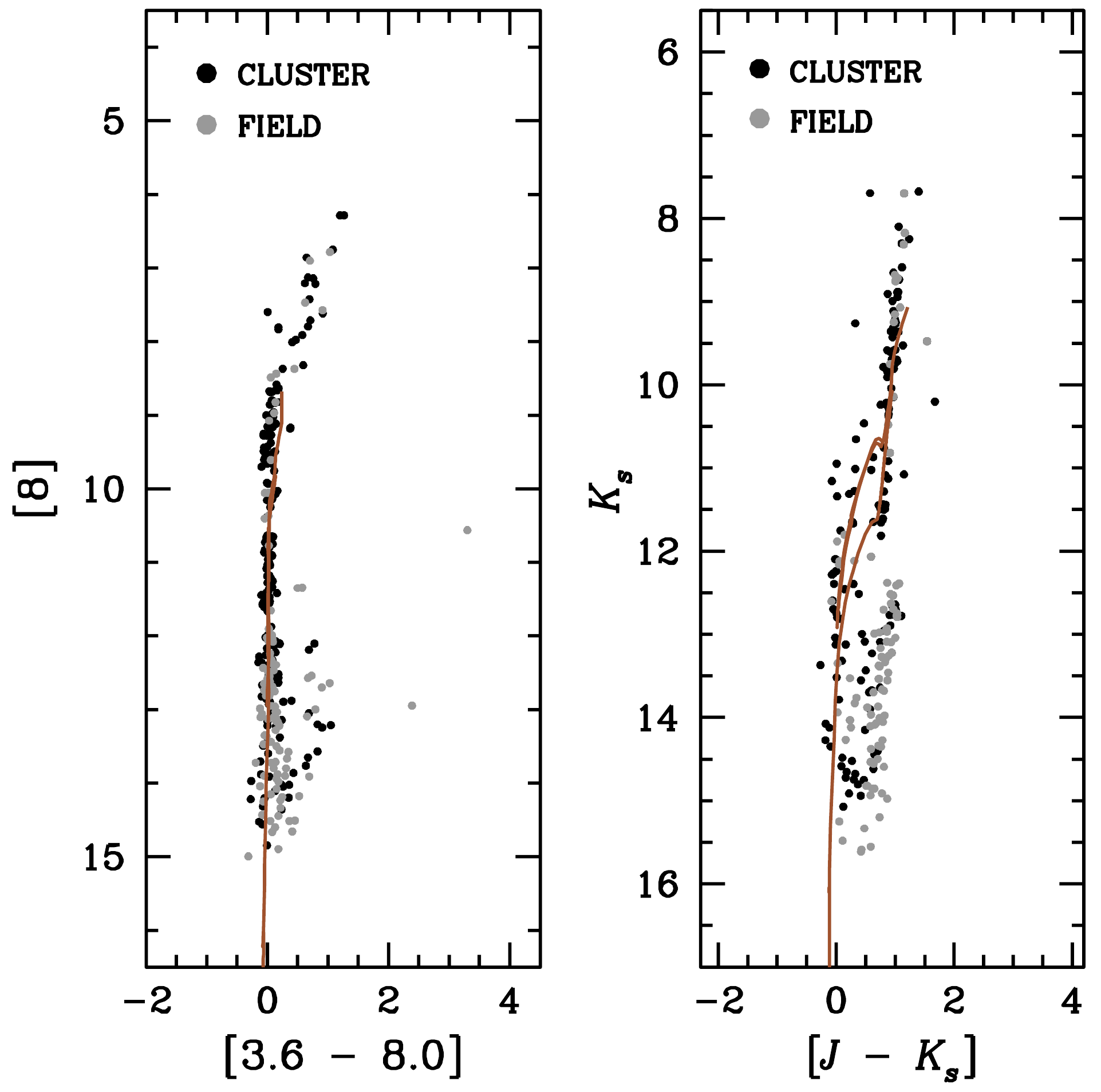}
             }\\
     \vspace{.3in}
     \subfigure[ SWB II. Isochrone $Z =$ 0.008; age = 76 Myr. 
     ]{
           \label{cmdtwo}
           \includegraphics[width=.45\textwidth] {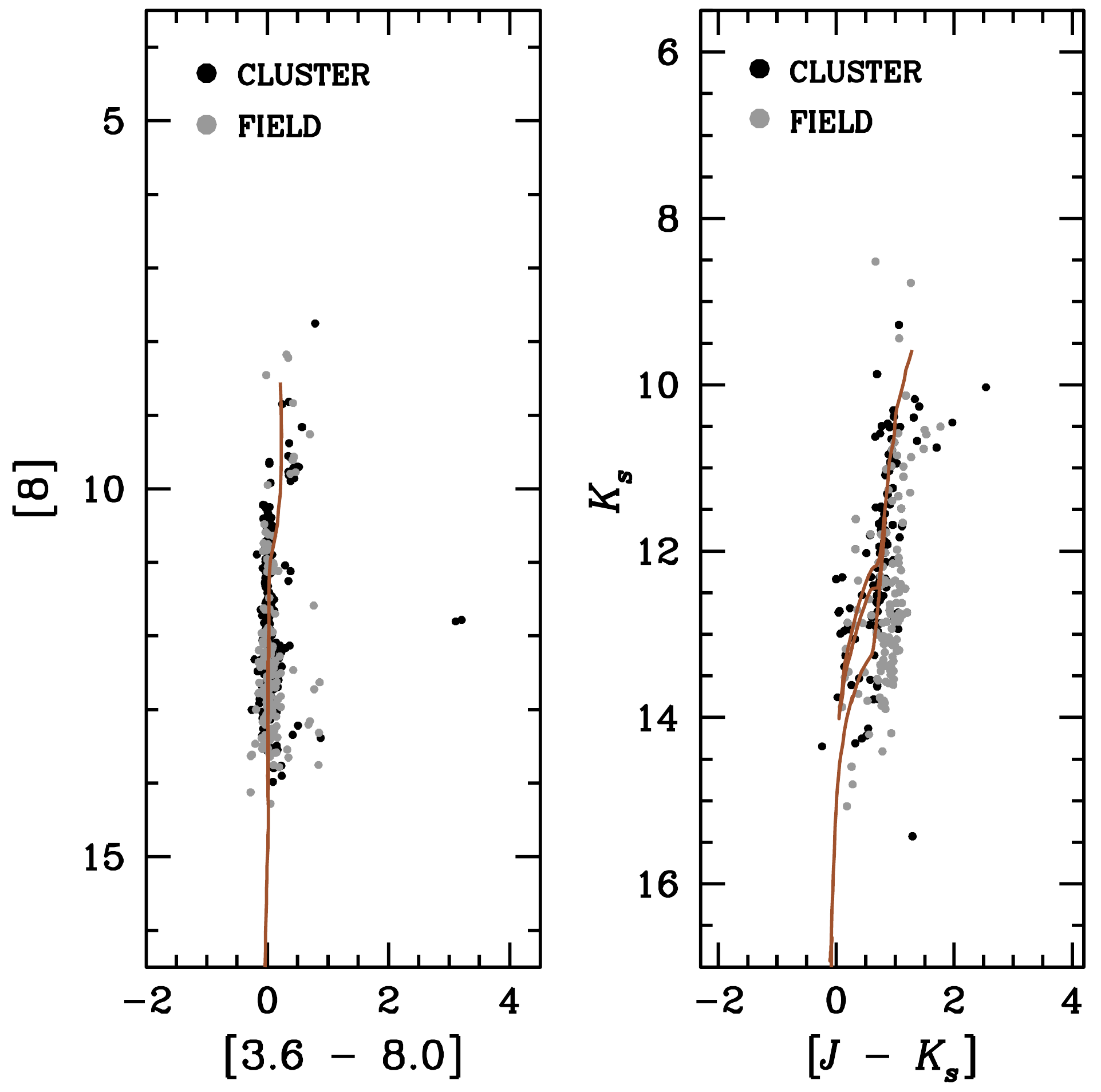}
            }
     \subfigure[SWB III. Isochrone $Z =$ 0.008; age = 160 Myr. ]{
           \label{cmdthree}
          \includegraphics[width=.45\textwidth]{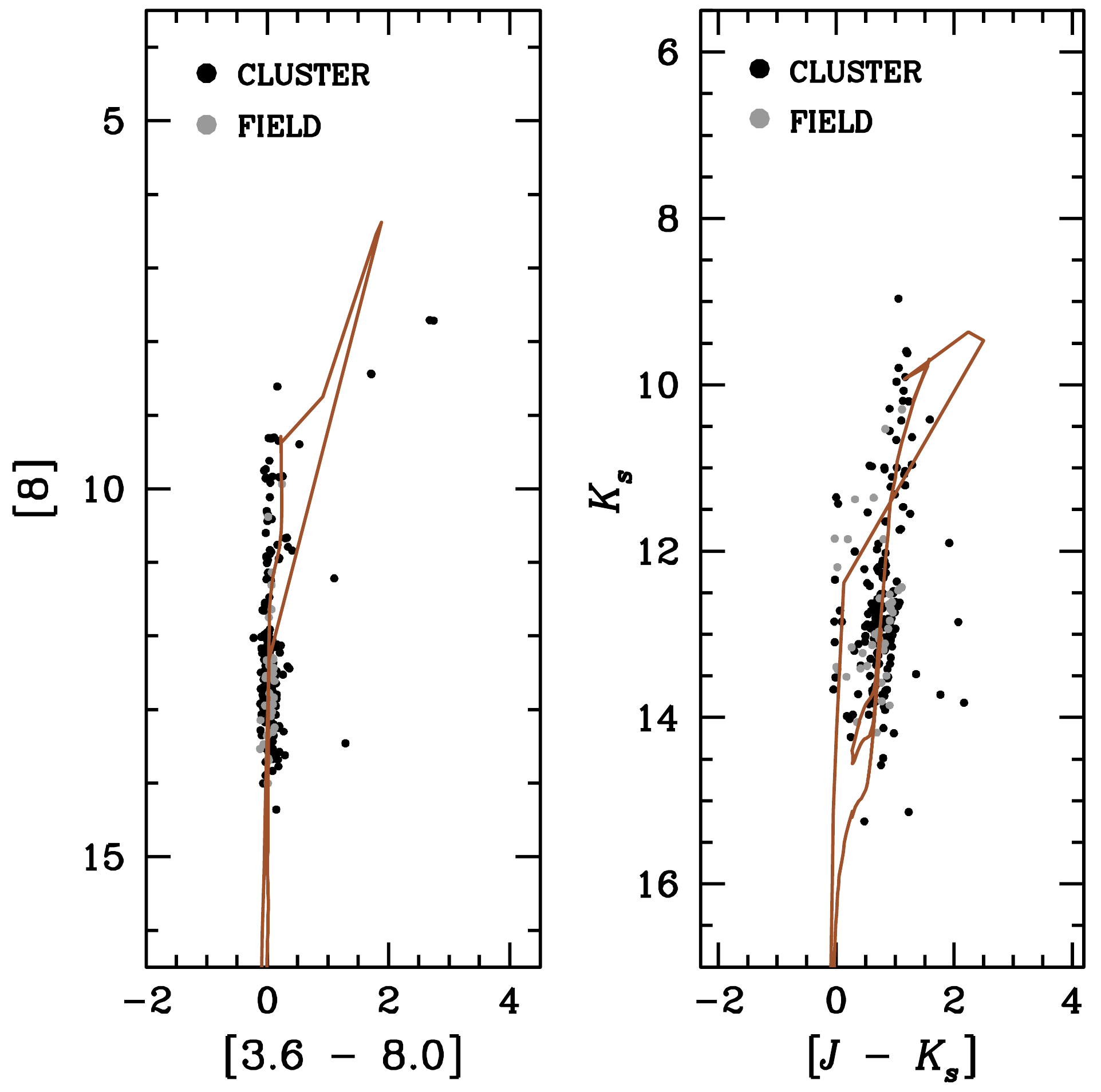}
          }
     \caption{ Color-magnitude diagrams of MC superclusters, SWB classes pre-SWB (top left), 
I (top right), II (bottom left), and  III (bottom right). Left panels: apparent 8 $\mu$m mag, 
[8], vs.\ [3.6 - 8] color;
right panels: apparent $K_s$-band mag vs.\ [$J$ - $K_s$] color. 
Solid black dots:
Supercluster decontaminated stars within 60$\arcsec$ from the center (at the distance of the LMC); 
solid gray dots: contaminating field stars (see the text). 
Solid brown lines: theoretical isochrones from CB$^*$ models, with indicated ages and metallicities. 
These are approximately the same as those estimated for the superclusters (Table \ref{tabclust}), except for the pre-SWB supercluster, for which
the isochrone is a few Myr older (10 Myr versus 6 Myr; see the text). 
           }
     \label{cmdsI}
\end{figure}

\begin{figure}
     \centering
     \subfigure[ SWB IV. Isochrone $Z =$ 0.004; age = 450 Myr. 
     ]{
          \label{cmdfour}
          \includegraphics[width=.45\textwidth]{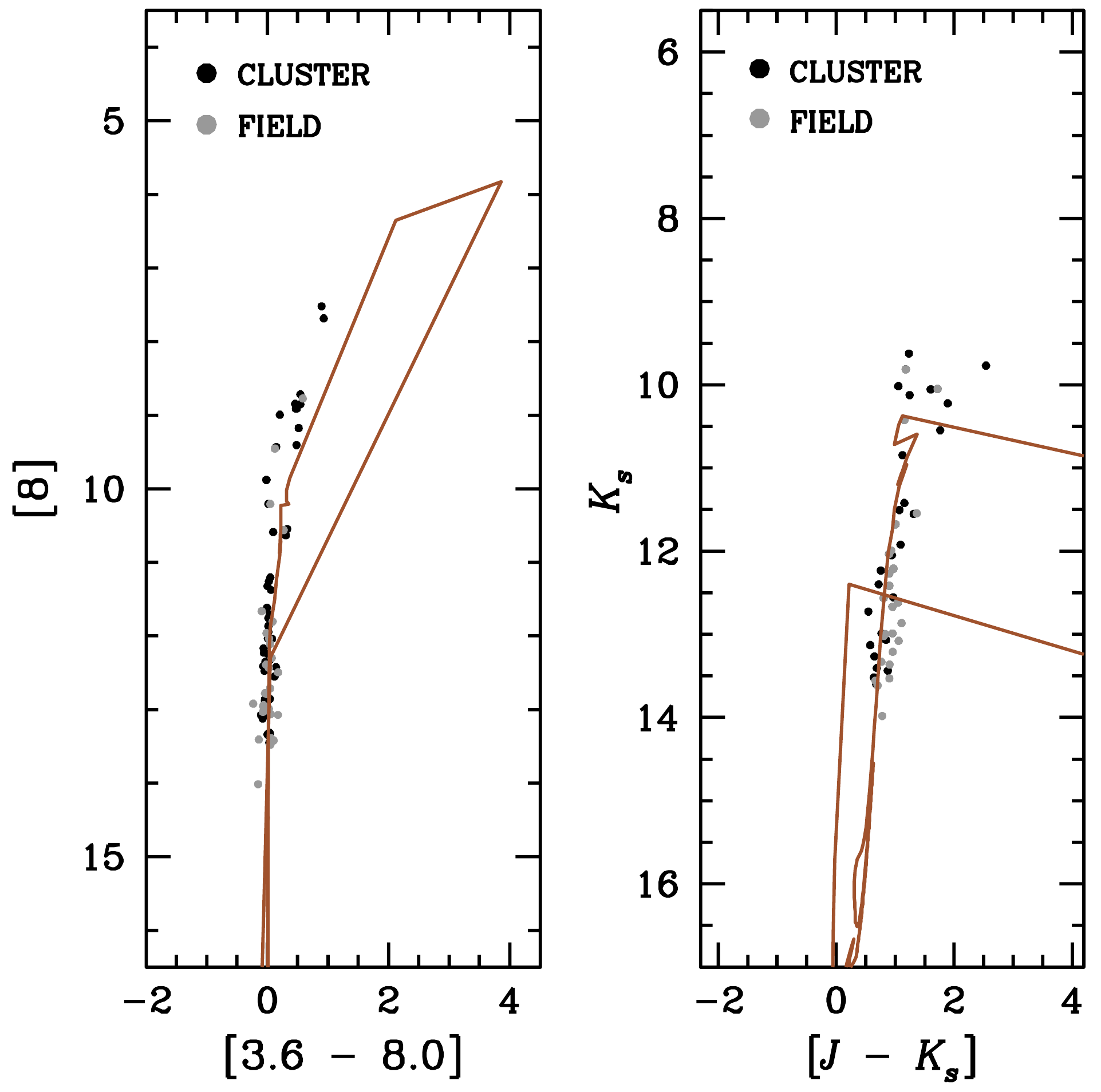}
           }
     \subfigure[ SWB V. Isochrone $Z =$ 0.004; age = 1.2 Gyr.
         ]{
          \label{cmdfive}
          \includegraphics[width=.45\textwidth]{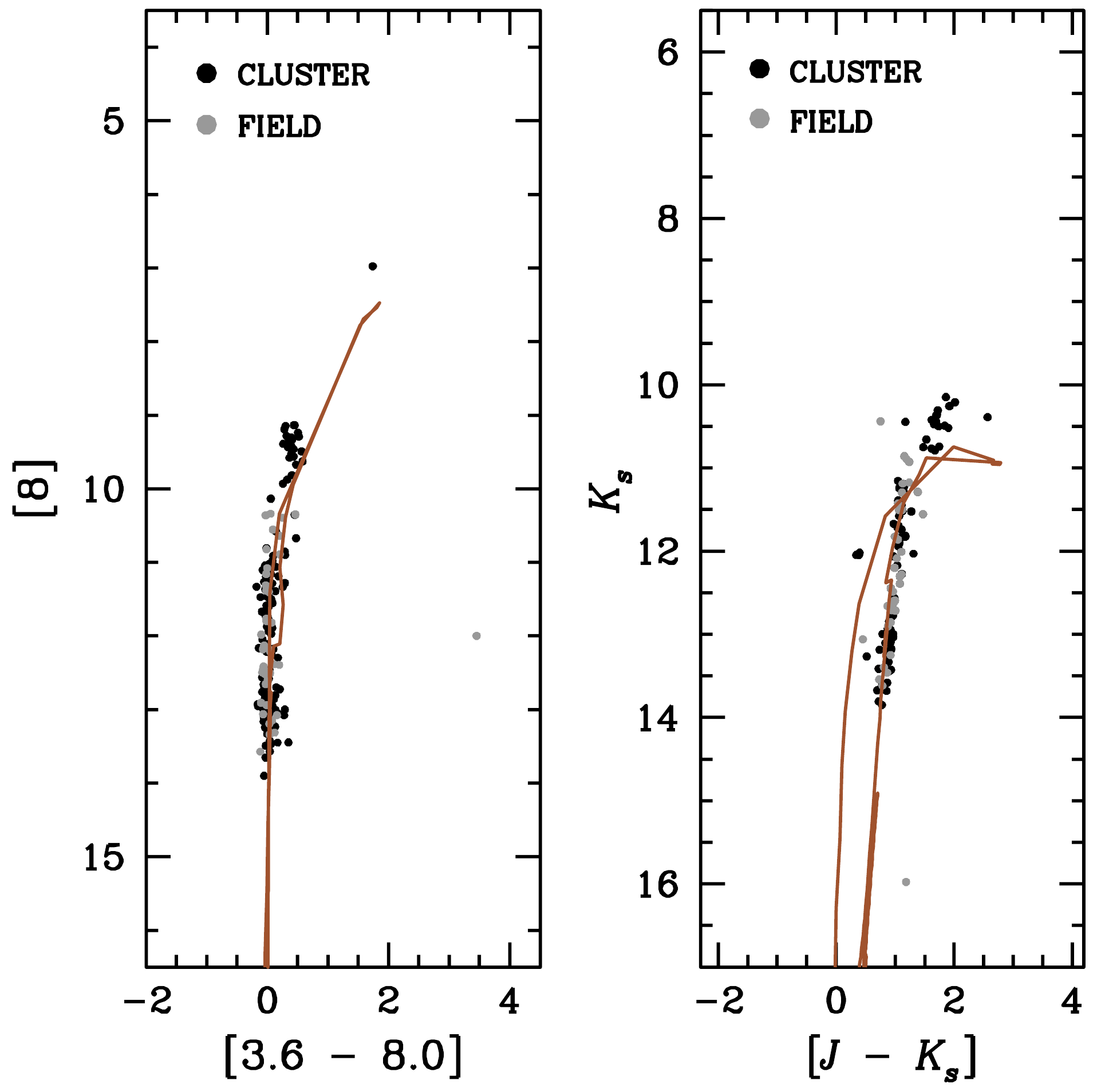}
             }\\
     \vspace{.3in}
     \subfigure[ SWB VI.  Isochrone $Z =$ 0.002; age = 2.8 Gyr.
     ]{
           \label{cmdsix}
          \includegraphics[width=.45\textwidth] {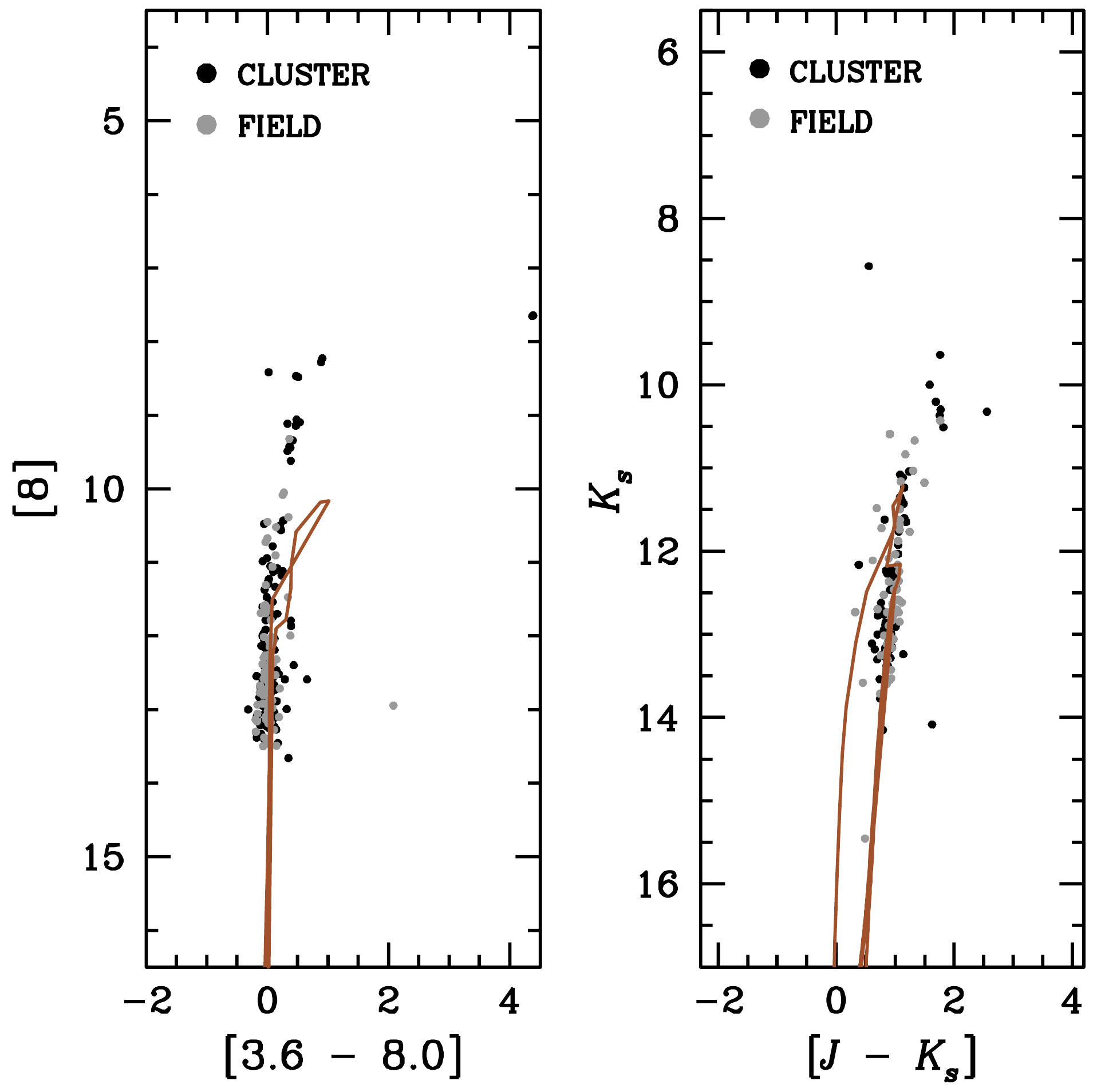}
            }
     \subfigure[SWB VII. Isochrone $Z =$ 0.001; age = 6.5 Gyr.
     ]{
           \label{cmdseven}
          \includegraphics[width=.45\textwidth]{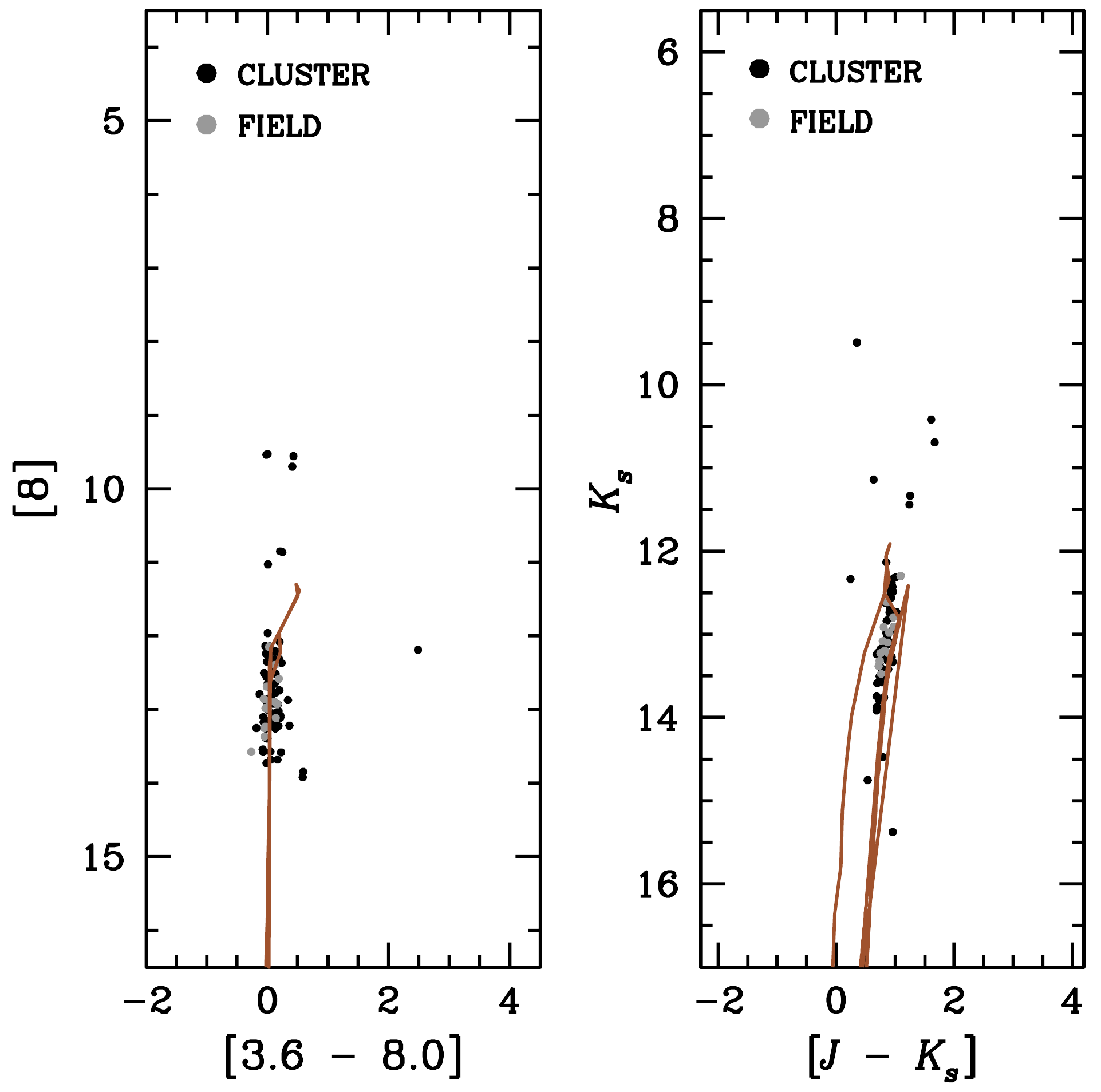}
          }
     \caption{
Color-magnitude diagrams of MC superclusters, SWB classes IV (top left), 
V (top right),  VI (bottom left), and  VII (bottom right). 
The models and symbols have the same meanings as in Figure \ref{cmdsI}.
}
     \label{cmdsII}
\end{figure}

\begin{figure}
\includegraphics[width=.75\textwidth] {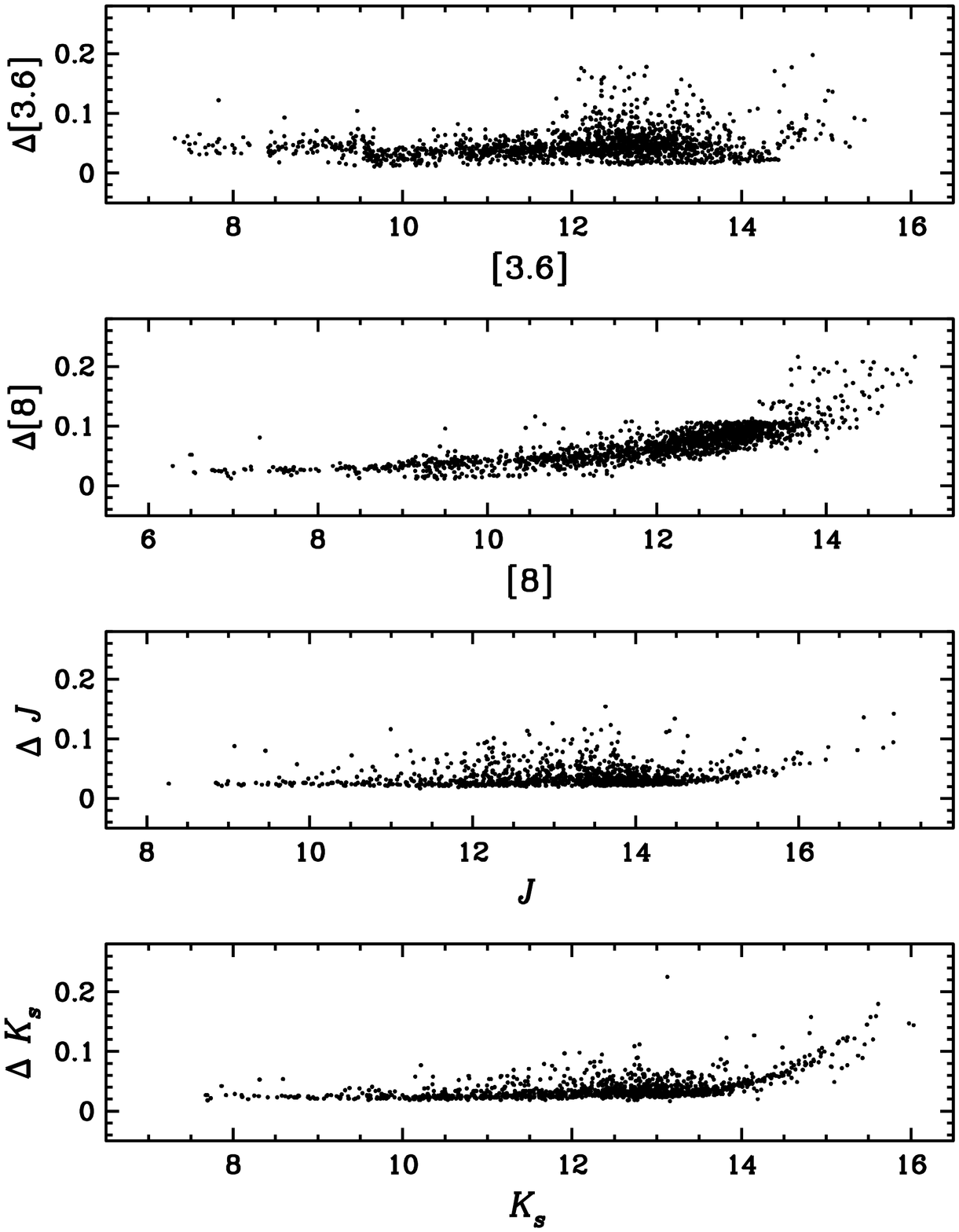}
\caption{
Photometric errors as a function of magnitude for individual stars. 
Top row: $\Delta$ [3.6]; second row: $\Delta$ [8]; 
third row: $\Delta\ J$; bottom row: $\Delta\ K_s$.
}
\label{photerrind}
\end{figure}

I have used the models to check that
the stars that have been detected as point sources in the SAGE mosaics 
are enough to obtain a reliable estimate of the mid-IR SBFs of the
Magellanic star clusters. In this experiment, SBF magnitudes are calculated 
from models as done for observations, that is, the denominator of eq.~\ref{theequation} 
is always the sum of all the stars in the isochrone, while the 
stars considered in the numerator are only those brighter than the data detection limit.
Figure~\ref{convtestirac} shows, for $Z =$
0.0005, 0.004, 0.008, 0.017, and 0.04, the difference between the 3.6, 4.5,
5.8, and 8 $\mu$m integrated and fluctuation magnitudes calculated with all the
stars, and only from those with $M_{5.8} \leq -4.5$ (or [5.8] = 
14 mag at the LMC, and 14.5 mag at the SMC).\footnote{Stars in the isochrones fainter than this limit are
excluded from the calculation of the integrated magnitudes, and from the numerator of eq.~\ref{theequation} 
when computing SBFs.} 
This difference is in fact an overestimate, since  
fainter stars are detected. Barring ages younger than $\sim$ 2 Myr, which
are hardly relevant to the clusters in this work, 
at all times 
the anticipated differences in the derived fluctuation magnitudes for all IRAC bands
are smaller than the expected errors, due mainly to stochastic fluctuations in the number of stars 
(see Table~\ref{latabla}); in fact,  
for ages younger than a few Gyr, the differences are barely noticeable. 
This is true even for the oldest clusters, 
in which the contribution from bright stars to the integrated luminosity is of the order of
30\% -- 50\%, depending on metallicity.

\begin{figure}
\includegraphics[angle=0.,width=0.84\hsize,clip=]{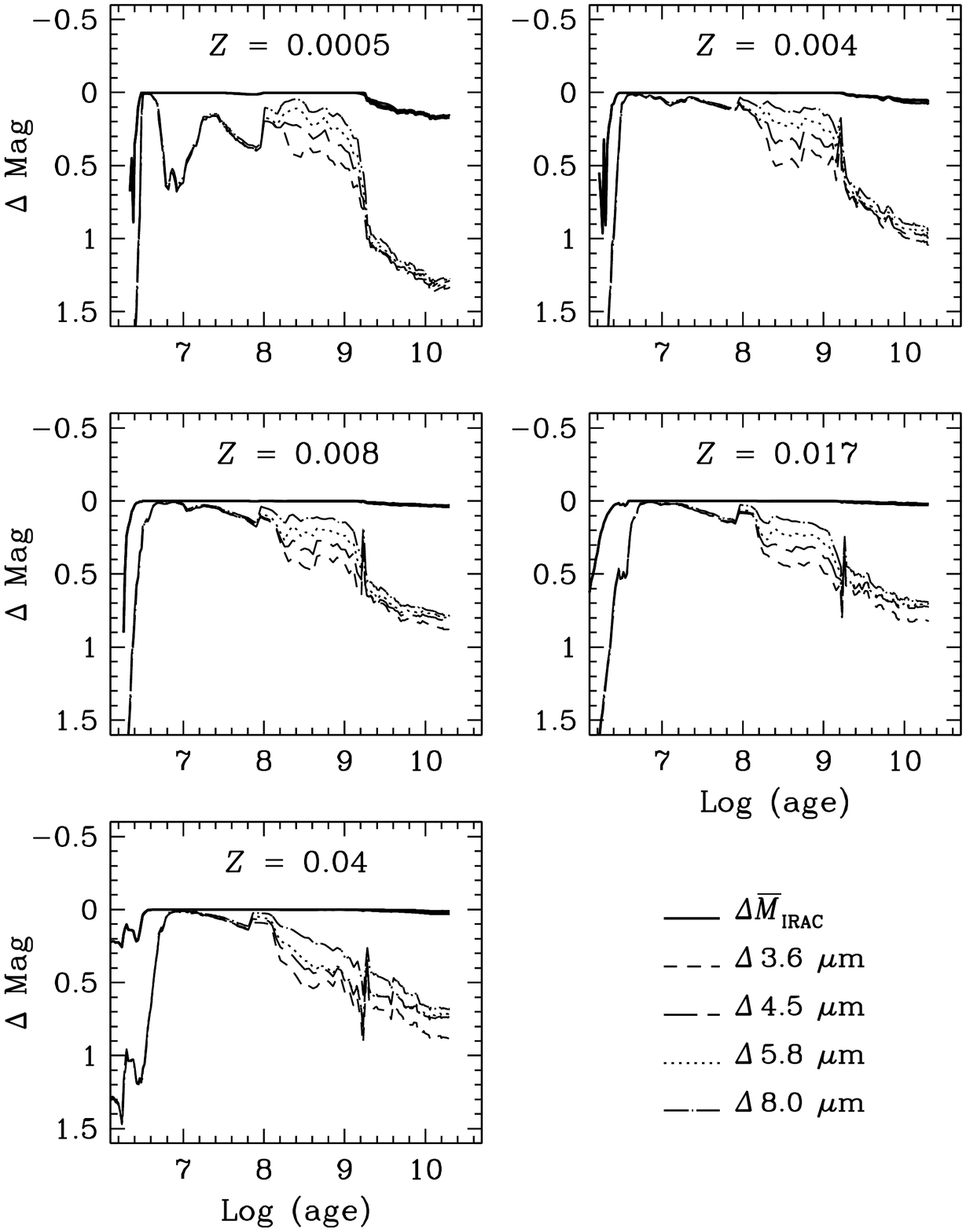}
\caption{
Comparison between contributions to integrated and
fluctuation magnitudes from all stars in the
isochrone ($\Delta$ mag = 0), and only from stars brighter than $M_{5.8}$ = -4.5
(or [5.8] = 14.0 mag at the LMC, and 14.5 mag at the SMC).
Top left panel: $Z$ = 0.0005;  top right panel:
$Z$ = 0.004; middle left panel: $Z$ = 0.008; middle right panel:
$Z$ = 0.017; bottom left panel: $Z$ = 0.04.
Displayed Within each panel is the
difference in integrated magnitudes
at [3.6] (short--dashed line),
[4.5] (long--dashed line), [5.8] (dotted line), [8] (dotted--long--dashed line),
and for fluctuation magnitudes in all the IRAC bands (thick solid line).
}
\label{convtestirac}
\end{figure}

Figure~\ref{convtestmips} displays a similar plot for the MIPS 24 $\mu$m band, including only stars with 
$M_{24} \leq -9.0$ (or [24] = 9.5 and 10.0 mag at the
LMC and SMC, respectively). This graph shows that there are not enough stars to 
calculate SBFs for populations with ages between $\sim$ 15 and 100 Myr, regardless of 
metallicity. Also, SBFs cannot be accurately derived after 1.5 Gyr for $Z$ between 0.0005 and 0.004, 
and after 4 Gyr for $Z$ = 0.008.  

\begin{figure}
\includegraphics[angle=0.,width=0.85\hsize,clip=]{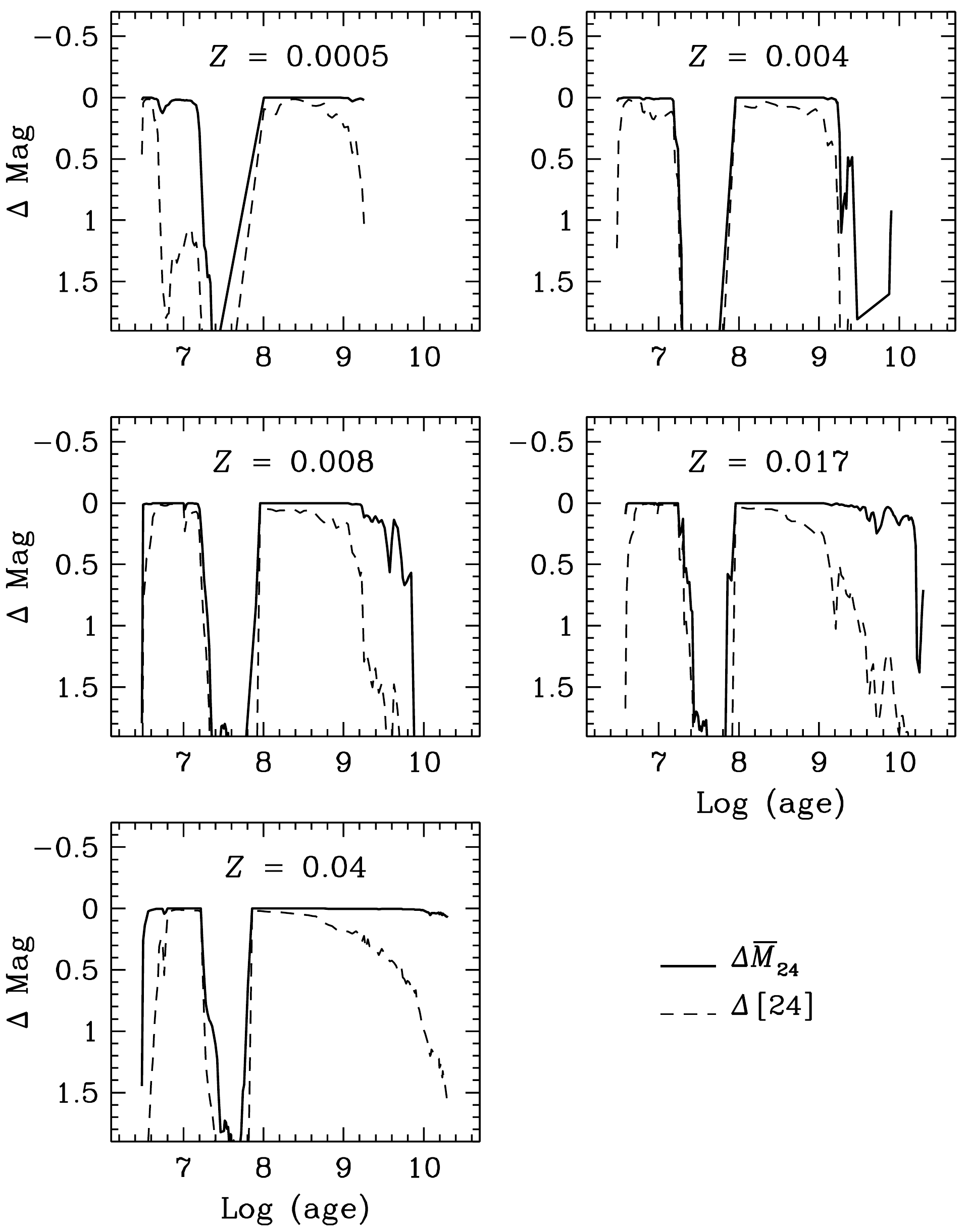}
\caption{
Comparison between contributions to 24 $\mu$m integrated and
fluctuation magnitudes from all stars in the
isochrone ($\Delta$ mag = 0), and only from stars brighter than $M_{24}$ = -9.
(or [24] = 9.5 mag at the LMC, and 10.0 mag at the SMC).
Dashed line: difference in integrated magnitudes; solid line: difference
in fluctuation magnitudes.
Metallicities are the same as in Figure~\ref{convtestirac}.
}
\label{convtestmips}
\end{figure}

The SBF magnitudes of the Magellanic superclusters in the IRAC bands and at [24] are presented in
Table\ \ref{latabla}. Unreliable values, per Figure~\ref{convtestmips}, are obviously omitted.

For comparison with the data, I compute the time evolution of SBF magnitudes
of single-burst stellar populations in the IRAC and 24 $\mu$m bands, with the metallicities and helium
contents available in the CB$^*$ models.
Figure\ \ref{fig_allZ} shows absolute fluctuation magnitudes versus log (age) for CB$^*$ 
models with fiducial mass-loss rate and different metallicities, from $Z$ = 0.0005 ($\sim$ 1/34 solar) to
$Z =$ 0.04 ($\sim$ 2.4 times solar).
Colored regions delimit
expected $\pm$ 1 $\sigma$ stochastic errors for a stellar population with
$5 \times 10^5~M_\odot$.
At very young ages (around 10 Myr), the red supergiants produced by the models with
the lowest $Z = 0.0005$ are fewer and fainter in the mid-IR;  
hence the SBF magnitudes of these populations are also fainter, and suffer from a larger stochastic error.
In contrast, at intermediate ages ($\sim$ 200 Myr to 1 Gyr), when TP-AGB stars are predominant,  
the SBF magnitudes of the population with $Z = 0.04$ are faintest between [4.5] and [8]: 
its TP-AGB stars are also faintest; here, opacity is seemingly more important than 
temperature. 

Finally, after 1 Gyr, when cluster luminosities are dominated by red giant branch (RGB) stars,
the opposite is true, and we get the more intuitive result that
there is a strong trend with metallicity and wavelength: mid-IR SBF magnitudes will be brighter
for higher metallicities, and the difference in SBF brightness between the lowest and  
highest $Z$ will increase with $\lambda$.

\begin{figure*}
\includegraphics[width=1.\hsize,clip=]{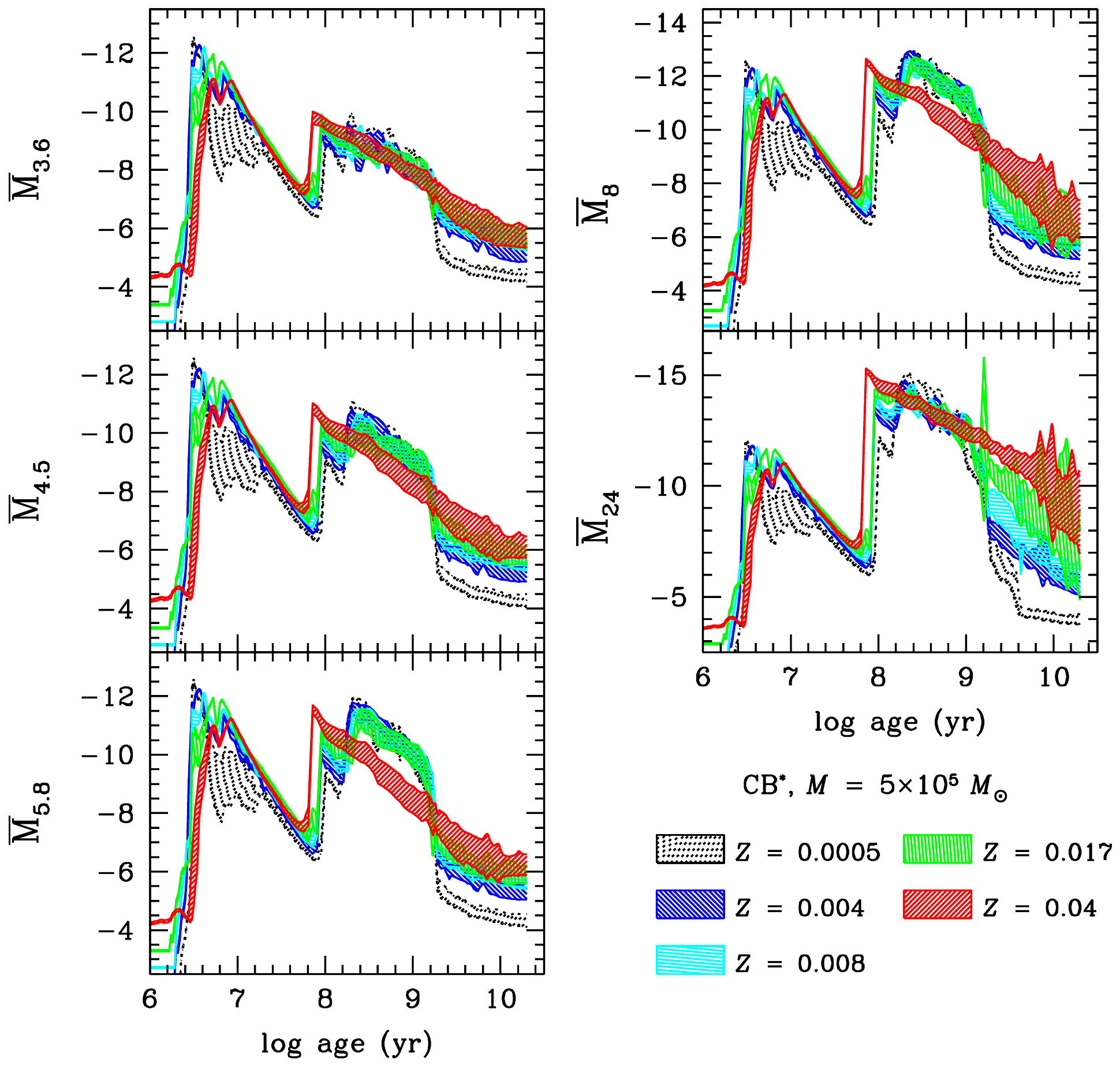}
\caption{Absolute fluctuation magnitudes versus log (age) for standard
CB$^*$ models with different metallicities.  Colored regions delimit
expected  $\pm 1 \sigma$ stochastic errors for stellar populations with $5
\times 10^5~M_\odot$. {\it Black dotted:} $Z$ = 0.0005;
blue left-hatched: $Z$ = 0.004;
cyan horizontal-hatched: $Z$ = 0.008;
green vertical-hatched: $Z$ = 0.017;
red right-hatched: $Z$ = 0.04.}
\label{fig_allZ}
\end{figure*}

I now compare the models to our MC supercluster data. 
Figure\ \ref{figclust52} presents mid-IR (3.6 to 24 $\mu$m) absolute SBF magnitudes
versus log (age) for MC clusters younger than $\sim$ 160 Myr, 
plotted together with models with $Z = 0.008$, and 3 different mass-loss rates:
fiducial $\dot M$ (cyan), fiducial $\dot M / 5$ (blue), and fiducial 
$\dot M \times 5$ (red); again, colored regions 
delimit expected $\pm 1 \sigma$ stochastic errors for a stellar population
with $5 \times 10^5 M_\odot$. 
Before the onset of the TP-AGB phase at $\sim$ 100 Myr, 
models are degenerate, but both models and data show the same trend of diminishing
SBF strength with time. 
(Supercluster type SWB I is omitted at 24 $\mu$, on account of insufficient statistics
to obtain its SBF magnitude in this filter.) 

I note that SBF magnitudes at 4.5, 5.8, and 8 $\mu$m can clearly discern between global changes in the mass-loss
rates for intermediate-age populations, in particular between $\sim 10^8$ yr --the onset of the TP-AGB phase-- and
1 Gyr. During most of this time span, the SBF magnitudes of a population with fiducial $\dot M$ will be brighter
than those for stars with a reduced mass-loss rate. A population with 5$\times \dot M$, on the other hand,  
will harbor more bright stars at the beginning and hence display SBF magnitudes 
with a smaller dispersion. However,  TP-AGB lifetimes
will be reduced according to the fuel-consumption theorem \citep{renz86}, and this will cause the 
SBF values, first to oscillate, and after age $\sim$ 400 Myr to remain lower than those for the population 
with $\dot M / 5$.  Then, between 2 and 5 Gyr --once the RGB dominates--, and from 3.6 to 5.8 $\mu$m, 
the dispersion will grow,
due to the appearance of a few bright red giant stars; only at 8 and 24 $\mu$m is there also
an increase of the SBF brightness at these times. 

In Figure\ \ref{figclust42}, data of superclusters types SWB IV and V are compared to 
models with $Z = 0.004$. The fit is quite good, and overall consistent with the fiducial mass-loss rate. 
Finally, Figures\ \ref{figclust32} and\ \ref{figclust22} show cluster types SWB VI and VII together with,
respectively, models with $Z = 0.002$ and $Z = 0.001$, from 3.6 to 8 $\mu$m (as the data
have insufficient statistics at 24 $\mu$m). At the age of cluster type VI the 
TP-AGB phase shuts off, hence the SBF magnitudes change abruptly and sensitivity to $\dot M$ is poor.
The fit between models and data points, however, is fair within the errors.  
On the other hand, the derived SBF magnitudes of cluster type VII are too bright at all 
wavelengths, compared to the model. This problem could be explained if the handful of stars brighter than 
[8] = 11.5 mag (see Figure\ \ref{cmdsII}, bottom right panel) are actually foreground in the
Milky Way halo, for example,
and thus our decontamination scheme did not work properly.

\begin{figure*}
\includegraphics[width=1.\hsize]{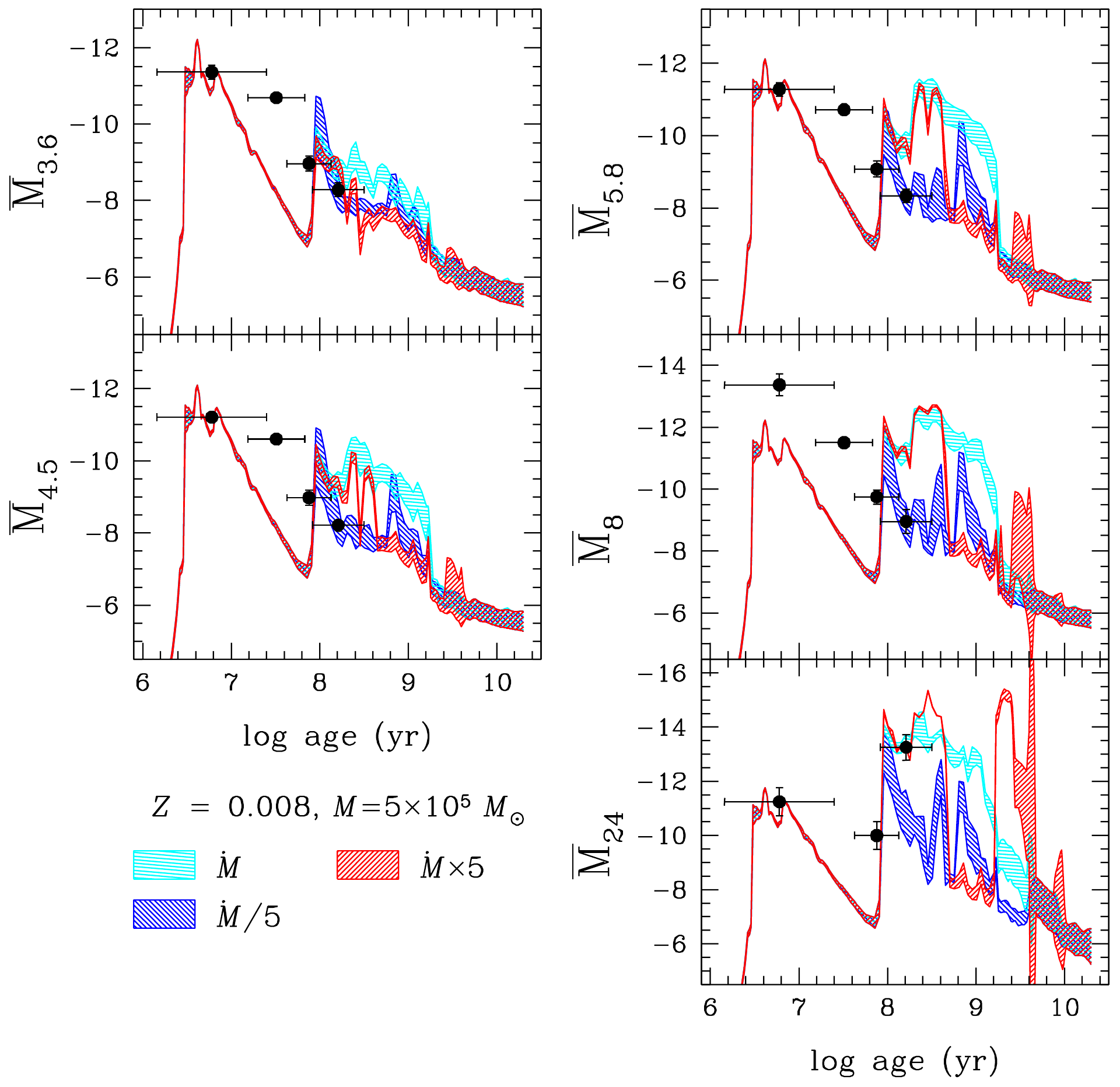}
\caption{SBFs of Magellanic star clusters I. Observations of young and
intermediate-age MC superclusters (pre-SWB to SWB III) are compared to models with $Z$ =
0.008. $\pm 1 \sigma$ stochastic uncertainties are shown for stellar
populations with 5 $\times 10^5~M_\odot$.
Cyan horizontal-hatched: fiducial $\dot M$;
blue left-hatched: fiducial $\dot M$/5;
red right-hatched: fiducial $\dot M \times$5.
Solid circles: mid-IR SBF measurements. Statistics are insufficient to obtain an SBF magnitude at 24 $\mu$m for the
supercluster type SWB I.
}
\label{figclust52}
\end{figure*}

\begin{figure*}
\includegraphics[width=1.\hsize]{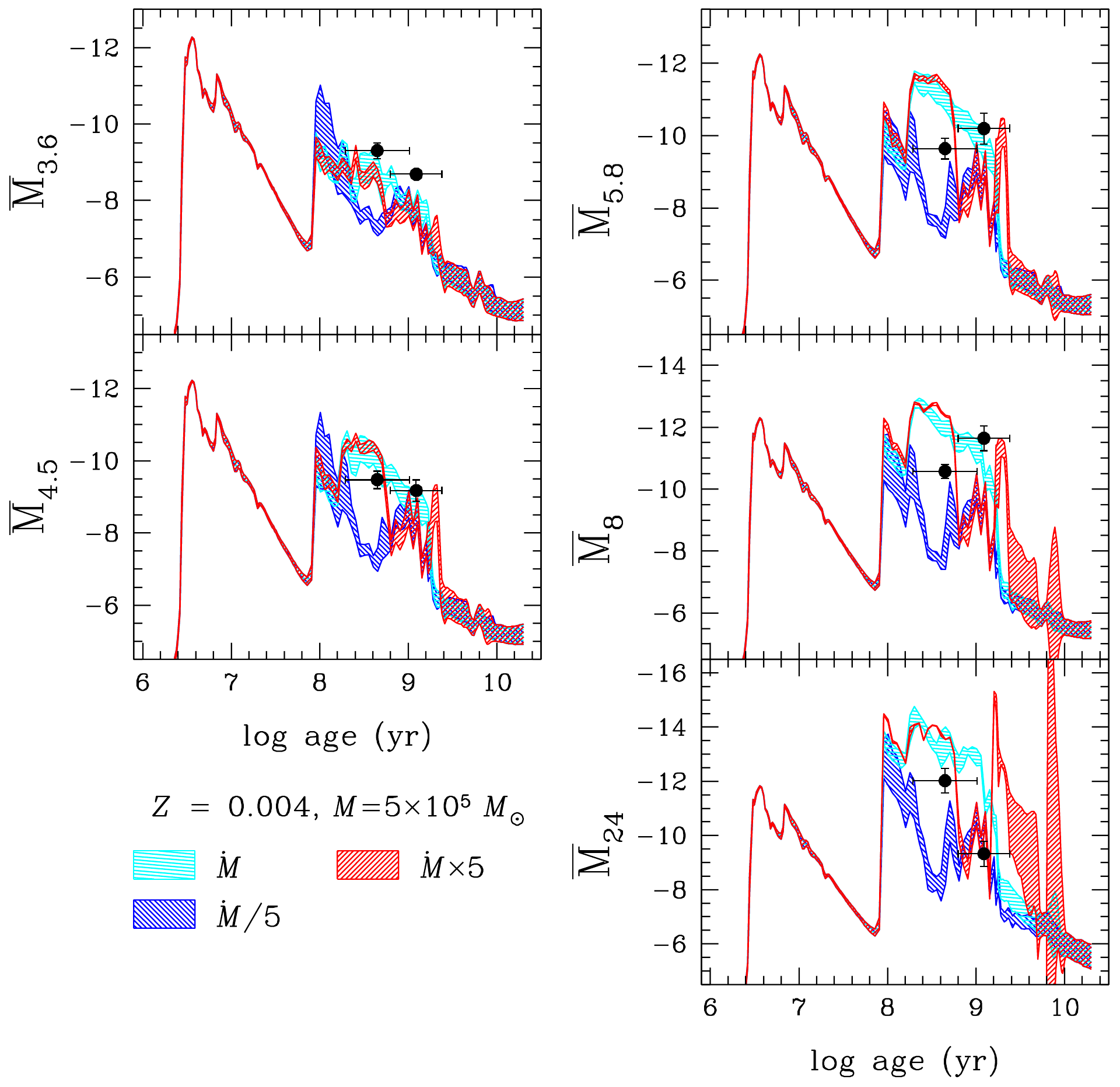}
\caption{SBFs of Magellanic star clusters II. Observations of
intermediate-age MC superclusters (SWB IV and SWB V) are compared to models with $Z$ =
0.004. $\pm 1 \sigma$ stochastic uncertainties are shown for stellar
populations with $5 \times 10^5~M_\odot$, coded as in Figure
\ref{figclust52}.  }
\label{figclust42}
\end{figure*}

\begin{figure*}
\includegraphics[width=1.\hsize]{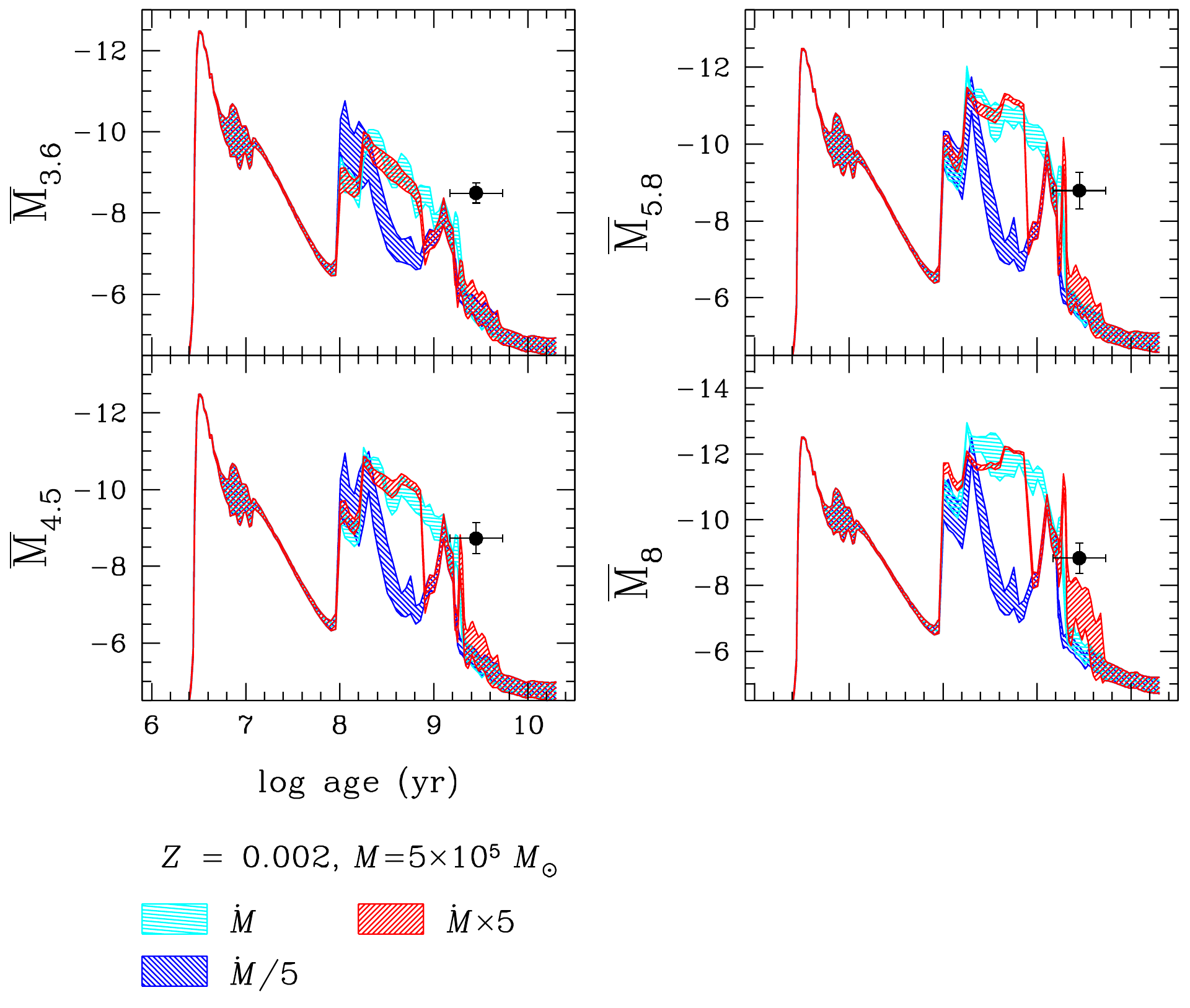}
\caption{SBFs of Magellanic star clusters III.  Observations of the
old MC
supercluster SWB VI are compared to models with $Z$ = 0.002. $\pm 1 \sigma$
stochastic uncertainties are shown for stellar populations with $5
\times 10^5~M_\odot$, coded as in Figure \ref{figclust52}.
}
\label{figclust32}
\end{figure*}

\begin{figure*}
\includegraphics[width=1.\hsize]{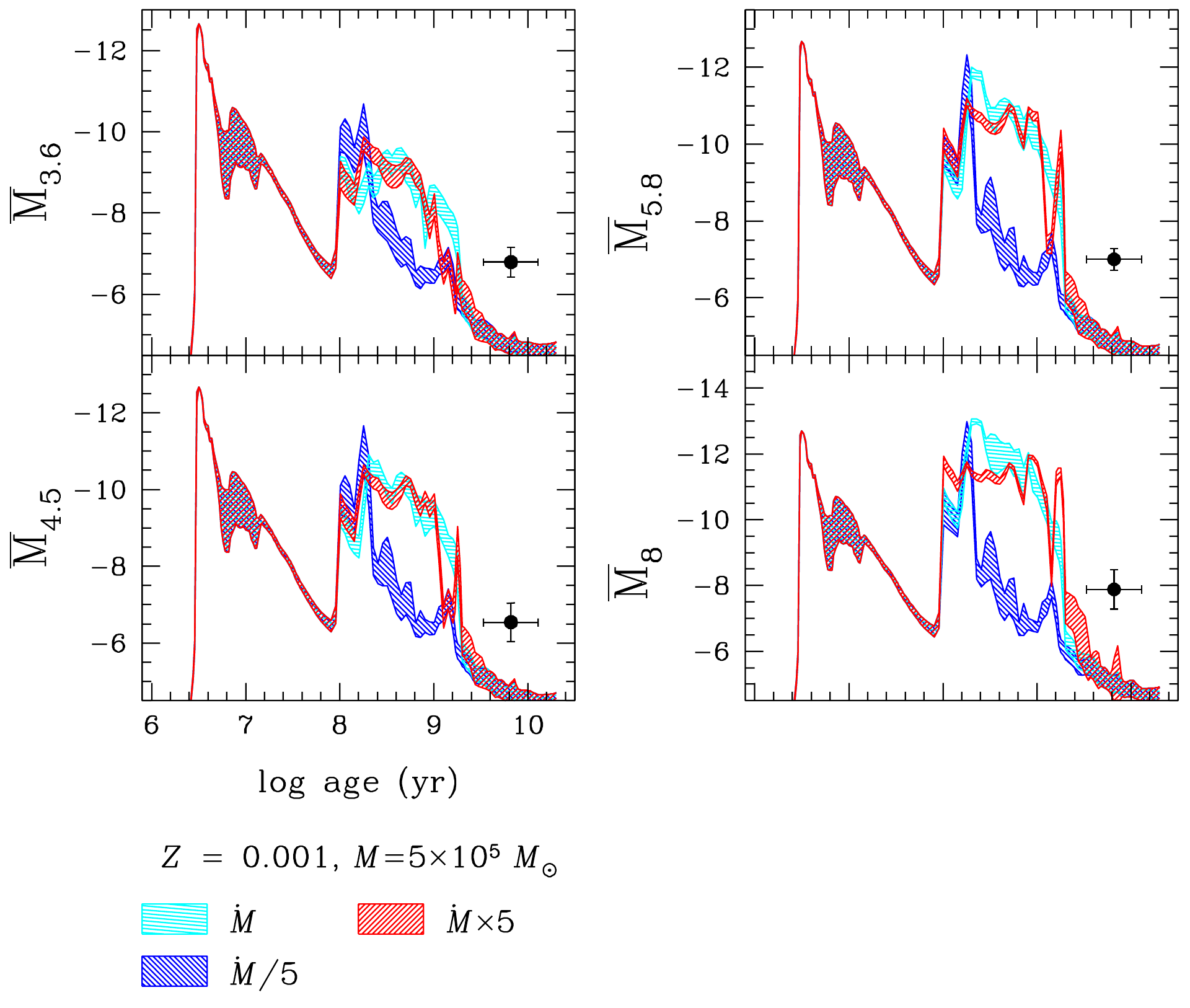}
\caption{SBFs of Magellanic star clusters IV. Observations of the old MC
supercluster SWB VII are compared to models with $Z$ = 0.001.  $\pm 1 \sigma$
stochastic uncertainties are shown for stellar populations with $5
\times 10^5~M_\odot$, coded as in Figure \ref{figclust52}.
}
\label{figclust22}
\end{figure*}

\section{Summary and Conclusions.} \label{sec:summconcl}

I have presented mid-IR broadband colors and fluctuation magnitudes computed
from SSP models, with the main goal of exploring their ability to detect 
changes in the global mass-loss rate of stars undergoing the TP-AGB phase.
To this end, I have used the CB$^*$ evolutionary tracks to produce spectra of
TP-AGB stars, considering the radiative transfer in their circumstellar envelopes.
I have processed SEDs for fiducial $\dot M$, as well as for twice, 5$\times$, 10$\times$ fiducial,
$\dot M/2$, $\dot M/5$, and $\dot M/10$. In all cases, the stellar parameters, mass-loss rate, and 
length of the superwind evolutionary phase 
have been varied simultaneously and consistently, as described in \citet{gonz10}. 
The model colors and SBF magnitudes have then been compared to mid-IR data of single AGB stars
and star clusters in the Magellanic Clouds.
My conclusions are as follows:

\begin{itemize}

\item[1.] Models with different mass-loss rates and metallicities differ significantly in their predicted mid-IR 
      colors and SBF magnitudes.

\item[2.] Models with a higher than fiducial mass-loss rate are needed to fit the mid-IR colors of ``extreme" single AGB stars in the LMC.

\item[3.] The range of mid-IR colors of individual MC clusters is consistent with models with $Z = 0.008$, a mass-loss rate between fiducial and 5 $\times$ fiducial, and the stochastic errors expected for a cluster population between 5 $\times 10^3$ and 5 $\times 10^4 M_\odot$.  

\item[4.] In the case of artificial ``superclusters," built by grouping clusters in the MCs by age and metallicity,  
although models are compatible with the observations, integrated colors cannot strongly constrain
the mass-loss rate, given the present data and theoretical uncertainties. The colors of the SWB VI cluster (3 Gyr), 
however, suggest a higher than fiducial mass-loss rate.

\item[5.] Model SBF magnitudes are quite sensitive to metallicity for 4.5 $\mu$m and longer wavelengths, basically at all
      stellar population ages. For metallicities between $Z = 0.004$ and $Z = 0.04$ and populations 
      younger than 1-2 Gyr, fluctuations grow fainter as $Z$ increases; 
      for older populations the trend reverses, and SBFs are brighter for higher $Z$.
      For $Z = 0.0005$, fluctuations are faintest for ages $<$ 100 Myr and $>$ 1-2 Gyr; in between, SBF magnitudes are as 
      bright as for 0.004 $\la Z \la$ 0.017.

\item[6.] Fluctuation magnitudes are powerful diagnostics of mass-loss rate in the TP-AGB. 

\item[7.] The SBF measurements of the MC superclusters suggest a mass-loss rate close to fiducial. Although consistent with the models within the errors, it is possible that the oldest SWB VII cluster (7 Gyr, and hence with the faintest RGB stars) is contaminated by foreground stars.  

\end{itemize}

The present work has in fact performed an independent calibration of the CB$^*$ models, and confirmed that colors and
SBF magnitudes in the mid-IR are sensitive to global changes in mass-loss rate, in particular during the TP-AGB phase.
For this kind of proof of concept work, the MC clusters offer the advantages of their known distances and wide range of
metallicities. However, 
star clusters with masses lower than $ \sim 5\times 10^5 M_\odot$ may be subject to 
stochastic fluctuations in their population properties that can be significant.  
Colors and fluctuation magnitudes of whole galaxies obtained with the upcoming James Webb Space Telescope
should be much less affected by these systematics, and
hence able to measure with a high degree of confidence global mass-loss rates and their correlation (or not) with metallicity.

\acknowledgments
I thank G.\ Bruzual for providing his models in the format required for this work, and 
L.\ Loinard for a critical reading of the manuscript.
I am grateful for the very thorough and constructive comments of the anonymous referee.

This research was supported by a grant from the program PASPA--DGAPA, UNAM.

This research has made use of the VizieR Service
and the SIMBAD database at
the Centre de Donn\'ees Astronomiques de Strasbourg, as well as NASA's
Astrophysics Data System Abstract Service.

This research has made use of the NASA/ IPAC Infrared Science Archive, which is operated by the Jet Propulsion Laboratory, California Institute of Technology, under contract with the National Aeronautics and Space Administration.

\facilities{IRSA, {\it Spitzer}}

\software{DUSTY \citep{ivez99}}

\end{document}

%% file: table1.tex
\begin{deluxetable}{l|c|c}
\tablecolumns{5}
%\tablewidth{150pt}
\tablewidth{0pt}
\tablecaption{Stellar Libraries Used in the CB$^*$ Models\label{cbstar}}
\tablehead{
\multicolumn{1}{c|}{Stellar} & \multicolumn{1}{c|}{Stellar} & \multicolumn{1}{c}{Wavelength} \\ 
\multicolumn{1}{c|}{Library} & \multicolumn{1}{c|}{Type}    & \multicolumn{1}{c}{Range}       }
\startdata
TLUSTY (a)         & O stars          &    45\AA - 300$\mu$m        \\
TLUSTY (b)         & B stars          &    54\AA - 300$\mu$m        \\
Martins et al.\ (c)  & A stars          &     3000 - 7000\AA              \\
UVBlue (d)          & F,G,K stars   &    \ 850 - 4700\AA            \\
Rauch (e)            & T$>$55kK       &\ \ \ \ 5 - 2000\AA         \\
Miles (f)              & A-M stars     &      3540 - 7351\AA     \\
Stelib (g)             & A-M stars     &      7351 - 8750\AA     \\
BaSeL 3.1 (h)     & A-M stars     &       8750\AA - 36000$\mu$m   \\
BaSeL 3.1, Aringer et al.\ (i), IRTF (j), Dusty models (k)    & TP-AGB stars &       8750\AA - 36000$\mu$m   \\
\enddata
\tablerefs{
(a) Lanz \& Hubeny (2003a, 2003b), 
(b) \citet{lanz2007},
(c) \citet{martins2005},
(d) \citet{rodmer2005},
(e) \citet{rauch2003},
(f) S\'anchez-Bl\'azquez et al.\ (2006), Falc\'on-Barroso et al.\ (2011), Prugniel et al.\ (2011), 
(g) \citet{jfl03},
(h) \citet{pw02},
(i) \citet{ba09},
(j) \citet{jr09},
(k) Nenkova et al.\ (2000), Gonz\'alez-L\'opezlira et al.\ (2010).
}
\end{deluxetable}

%% file: table2.tex
\begin{table*}
 \centering
  \caption{IR Colors for CB$^*$ Models with Fiducial Mass-loss}
 \begin{minipage}{280mm}
  \begin{scriptsize}
  \begin{tabular}{@{}rrrrrr@{}}
\hline
 Age (Gyr)  & $[V-3.6]$  & $[3.6-4.5]$ & $[4.5-5.8]$  & $[5.8-8]$  & $[8-24]$\\
\hline
\multicolumn {6}{|c|}{$Z = 0.017 \ \ Y = 0.279$}\\
\hline
     0.005  &     2.110  &    -0.068  &     0.045  &     0.110  &    -0.428  \\
     0.006  &     2.541  &    -0.065  &     0.048  &     0.118  &    -0.419  \\
     0.007  &     3.674  &     0.095  &     0.079  &     0.086  &    -0.330  \\
     0.008  &     3.515  &     0.094  &     0.079  &     0.085  &    -0.331  \\
     0.009  &     3.192  &     0.091  &     0.077  &     0.084  &    -0.334  \\
     0.010  &     2.389  &     0.084  &     0.073  &     0.081  &    -0.340  \\
     0.020  &     2.590  &     0.060  &     0.070  &     0.085  &    -0.353  \\
     0.030  &     1.935  &     0.002  &     0.059  &     0.092  &    -0.387  \\
     0.040  &     1.707  &    -0.033  &     0.050  &     0.094  &    -0.413  \\
     0.050  &     1.614  &    -0.056  &     0.045  &     0.095  &    -0.429  \\
     0.060  &     1.602  &    -0.068  &     0.042  &     0.096  &    -0.438  \\
     0.070  &     1.699  &    -0.062  &     0.045  &     0.094  &    -0.431  \\
     0.080  &     1.666  &    -0.066  &     0.042  &     0.091  &    -0.437  \\
     0.090  &     2.309  &     0.149  &     0.213  &     0.755  &     1.697  \\
     0.100  &     2.370  &     0.109  &     0.187  &     0.725  &     1.600  \\
     0.200  &     2.308  &     0.084  &     0.184  &     0.571  &     1.520  \\
     0.300  &     2.250  &     0.188  &     0.394  &     0.639  &     0.940  \\
     0.400  &     2.287  &     0.176  &     0.370  &     0.589  &     0.872  \\
     0.500  &     2.395  &     0.203  &     0.339  &     0.500  &     0.703  \\
     0.600  &     2.412  &     0.196  &     0.337  &     0.512  &     0.753  \\
     0.700  &     2.557  &     0.194  &     0.312  &     0.450  &     0.544  \\
     0.800  &     2.594  &     0.180  &     0.291  &     0.418  &     0.476  \\
     0.900  &     2.626  &     0.164  &     0.269  &     0.382  &     0.394  \\
     1.000  &     2.603  &     0.137  &     0.283  &     0.444  &     0.640  \\
     1.500  &     2.675  &     0.057  &     0.180  &     0.232  &     0.096  \\
     2.000  &     2.926  &    -0.058  &     0.059  &     0.227  &     0.429  \\
     3.000  &     3.145  &    -0.025  &     0.045  &     0.180  &     0.202  \\
     4.000  &     3.036  &    -0.030  &     0.044  &     0.157  &     0.067  \\
     5.000  &     3.111  &    -0.029  &     0.049  &     0.147  &     0.000  \\
     6.000  &     3.071  &    -0.026  &     0.044  &     0.143  &    -0.019  \\
     7.000  &     3.091  &    -0.025  &     0.044  &     0.144  &    -0.004  \\
     8.000  &     3.116  &    -0.023  &     0.052  &     0.140  &    -0.032  \\
     9.000  &     3.130  &    -0.023  &     0.054  &     0.128  &    -0.121  \\
    10.000  &     3.170  &    -0.023  &     0.056  &     0.123  &    -0.156  \\
    11.000  &     3.190  &    -0.018  &     0.055  &     0.133  &    -0.086  \\
    12.000  &     3.205  &    -0.016  &     0.056  &     0.128  &    -0.123  \\
    13.000  &     3.231  &    -0.016  &     0.058  &     0.126  &    -0.137  \\
    13.500  &     3.243  &    -0.016  &     0.059  &     0.123  &    -0.153  \\
\hline
\end{tabular}
\end{scriptsize}
\end{minipage}
{{\bf Note.} Values for solar metallicity and helium content are shown here for guidance regarding the table's form and content.
(This table is available in its entirety in machine-readable form.)}
\label{tabla2}
\end{table*}

%% file: table3.tex
%\begin{landscape}
\begin{table*}
  \caption{SBF Amplitudes for CB$^*$ Models with Fiducial Mass-loss }
 \centering
 \begin{minipage}{280mm}
  \begin{scriptsize}
  \begin{tabular}{@{}rrrrrr@{}}
\hline
 Age (Gyr)  & $\barM_{3.6}$  & $\barM_{4.5}$  & $\barM_{5.8}$  & $\barM_{8}$  & $\barM_{24}$  \\
\hline
\multicolumn {6}{|c|}{$Z = 0.017 \ \ Y = 0.279$}\\
\hline
     0.005  &   -11.795  &   -11.722  &   -11.775  &   -11.899  &   -11.484  \\
     0.006  &   -11.281  &   -11.238  &   -11.295  &   -11.412  &   -11.015  \\
     0.007  &   -11.596  &   -11.693  &   -11.776  &   -11.864  &   -11.539  \\
     0.008  &   -11.105  &   -11.203  &   -11.285  &   -11.373  &   -11.048  \\
     0.009  &   -10.740  &   -10.836  &   -10.917  &   -11.006  &   -10.680  \\
     0.010  &   -10.402  &   -10.497  &   -10.582  &   -10.674  &   -10.353  \\
     0.020  &    -9.231  &    -9.315  &    -9.396  &    -9.488  &    -9.159  \\
     0.030  &    -8.408  &    -8.488  &    -8.573  &    -8.670  &    -8.349  \\
     0.040  &    -7.839  &    -7.899  &    -7.987  &    -8.092  &    -7.771  \\
     0.050  &    -7.448  &    -7.494  &    -7.582  &    -7.694  &    -7.372  \\
     0.060  &    -7.221  &    -7.260  &    -7.349  &    -7.464  &    -7.141  \\
     0.070  &    -7.622  &    -7.758  &    -7.875  &    -7.983  &    -7.740  \\
     0.080  &    -7.353  &    -7.484  &    -7.599  &    -7.706  &    -7.459  \\
     0.090  &    -9.595  &   -10.227  &   -10.699  &   -11.978  &   -14.212  \\
     0.100  &    -9.534  &   -10.107  &   -10.564  &   -11.863  &   -14.102  \\
     0.200  &    -8.873  &    -9.510  &   -10.035  &   -11.379  &   -13.920  \\
     0.300  &    -8.611  &    -9.851  &   -11.128  &   -12.366  &   -13.695  \\
     0.400  &    -8.363  &    -9.474  &   -10.721  &   -12.004  &   -13.418  \\
     0.500  &    -8.533  &    -9.549  &   -10.535  &   -11.705  &   -13.227  \\
     0.600  &    -8.419  &    -9.474  &   -10.475  &   -11.670  &   -13.304  \\
     0.700  &    -8.319  &    -9.430  &   -10.371  &   -11.386  &   -12.589  \\
     0.800  &    -8.275  &    -9.354  &   -10.246  &   -11.206  &   -12.333  \\
     0.900  &    -8.229  &    -9.273  &   -10.110  &   -11.009  &   -12.044  \\
     1.000  &    -7.985  &    -8.968  &    -9.941  &   -11.165  &   -12.800  \\
     1.500  &    -7.771  &    -8.634  &    -9.265  &    -9.945  &   -10.703  \\
     2.000  &    -6.939  &    -7.008  &    -7.095  &    -8.215  &   -10.699  \\
     3.000  &    -6.482  &    -6.568  &    -6.595  &    -7.569  &   -10.195  \\
     4.000  &    -6.281  &    -6.342  &    -6.335  &    -7.059  &    -9.344  \\
     5.000  &    -6.146  &    -6.203  &    -6.223  &    -6.859  &    -9.013  \\
     6.000  &    -6.096  &    -6.168  &    -6.148  &    -6.760  &    -8.872  \\
     7.000  &    -6.076  &    -6.178  &    -6.178  &    -6.953  &    -9.580  \\
     8.000  &    -5.931  &    -6.034  &    -6.070  &    -6.726  &    -9.082  \\
     9.000  &    -5.810  &    -5.907  &    -5.957  &    -6.437  &    -8.384  \\
    10.000  &    -5.767  &    -5.868  &    -5.929  &    -6.327  &    -8.041  \\
    11.000  &    -5.799  &    -5.930  &    -5.989  &    -6.566  &    -8.865  \\
    12.000  &    -5.761  &    -5.910  &    -5.975  &    -6.574  &    -9.040  \\
    13.000  &    -5.743  &    -5.894  &    -5.970  &    -6.513  &    -8.889  \\
    13.500  &    -5.734  &    -5.885  &    -5.964  &    -6.476  &    -8.814  \\
\hline
\end{tabular}
\end{scriptsize}
\end{minipage}
{{\bf Note.} Values for solar metallicity and helium content are shown here for guidance regarding the table's form and content.
(This table is available in its entirety in machine-readable form.)}
\label{tabla4}
\end{table*}
%\end{landscape}

%% file: table4.tex
%\documentclass[preprint]{aastex}
%\usepackage{epsfig,lscape}
%\begin{document}
\startlongtable
\begin{deluxetable}{llcrrrrr}
\tablewidth{6.5in}
\tablecaption{Individual Cluster Photometry}
\tablehead{
\colhead{} &
\colhead{} &
\colhead{} &
\colhead {[3.6]} &
\colhead {[4.5]} &
\colhead {[5.8]} &
\colhead {[8]} &
\colhead {[24]}\\[-0.2cm]
\colhead {Supercluster} &
\colhead {Name} &
\colhead {Cloud} &
\colhead { mag}& 
\colhead { mag}& 
\colhead { mag}& 
\colhead { mag}& 
\colhead { mag}\\[-0.3cm]
}
\colnumbers
\startdata
Pre-SWB\dotfill&     L~84 & SMC  & 11.30 $\pm$ 0.10 & 11.03 $\pm$ 0.09 &  9.83 $\pm$ 0.07 &  8.38 $\pm$ 0.04& ...\ \ \ \ \ \ \ \\ 
&    L~107 & SMC  & 11.51 $\pm$ 0.09 & 11.63 $\pm$ 0.10 & 11.25 $\pm$ 0.11 & 10.80 $\pm$ 0.12&...\ \ \ \ \ \ \  \\ 
&  NGC~602 & SMC  & 10.86 $\pm$ 0.05 & 10.39 $\pm$ 0.06 &  9.36 $\pm$ 0.05 &  7.99 $\pm$ 0.03&...\ \ \ \ \ \ \  \\ 
& NGC~1983 & LMC  &  7.39 $\pm$ 0.01 &  7.23 $\pm$ 0.01 &  6.49 $\pm$ 0.01 &  6.00 $\pm$ 0.01& 4.18 $\pm$  0.04 \\ 
& NGC~1984 & LMC  &  7.13 $\pm$ 0.01 &  6.73 $\pm$ 0.01 &  5.44 $\pm$ 0.01 &  4.10 $\pm$ 0.01&  -0.669 $\pm$0.004 \\ 
& NGC~2001 & LMC  &  8.37 $\pm$ 0.02 &  8.46 $\pm$ 0.02 &  7.34 $\pm$ 0.02 &  7.49 $\pm$ 0.02& 5.98 $\pm$   0.08 \\ 
& NGC~2006 & LMC  &  8.70 $\pm$ 0.02 &  8.73 $\pm$ 0.03 &  8.58 $\pm$ 0.03 &  8.69 $\pm$ 0.05&...\ \ \ \ \ \ \   \\ 
& NGC~2011 & LMC  &  7.48 $\pm$ 0.01 &  7.44 $\pm$ 0.01 &  6.86 $\pm$ 0.02 &       ...\ \ \ \ \ \ \ \        &...\ \ \ \ \ \ \   \\ 
& NGC~2014 & LMC  &  7.14 $\pm$ 0.01 &  7.01 $\pm$ 0.01 &  5.05 $\pm$ 0.01 &  3.515 $\pm$ 0.005&...\ \ \ \ \ \ \   \\ 
& NGC~2027 & LMC  &  8.33 $\pm$ 0.02 &  7.83 $\pm$ 0.02 &  7.08 $\pm$ 0.02 &  6.80 $\pm$ 0.02& 6.23 $\pm$  0.08 \\ 
&   SL~114 & LMC  &  9.40 $\pm$ 0.03 &  9.19 $\pm$ 0.03 &  8.53 $\pm$ 0.04 &  7.31 $\pm$ 0.03& 5.57 $\pm$  0.07   \\ 
%---------------pre swb up to here--------------------------
SWB I\dotfill& L~45    & SMC  &  9.63 $\pm$ 0.02 &  9.63 $\pm$ 0.03 &  9.13 $\pm$ 0.04 &  8.57 $\pm$ 0.04 &...\ \ \ \ \ \ \  \\ 
& L~51    & SMC  &  8.98 $\pm$ 0.02 &  8.84 $\pm$ 0.02 &  8.57 $\pm$ 0.04 &  8.59 $\pm$ 0.04 &...\ \ \ \ \ \ \  \\ 
& L~56    & SMC  & 11.15 $\pm$ 0.11 & 11.06 $\pm$ 0.12 & 10.28 $\pm$ 0.06 &  9.95 $\pm$ 0.08 &...\ \ \ \ \ \ \  \\ 
& L~66    & SMC  & 10.64 $\pm$ 0.02 & 10.74 $\pm$ 0.04 & 10.64 $\pm$ 0.04 & 10.80 $\pm$ 0.09 &...\ \ \ \ \ \ \  \\ 
& NGC~290 & SMC  & 10.13 $\pm$ 0.05 & 10.37 $\pm$ 0.07 &  9.72 $\pm$ 0.05 &       ...\ \ \ \ \ \ \ \         &...\ \ \ \ \ \ \  \\ 
& NGC~299 & SMC  &  9.07 $\pm$ 0.02 &  9.21 $\pm$ 0.02 &  8.88 $\pm$ 0.03 &  8.87 $\pm$ 0.05 &...\ \ \ \ \ \ \  \\ 
& NGC~330 & SMC  &  7.95 $\pm$ 0.02 &  8.09 $\pm$ 0.02 &  7.75 $\pm$ 0.02 &  7.66 $\pm$ 0.03 &...\ \ \ \ \ \ \  \\ 
& NGC~376 & SMC  &  8.76 $\pm$ 0.02 &  8.92 $\pm$ 0.02 &  8.61 $\pm$ 0.03 &  8.63 $\pm$ 0.05 &...\ \ \ \ \ \ \  \\ 
& NGC~1704& LMC  &  9.06 $\pm$ 0.02 &  9.23 $\pm$ 0.03 &  9.16 $\pm$ 0.04 &       ...\ \ \ \ \ \ \ \         &...\ \ \ \ \ \ \  \\ 
& NGC~1711& LMC  &  8.13 $\pm$ 0.02 &  8.18 $\pm$ 0.02 &  7.96 $\pm$ 0.02 &  7.96 $\pm$ 0.03 & 7.99 $\pm$ 0.31\\ 
& NGC~1787& LMC  &  8.06 $\pm$ 0.02 &  8.19 $\pm$ 0.02 &  7.91 $\pm$ 0.02 &  7.81 $\pm$ 0.03 & 7.50 $\pm$ 0.16\\ 
& NGC~1805& LMC  &  8.03 $\pm$ 0.02 &  7.51 $\pm$ 0.01 &  6.87 $\pm$ 0.01 &  6.47 $\pm$ 0.02 & 4.73 $\pm$ 0.04\\ 
& NGC~1810& LMC  &  8.88 $\pm$ 0.02 &  9.10 $\pm$ 0.03 &  9.10 $\pm$ 0.04 &  9.24 $\pm$ 0.08 & ...\ \ \ \ \ \ \  \\ 
& NGC~1818& LMC  &  7.12 $\pm$ 0.01 &  7.26 $\pm$ 0.01 &  6.98 $\pm$ 0.01 &  6.97 $\pm$ 0.02 & ...\ \ \ \ \ \ \  \\ 
& NGC~2002& LMC  &  6.89 $\pm$ 0.01 &  6.86 $\pm$ 0.01 &  6.09 $\pm$ 0.01 &  5.83 $\pm$ 0.01 & 4.32 $\pm$ 0.04\\ 
& NGC~2003& LMC  &  9.05 $\pm$ 0.02 &  9.27 $\pm$ 0.03 &  9.00 $\pm$ 0.04 &  9.06 $\pm$ 0.05 & 8.73 $\pm$ 0.42 \\ 
& NGC~2004& LMC  &  6.30 $\pm$ 0.01 &  6.41 $\pm$ 0.01 &  5.94 $\pm$ 0.01 &  5.72 $\pm$ 0.01 & ...\ \ \ \ \ \ \  \\ 
& NGC~2009& LMC  &  7.36 $\pm$ 0.01 &  7.02 $\pm$ 0.01 &  6.13 $\pm$ 0.01 &  5.39 $\pm$ 0.01 & ...\ \ \ \ \ \ \  \\ 
& NGC~2098& LMC  &  8.11 $\pm$ 0.02 &  8.29 $\pm$ 0.02 &  8.04 $\pm$ 0.03 &  8.33 $\pm$ 0.05 & ...\ \ \ \ \ \ \  \\ 
& NGC~2100& LMC  &  6.11 $\pm$ 0.01 &  6.20 $\pm$ 0.01 &  5.73 $\pm$ 0.01 &  5.74 $\pm$ 0.01 & ...\ \ \ \ \ \ \  \\ 
& NGC~l477& LMC  &  8.94 $\pm$ 0.02 &  9.18 $\pm$ 0.03 &  9.19 $\pm$ 0.05 &  9.78 $\pm$ 0.15 & ...\ \ \ \ \ \ \  \\ 
& NGC~l538& LMC  &  8.69 $\pm$ 0.02 &  8.78 $\pm$ 0.03 &  8.36 $\pm$ 0.03 &  8.39 $\pm$ 0.04 & ...\ \ \ \ \ \ \  \\ 
%--------------------swb I up to here----------------------------
SWB II\dotfill&IC~1624 &  SMC   & 9.22 $\pm$ 0.02 &  9.31 $\pm$ 0.02 &  8.97 $\pm$ 0.03 &  8.92 $\pm$ 0.02 & ...\ \ \ \ \ \ \  \\ 
&IC~1655 &  SMC   & 13.04 $\pm$ 0.23 &  13.14 $\pm$ 0.24 &      ...\ \ \ \ \ \ \ \          &  12.53 $\pm$ 0.33 & ...\ \ \ \ \ \ \  \\ 
&NGC~220 &  SMC   & 10.88 $\pm$ 0.06 &  11.05 $\pm$ 0.10 &  10.74 $\pm$ 0.10&  10.75 $\pm$ 0.13 & ...\ \ \ \ \ \ \  \\ 
&NGC~222 &  SMC   & 9.42 $\pm$ 0.03 &  9.51 $\pm$ 0.03 &  9.23 $\pm$ 0.04 &  9.33 $\pm$ 0.04 & ...\ \ \ \ \ \ \  \\ 
&NGC~231 &  SMC   & 11.07 $\pm$ 0.06 &  11.19 $\pm$ 0.11 &  11.29 $\pm$ 0.11&  11.04 $\pm$ 0.14 & ...\ \ \ \ \ \ \  \\ 
&NGC~242 &  SMC   & 9.60 $\pm$ 0.03 &  9.72 $\pm$ 0.05 &  9.55 $\pm$ 0.05 &  9.30 $\pm$ 0.07 & ...\ \ \ \ \ \ \  \\ 
&NGC~422 &  SMC   & 11.91 $\pm$ 0.12 &  12.22 $\pm$ 0.17 &  11.90 $\pm$ 0.17&     ...\ \ \ \ \ \ \ \         & ...\ \ \ \ \ \ \   \\ 
&NGC~1732 &  LMC   & 9.67 $\pm$ 0.04 &  9.39 $\pm$ 0.04 &  8.82 $\pm$ 0.04 &  7.60 $\pm$ 0.03 & 6.02 $\pm$ 0.07 \\ 
&NGC~1735 &  LMC   & 8.68 $\pm$ 0.02 &  8.70 $\pm$ 0.03 &  8.13 $\pm$ 0.03 &  7.30 $\pm$ 0.03 & ...\ \ \ \ \ \ \  \\ 
&NGC~1755 &  LMC   & 8.05 $\pm$ 0.02 &  8.20 $\pm$ 0.02 &  7.91 $\pm$ 0.02 &  7.87 $\pm$ 0.04 & 6.89 $\pm$ 0.17 \\ 
&NGC~1774 &  LMC   & 8.27 $\pm$ 0.02 &  8.55 $\pm$ 0.02 &  8.29 $\pm$ 0.03 &  9.36 $\pm$ 0.07 & ...\ \ \ \ \ \ \  \\ 
&NGC~1782 &  LMC   & 7.45 $\pm$ 0.01 &  7.60 $\pm$ 0.02 &  7.28 $\pm$ 0.02 &  7.20 $\pm$ 0.03 & ...\ \ \ \ \ \ \  \\ 
&NGC~1793 &  LMC   & 9.34 $\pm$ 0.03 &  9.38 $\pm$ 0.04 &  9.37 $\pm$ 0.06 &       ...\ \ \ \ \ \ \ \        &  6.65 $\pm$  0.15 \\ 
&NGC~1834 &  LMC   & 8.57 $\pm$ 0.02 &  8.74 $\pm$ 0.03 &  8.73 $\pm$ 0.04 &                  & 7.60 $\pm$  0.20 \\ 
&NGC~1847 &  LMC   & 7.75 $\pm$ 0.01 &  7.70 $\pm$ 0.02 &  7.29 $\pm$ 0.02 &  6.90 $\pm$ 0.03 & ...\ \ \ \ \ \ \  \\ 
&NGC~1854 &  LMC   & 7.90 $\pm$ 0.02 &  8.05 $\pm$ 0.02 &       ...\ \ \ \ \ \ \ \         &       ...\ \ \ \ \ \ \ \         & ...\ \ \ \ \ \ \  \\
&NGC~1863 &  LMC   & 9.12 $\pm$ 0.03 &       ...\ \ \ \ \ \ \ \         &       ...\ \ \ \ \ \ \ \         &       ...\ \ \ \ \ \ \ \         & ...\ \ \ \ \ \ \  \\
&NGC~1870 &  LMC   & 8.97 $\pm$ 0.03 &  9.15 $\pm$ 0.03 &  8.61 $\pm$ 0.04 &  7.45 $\pm$ 0.04  & ...\ \ \ \ \ \ \    \\ 
&NGC~1928 &  LMC   & 8.13 $\pm$ 0.02 &  8.37 $\pm$ 0.02 &  8.45 $\pm$ 0.04 &       ...\ \ \ \ \ \ \ \          & ...\ \ \ \ \ \ \    \\ 
&NGC~1951 &  LMC   & 8.52 $\pm$ 0.02 &  8.53 $\pm$ 0.02 &  8.09 $\pm$ 0.02 &  8.04 $\pm$ 0.04  & 6.74 $\pm$  0.15   \\ 
&NGC~2118 &  LMC   & 8.87 $\pm$ 0.02 &  9.04 $\pm$ 0.03 &  8.77 $\pm$ 0.04 & 8.00 $\pm$ 0.05   & ...\ \ \ \ \ \ \    \\ 
&NGC~2164 &  LMC   & 9.00 $\pm$ 0.02 &  8.90 $\pm$ 0.03 &  8.70 $\pm$ 0.03   & 8.67 $\pm$ 0.04  & 8.13 $\pm$ 0.30   \\ 
&SL~56 &  LMC   & 10.07 $\pm$ 0.04 &  9.80 $\pm$ 0.04 &  9.38 $\pm$ 0.05   & 9.60 $\pm$ 0.07  & 9.37 $\pm$  0.82   \\ 
&SL~106 &  LMC   & 9.44 $\pm$ 0.03 &  9.46 $\pm$ 0.04 & 9.07 $\pm$ 0.04  &        ...\ \ \ \ \ \ \ \         &  ...\ \ \ \ \ \ \   \\ 
%----------------------swb II up to here--------------------------
SWB III\dotfill& IC~1611 & SMC    & 10.37 $\pm$ 0.04 & 10.51 $\pm$ 0.04 & 10.08 $\pm$ 0.09 & 10.17 $\pm$ 0.16 &  ...\ \ \ \ \ \ \ \\ 
& L~40    & SMC    &  9.55 $\pm$ 0.03 &  9.72 $\pm$ 0.04 &  9.54 $\pm$ 0.05 &  9.82 $\pm$ 0.08 & ...\ \ \ \ \ \ \  \\ 
& L~44    & SMC    &  9.56 $\pm$ 0.02 &  9.59 $\pm$ 0.03 &  9.28 $\pm$ 0.06 &  9.03 $\pm$ 0.05 & ...\ \ \ \ \ \ \  \\ 
& L~63    & SMC    &       ...\ \ \ \ \ \ \ \         &       ...\ \ \ \ \ \ \ \         & 10.84 $\pm$ 0.12 & 11.09 $\pm$ 0.33 & ...\ \ \ \ \ \ \  \\ 
& L~114   & SMC    & 10.77 $\pm$ 0.06 & 10.88 $\pm$ 0.07 & 10.61 $\pm$ 0.09 & 10.80 $\pm$ 0.19 & ...\ \ \ \ \ \ \  \\ 
& NGC~265 & SMC    & 10.23 $\pm$ 0.03 & 10.32 $\pm$ 0.03 &  9.78 $\pm$ 0.07 &  8.89 $\pm$ 0.05 & ...\ \ \ \ \ \ \  \\ 
& NGC~458 & SMC    & 10.69 $\pm$ 0.05 & 10.81 $\pm$ 0.07 & 10.65 $\pm$ 0.09 & 10.53 $\pm$ 0.11 & ...\ \ \ \ \ \ \  \\ 
& NGC~1844& LMC    &  9.94 $\pm$ 0.04 & 10.03 $\pm$ 0.05 &  9.63 $\pm$ 0.04 &  9.22 $\pm$ 0.06 & 7.73 $\pm$ 0.02\\ 
& NGC~1866& LMC    &  7.29 $\pm$ 0.01 &  7.47 $\pm$ 0.01 &  7.12 $\pm$ 0.02 &  7.13 $\pm$ 0.02 & 6.90 $\pm$ 0.14\\ 
& NGC~1895& LMC    &  8.79 $\pm$ 0.02 &  8.50 $\pm$ 0.02 &  6.59 $\pm$ 0.01 &  5.03 $\pm$ 0.01 & 2.33 $\pm$ 0.01\\ 
& NGC~1953& LMC    &  8.91 $\pm$ 0.02 &  8.90 $\pm$ 0.03 &  8.61 $\pm$ 0.05 &       ...\ \ \ \ \ \ \ \         & ...\ \ \ \ \ \ \  \\ 
& NGC~2000& LMC    &  9.43 $\pm$ 0.03 &  9.58 $\pm$ 0.04 &  9.56 $\pm$ 0.05 & 10.25 $\pm$ 0.08 & ...\ \ \ \ \ \ \  \\ 
& NGC~2025& LMC    &  9.15 $\pm$ 0.03 &  9.10 $\pm$ 0.03 &  8.93 $\pm$ 0.04 &  9.13 $\pm$ 0.06 & 8.51 $\pm$ 0.46 \\ 
& NGC~2031& LMC    &  8.15 $\pm$ 0.02 &  8.52 $\pm$ 0.02 &  7.69 $\pm$ 0.02 &  8.38 $\pm$ 0.03 & 7.62 $\pm$ 0.20\\
& NGC~2134& LMC    &  8.40 $\pm$ 0.02 &  8.54 $\pm$ 0.02 &  8.19 $\pm$ 0.03 &  8.26 $\pm$ 0.04 & 7.93 $\pm$ 0.08\\ 
& NGC~2136& LMC    &  8.14 $\pm$ 0.02 &  8.30 $\pm$ 0.02 &  7.94 $\pm$ 0.02 &  7.97 $\pm$ 0.03 & 7.20 $\pm$ 0.13\\
& NGC~2156& LMC    &  9.96 $\pm$ 0.04 & 10.07 $\pm$ 0.05 & 10.03 $\pm$ 0.08 &  9.73 $\pm$ 0.04 & 7.74 $\pm$ 0.18\\ 
& NGC~2157& LMC    &  8.11 $\pm$ 0.02 &  8.24 $\pm$ 0.02 &  7.98 $\pm$ 0.02 &  7.96 $\pm$ 0.03 & 7.66 $\pm$ 0.23\\ 
& NGC~2159& LMC    &  9.56 $\pm$ 0.03 &  9.64 $\pm$ 0.04 &  9.37 $\pm$ 0.04 &  9.56 $\pm$ 0.04 & 9.73 $\pm$ 0.85\\ 
& NGC~2172& LMC    & 10.02 $\pm$ 0.04 & 10.21 $\pm$ 0.05 &  9.78 $\pm$ 0.04 & 10.18 $\pm$ 0.05 & ...\ \ \ \ \ \ \  \\ 
& NGC~l539& LMC    &  7.64 $\pm$ 0.01 &  7.89 $\pm$ 0.02 &  7.56 $\pm$ 0.02 &  7.59 $\pm$ 0.03 & ...\ \ \ \ \ \ \  \\ 
%--------------------------swb III up to here-------------------------
SWB IV\dotfill& L~26&    SMC   & 11.22 $\pm$ 0.09 & 11.37 $\pm$ 0.10 & 11.36 $\pm$ 0.13 & 10.98 $\pm$ 0.15  &  ...\ \ \ \ \ \ \ \\ 
& L~53&    SMC   & 10.14 $\pm$ 0.04 & 10.20 $\pm$ 0.05 &  9.85 $\pm$ 0.06 &  9.66 $\pm$ 0.08  & ...\ \ \ \ \ \ \  \\ 
&NGC~294&  SMC    & 9.96 $\pm$ 0.04 & 10.09 $\pm$ 0.03 &  9.97 $\pm$ 0.08 &  9.99 $\pm$ 0.14  & ...\ \ \ \ \ \ \  \\ 
&NGC~1801& LMC    & 8.32 $\pm$ 0.02 &  8.48 $\pm$ 0.02 &  8.23 $\pm$ 0.03 &  8.13 $\pm$ 0.04  & ...\ \ \ \ \ \ \  \\ 
&NGC~1831& LMC    & 8.27 $\pm$ 0.02 &  8.40 $\pm$ 0.02 &  8.11 $\pm$ 0.02 &  7.92 $\pm$ 0.03  & 8.07  $\pm$   0.21 \\ 
&NGC~1849& LMC    & 8.47 $\pm$ 0.02 &  8.28 $\pm$ 0.02 &  7.83 $\pm$ 0.02 &  7.37 $\pm$ 0.02  & 6.60  $\pm$  0.12\\ 
&NGC~1868& LMC    & 9.17 $\pm$ 0.03 &  9.44 $\pm$ 0.04 &  9.25 $\pm$ 0.04 &  9.06 $\pm$ 0.05  & ...\ \ \ \ \ \ \  \\ 
&NGC~1987& LMC    & 8.43 $\pm$ 0.02 &  8.53 $\pm$ 0.02 &  8.33 $\pm$ 0.03 &  8.41 $\pm$ 0.04  & 9.65  $\pm$  0.61 \\ 
&NGC~2056& LMC    & 9.07 $\pm$ 0.03 &  9.12 $\pm$ 0.03 &  8.73 $\pm$ 0.04 &       ...\ \ \ \ \ \ \ \          &  ...\ \ \ \ \ \ \   \\ 
&NGC~2107& LMC    & 8.61 $\pm$ 0.02 &  8.63 $\pm$ 0.03 &  8.25 $\pm$ 0.03 &  7.88 $\pm$ 0.04  & ...\ \ \ \ \ \ \  \\ 
%-------------------------swb IV up to here-------------------------
%\
%\
%\tablebreak
SWB V\dotfill&NGC~152 & SMC    & 9.41 $\pm$ 0.03 &  9.60 $\pm$ 0.04 &  9.28 $\pm$ 0.04 &  9.05 $\pm$ 0.05 &  ...\ \ \ \ \ \ \ \\ 
&NGC~411 & SMC    & 9.55 $\pm$ 0.03 &  9.65 $\pm$ 0.04 &  9.34 $\pm$ 0.04 &  9.20 $\pm$ 0.05 & ...\ \ \ \ \ \ \  \\ 
&NGC~419 & SMC    & 7.51 $\pm$ 0.01 &  7.51 $\pm$ 0.02 &  6.90 $\pm$ 0.01 &  6.36 $\pm$ 0.01 & ...\ \ \ \ \ \ \  \\ 
&NGC~1651& LMC    & 8.75 $\pm$ 0.02 &  8.93 $\pm$ 0.03 &  8.61 $\pm$ 0.03 &  8.53 $\pm$ 0.05 & 8.21 $\pm$ 0.23\\ 
&NGC~1783& LMC    & 7.49 $\pm$ 0.01 &  7.65 $\pm$ 0.02 &  7.24 $\pm$ 0.02 &  6.99 $\pm$ 0.02 & ...\ \ \ \ \ \ \  \\ 
&NGC~1795& LMC    & 9.27 $\pm$ 0.03 &  9.55 $\pm$ 0.04 &  9.35 $\pm$ 0.05 &  9.13 $\pm$ 0.06 & 8.08 $\pm$ 0.29\\ 
&NGC~1846& LMC    & 6.91 $\pm$ 0.01 &  7.13 $\pm$ 0.01 &  6.84 $\pm$ 0.01 &  6.59 $\pm$ 0.02 & 6.56 $\pm$ 0.09\\ 
&NGC~1917& LMC    & 8.80 $\pm$ 0.02 &       ...\ \ \ \ \ \ \ \         &       ...\ \ \ \ \ \ \ \         &       ...\ \ \ \ \ \ \ \         &  ...\ \ \ \ \ \ \ \\
&NGC~2154& LMC    & 8.15 $\pm$ 0.02 &  8.39 $\pm$ 0.02 &  8.16 $\pm$ 0.03 &  7.85 $\pm$ 0.03 & 7.58 $\pm$ 0.25\\ 
%---------------------swb V up to here----------------------------------
SWB VI\dotfill&NGC~416   & SMC  &  9.29 $\pm$ 0.03 &  9.41 $\pm$ 0.03 &  9.13 $\pm$ 0.04 &  9.18 $\pm$ 0.05 & 8.94 $\pm$ 0.09 \\ 
&NGC~1751  & LMC  &  7.76 $\pm$ 0.01 &  7.96 $\pm$ 0.02 &  7.79 $\pm$ 0.02 &  7.55 $\pm$ 0.03 & ...\ \ \ \ \ \ \   \\ 
&NGC~1754  & LMC  &  7.93 $\pm$ 0.01 &  7.98 $\pm$ 0.02 &  7.65 $\pm$ 0.02 &  7.53 $\pm$ 0.03 & 6.90 $\pm$ 0.11 \\ 
&NGC~1852  & LMC  &  8.09 $\pm$ 0.02 &  8.18 $\pm$ 0.02 &  7.76 $\pm$ 0.02 &  6.96 $\pm$ 0.02 & 3.29 $\pm$ 0.02 \\ 
&NGC~1916  & LMC  &  6.86 $\pm$ 0.01 &  6.96 $\pm$ 0.01 &  5.80 $\pm$ 0.01 &  4.44 $\pm$ 0.01 & ...\ \ \ \ \ \ \   \\ 
&NGC~1978  & LMC  &  6.79 $\pm$ 0.01 &  6.72 $\pm$ 0.01 &  6.14 $\pm$ 0.01 &  5.72 $\pm$ 0.01 &  ...\ \ \ \ \ \ \  \\ 
&NGC~2005  & LMC  &  8.42 $\pm$ 0.02 &  8.63 $\pm$ 0.03 &  8.26 $\pm$ 0.03 &  8.29 $\pm$ 0.05 & ...\ \ \ \ \ \ \   \\ 
&NGC~2019  & LMC  &  7.94 $\pm$ 0.02 &  8.14 $\pm$ 0.02 &  7.90 $\pm$ 0.02 &  7.84 $\pm$ 0.05 & ...\ \ \ \ \ \ \   \\ 
&NGC~2121  & LMC  &  8.05 $\pm$ 0.02 &  8.08 $\pm$ 0.02 &  7.70 $\pm$ 0.02 &  7.51 $\pm$ 0.03 & 7.31 $\pm$ 0.19 \\ 
&SL~506    & LMC  & 10.56 $\pm$ 0.05 & 10.68 $\pm$ 0.07 & 10.42 $\pm$ 0.09 & 10.22 $\pm$ 0.04 & 9.16 $\pm$ 0.41 \\ 
%---------------------swb VI up to here----------------------------------
SWB VII\dotfill&L~8     &  SMC  &      ...\ \ \ \ \ \ \ \          &       ...\ \ \ \ \ \ \ \          &        ...\ \ \ \ \ \ \ \         &   9.60 $\pm$ 0.05 &  ...\ \ \ \ \ \ \       \\ 
&L~11    &  SMC  & 10.66 $\pm$ 0.05 &  10.81 $\pm$ 0.06 &  10.75 $\pm$ 0.09 &  10.26 $\pm$ 0.10 &  ...\ \ \ \ \ \ \       \\ 
&L~68    &  SMC  &      ...\ \ \ \ \ \ \ \          &        ...\ \ \ \ \ \ \ \         &  10.55 $\pm$ 0.07 &  10.66 $\pm$ 0.21 &   ...\ \ \ \ \ \ \      \\ 
&L~113   &  SMC  & 10.37 $\pm$ 0.03 &  10.51 $\pm$ 0.05 &  10.22 $\pm$ 0.05 &   9.97 $\pm$ 0.08 &   ...\ \ \ \ \ \ \      \\ 
&NGC~339 &  SMC  & 10.40 $\pm$ 0.06 &  10.44 $\pm$ 0.06 &  10.26 $\pm$ 0.06 &  10.19 $\pm$ 0.10 &   ...\ \ \ \ \ \ \      \\ 
&NGC~361 &  SMC  &  9.85 $\pm$ 0.04 &  10.00 $\pm$ 0.05 &   9.73 $\pm$ 0.05 &   9.71 $\pm$ 0.08 &   ...\ \ \ \ \ \ \      \\ 
&NGC~1786&  LMC  &  7.74 $\pm$ 0.01 &   7.76 $\pm$ 0.02 &   7.48 $\pm$ 0.02 &   7.32 $\pm$ 0.02 & 6.75 $\pm$ 0.14  \\ 
&NGC~1835&  LMC  &  7.33 $\pm$ 0.01 &   7.38 $\pm$ 0.01 &   7.14 $\pm$ 0.02 &   7.14 $\pm$ 0.02 & 6.60 $\pm$ 0.13  \\ 
%-------------------  swb VII up to here---------------------------------
\enddata
\tablecomments{(This table is available in machine-readable form.)
}
\label{tabclust}
\end{deluxetable}